\RequirePackage{lineno}
\documentclass[aps,prb,10pt,twocolumn,superscriptaddress,longbibliography,nofootinbib]{revtex4-2}

\usepackage{graphicx}
\usepackage{amssymb,amsmath,amsfonts,gensymb}
\usepackage{bm,hyperref}
\usepackage{color}
\usepackage{ulem}
\usepackage{bbm}
\usepackage{subfigure}
\usepackage{dcolumn}
\usepackage{mathtools}
\usepackage{pifont}
\usepackage[symbol]{footmisc}
\usepackage{mhchem}

\newcommand{\ket}[1]{\vert #1 \rangle}

\newcommand{\eV}{{\text{eV}}}
\newcommand{\eVA}{{\text{eV}\cdot \text{\AA}}}
\newcommand{\bk}{\mathbf{k}}
\newcommand{\bq}{\mathbf{q}}

\newcommand{\bd}{\mathbf{d}}
\newcommand{\btk}{\widetilde{\mathbf{k}}}
\newcommand{\btq}{\widetilde{\mathbf{q}}}
\newcommand{\br}{\mathbf{r}}

\def\I{\uppercase\expandafter{\romannumeral 1}}
\def\II{\uppercase\expandafter{\romannumeral 2}}
\def\III{{\uppercase\expandafter{\romannumeral 3}}}
\def\IV{{\uppercase\expandafter{\romannumeral 4}}}
\def\V{{\uppercase\expandafter{\romannumeral 5}}}
\def\VI{{\uppercase\expandafter{\romannumeral 6}}}
\def\VII{{\uppercase\expandafter{\romannumeral 7}}}

\def\a{\mathbf{a}}
\def\b{\mathbf{b}}
\def\p{\mathbf{p}}

\def\k{\mathbf{k}}
\def\vr{\mathbf{r}}
\def\G{\mathbf{G}}
\def\Q{\mathbf{Q}}
\def\br{\mathbf{r}}
\def\bd{\mathbf{d}}
\def\kt{\widetilde{\mathbf{k}}}
\def\q{\mathbf{q}}

\def\qt{\widetilde{\mathbf{q}}}
\def\nn{\nonumber \\}
\def\hc{\hat{c}}

\def\hd{\hat{d}}

\def\R{\mathbf{R}}

\makeatletter
\def\@ssect@ltx#1#2#3#4#5#6[#7]#8{%
	\def\H@svsec{\phantomsection}%
	\@tempskipa #5\relax
	\@ifdim{\@tempskipa>\z@}{%
		\begingroup
		\interlinepenalty \@M
		#6{%
			\@ifundefined{@hangfroms@#1}{\@hang@froms}{\csname @hangfroms@#1\endcsname}%
			{\hskip#3\relax\H@svsec}{#8}%
		}%
		\@@par
		\endgroup
		\@ifundefined{#1smark}{\@gobble}{\csname #1smark\endcsname}{#7}%
	}{%
		\def\@svsechd{%
			#6{%
				\@ifundefined{@runin@tos@#1}{\@runin@tos}{\csname @runin@tos@#1\endcsname}%
				{\hskip#3\relax\H@svsec}{#8}%
			}%
			\@ifundefined{#1smark}{\@gobble}{\csname #1smark\endcsname}{#7}%
			\addcontentsline{toc}{#1}{\protect\numberline{}#8}%
		}%
	}%
	\@xsect{#5}%
}%
\makeatother

\begin{document}

	
\title{Fractional topological states in rhombohedral multilayer graphene modulated by kagome superlattice}

\author{Yanran Shi}
\affiliation{School of Physical Science and Technology, ShanghaiTech Laboratory for Topological Physics, ShanghaiTech University, Shanghai 201210, China}

\author{Bo Xie}
\affiliation{School of Physical Science and Technology, ShanghaiTech Laboratory for Topological Physics, ShanghaiTech University, Shanghai 201210, China}

\author{Fengfan Ren}
\affiliation{School of Physical Science and Technology, ShanghaiTech Laboratory for Topological Physics, ShanghaiTech University, Shanghai 201210, China}

\author{Xinyu Cai}
\affiliation{School of Physical Science and Technology, ShanghaiTech Laboratory for Topological Physics, ShanghaiTech University, Shanghai 201210, China}

\author{Zhongqing Guo}
\affiliation{School of Physical Science and Technology, ShanghaiTech Laboratory for Topological Physics, ShanghaiTech University, Shanghai 201210, China}

\author{Qiao Li}
\affiliation{School of Physical Science and Technology, ShanghaiTech Laboratory for Topological Physics, ShanghaiTech University, Shanghai 201210, China}

\author{Xin Lu}
\affiliation{School of Physical Science and Technology, ShanghaiTech Laboratory for Topological Physics, ShanghaiTech University, Shanghai 201210, China}

\author{Nicolas Regnault}
\affiliation{Center for Computational Quantum Physics, Flatiron Institute, 162 5th Avenue, New York, NY 10010, USA}
\affiliation{Department of Physics, Princeton University, Princeton, New Jersey 08544, USA}
\affiliation{Laboratoire de Physique de l'Ecole normale sup\'erieure, ENS, Universit\'e PSL, CNRS, Sorbonne Universit\'e}

\author{Zhongkai Liu}
\affiliation{School of Physical Science and Technology, ShanghaiTech Laboratory for Topological Physics, ShanghaiTech University, Shanghai 201210, China}

\author{Jianpeng Liu}
\email{liujp@shanghaitech.edu.cn}
\affiliation{School of Physical Science and Technology, ShanghaiTech Laboratory for Topological Physics, ShanghaiTech University, Shanghai 201210, China}
\affiliation{Liaoning Academy of Materials, Shenyang 110167, China}
	
\bibliographystyle{apsrev4-2}

\begin{abstract} 

Fractional quantum anomalous Hall effects realized in twisted bilayer MoTe$_2$ and multilayer-graphene-based  moir\'e heterostructures have captured a tremendous growth of interest. In this work, we propose that rhombohedral multilayer  graphene coupled with an artificial kagome superlattice potential is a new platform to realize various fractional topological phases. Taking Bernal bilayer graphene as the simplest example, when it is placed on top of a prepatterned SiO$_2$ substrate with periodic arrays of holes arranged into kagome lattice, the system would be subject to a tunable  kagome superlattice potential once an electrostatic voltage drop between the top and bottom gates is applied. Then, we theoretically study the electronic band structures, topological properties, and quantum geometric properties of the Bloch states of Bernal bilayer graphene coupled with a realistic kagome superlattice potential, which is well benchmarked by transport measurements in the weak superlattice-potential regime. We find that the system may exhibit nearly ideal topological flat bands in a substantial region of the parameter space spanned by superlattice constant and electrostatic potential strength. When these topological flat bands are fractionally filled, exact diagonalization calculations suggest that the system would exhibit rich fractional topological phases at 1/3, 2/3, 2/5, 3/5 and 1/2 fillings  including both fractional Chern insulators and  anomalous composite Fermi liquids under zero magnetic field.  
\end{abstract}

\maketitle

Recent experimental observations of fractional quantum anomalous Hall effects in both twisted MoTe$_2$\cite{fqah-nature23,fqah-prx23,fqah-optics-xu-nature23,fqah-mak-nature23} and multilayer graphene moir\'e heterostructures \cite{fqah-ju-nature24,lu-hexlayer-arxiv24,ju-eqah-nature25} have aroused significant research interest. The fractional quantum anomalous Hall effects characterized by fractionally quantized Hall plateaus under zero magnetic field originate from a type of intriguing many-body topological states known as fractional Chern insulator (FCI) states \cite{fci-prx11,sheng-fci-nc11,murdy-fci-prl11,wen-kagome-prl11,sarma-flatchern-prl11,cooper-fci-prl09,qi-fqah-prl11}. Such FCI states are the zero-field analogue of fractional quantum Hall states emerging from fractionally filled Landau levels \cite{fqhe-prl82,laughlin-prl83,jain-prl89,moore-read,fqhe-rmp99,qhe-book-2012}.  Different types of FCI states have been  extensively studied in  lattice models \cite{fqh-boson-prl11,fci-zoo-prb12,liuzhao-fci-prl12,vanderbos-fci-prl12,liu-fci-prb13,fiete-ruby-prb11} as well as in realistic moir\'e superlattice systems \cite{liangfu-fqah-tmd-prb23,xiao-fqah-arxiv23,zhangyang-fqah-tmd-arxiv23,regnault-fci-mote2-prb24,senthil-fci-prl24,zhangyh-fci-prl24,ashvin-fci-prl24,guo-prb24,bernevig-fci-iii-arxiv23,bernevig-fci-iv-arxiv24,lixiao-fci-prb24}.

Moir\'e superlattices have been an excellent platform to realize FCI states because of the presence of isolated topological flat bands with desirable quantum geometric properties in these systems \cite{ashvin-fci-tbg-prr20,wang-geometry-prl21,eslam-tmg-prl22,wang-tmg-prl22}. For example, Landau-level-like flat-band wavefunctions have been proposed to exist in both magic-angle twisted bilayer graphene \cite{origin-magic-angle-prl19,jpliu-prb19,ashvin-fci-tbg-prr20,zaletel-tbg-2019} and twisted MoTe$_2$ \cite{macdonald-tmd-ll-prl24,wu-mote2-arxiv24,lin-gauge-prb24}. 
However,  the appearance of flat band in moir\'e superlattice also requires a precise control of twist angle, which is still challenging using state-of-art techniques due to unavoidable twist angle inhomogeneities \cite{zeldov-disorder-np20} and uncontrolled lattice relaxations \cite{kazmierczak2021strain}.
 It is thus highly needed to find alternative material platforms which also host desirable topological flat bands, yet suffer less from unwanted inhomogeneities and disorder effects.  Recently, a new method of fabricating high-mobility superlattice device has been proposed, which is to integrate prepatterned dielectric substrate consisting of periodic arrays of holes with a van der Waals 2D material such as graphene \cite{dean-nn18}. Upon the application of electrostatic gate voltages, a superlattice potential  modulated by the patterned  dielectric substrate, with the period of tens of nanometers, would be exerted to the 2D material \cite{dean-nn18, chen-cp20, dean-nn21, ruiz-nc22,zeng-prl24, du-nanoletter24}. This folds the electronic band structures to the mini Brillouin zone of the superlattice, generating subbands. If the parent Bloch states from the 2D materials already exhibit non-vanishing Berry curvatures,  the subbands are likely to possess  nontrivial topological properties. Indeed, some of the previous works already suggest the presence of topological flat bands in both graphene systems \cite{cano-bilayer-prl23,lu-nc23, cano-multilayer-prb23,vanderbilt-bilayer-prb24,zhan-patterned-prb25}  and topological semimetals \cite{trithep-semiconductor-prl24,dai-artificial-arxiv24} coupled with electrostatic and even magnetostatic superlattice potentials. 

\begin{figure*}[hbtp!]
    \includegraphics[width=7in]{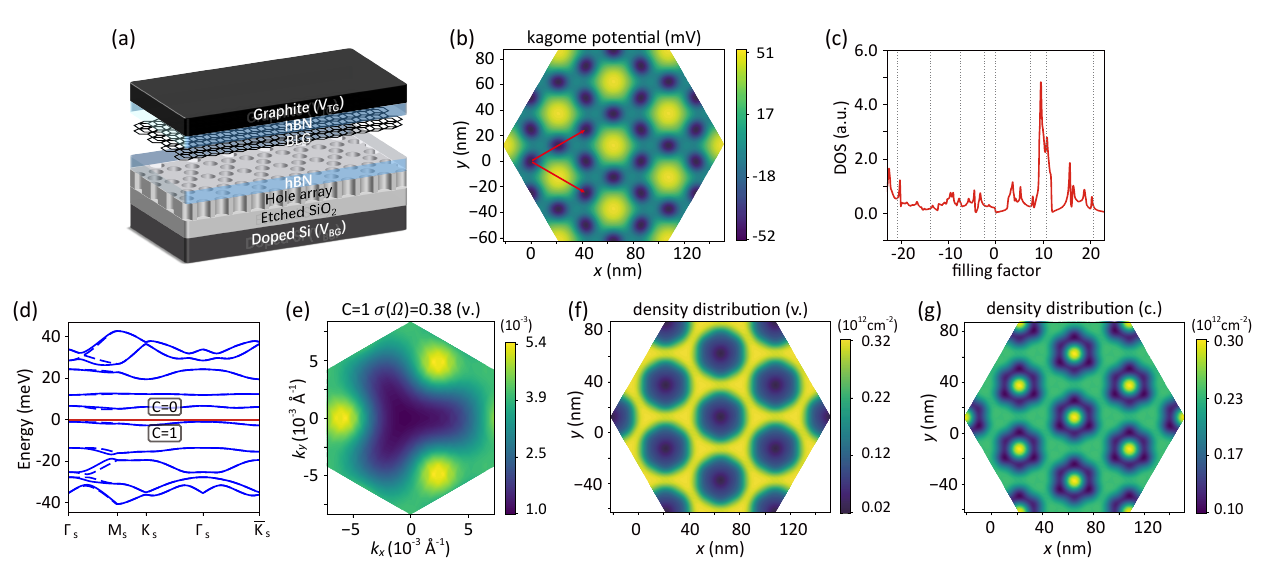}
\caption{Device set up, kagome superlattice potential distribution and single particle properties of BLG: (a) The device set up; (b) kagome superlattice potential distribution in real space; (c) the calculated density of states; (d) energy band structure; (e) Berry curvature distribution of highest valence band; (f) local charge density distribution in real space for highest valence band;  and (g) lowest conduction band.  
(b), (d), (e), (f), (g) are for the case of $L_s=50$\,nm, $\delta V=-20$\,V. (c) is the for case of $L_s=120$\,nm, $\delta V=3.5$\,V. The solid and dashed blue lines in (d) denote bands from $K$ and $K'$ valleys, respectively.}
\label{fig1}
\end{figure*}

Nevertheless, most of the previous studies have been focused on patterned triangular superlattice systems \cite{dean-nn18, du-nanoletter24,cano-bilayer-prl23,vanderbilt-bilayer-prb24,cano-multilayer-prb23,trithep-semiconductor-prl24,dai-artificial-arxiv24}. It is well known that flat band would necessarily emerge in a tight-binding model of kagome lattice due to destructive quantum interference effects \cite{balents-prb08}. It is then natural to ask what would happen if a patterned kagome superlattice potential is coupled with 2D materials.  It has been shown that a kagome superlattice potential would strongly modify the linear Dirac  dispersions of monolayer graphene, generating multiple Dirac bands \cite{zeng-prl24,zhan-patterned-prb25}. In this work, we consider integrating a patterned kagome superlattice with rhombohedral multilayer graphene (RMG).  We  study the electronic band structures, topological properties and quantum geometric properties of the low-energy Bloch states of RMG coupled with a realistic kagome superlattice potential. We take Bernal bilayer graphene (BLG) as the simplest example. 
Using on a realistic modelling of the system that is benchmarked by transport measurements, we find that the system may exhibit nearly ideal topological flat bands in a substantial region of the parameter space spanned by superlattice constant $L_s$ and  potential drop $\delta V$ between the bottom and top gate voltages. When these flat bands are fractionally filled, the system would exhibit rich fractional topological phases at 1/3, 2/3, 2/5, 3/5 and 1/2 fillings,  including both fractional Chern insulators and  composite Fermi liquids under zero magnetic field.


We first consider the experimental setup that  BLG encapsulated by hexagonal boron nitride is  placed on top of spatially periodic prepatterned SiO$_2$ substrate, as schematically shown in Fig.~\ref{fig1}(a). The artificially designed pattern with kagome-type lattice can be realized in practice by etching holes in the SiO$_2$ substrate \cite{supp_info}.  The tunable top gate ($V_{\text{tg}}$) and bottom gate potential ($V_{\text{bg}}$) synergistically controls the charge density and superlattice potential strength of BLG. We use the commercial software package MATLAB (Partial Differential Equation toolbox) to numerically simulate the spatial profile of superlattice potential of such patterned dielectric system with fixed top and bottom electrostatic voltages \cite{supp_info}. 
Specifically, the spatially periodic kagome lattice potential (denoted by $V_{\rm{SL}}(\mathbf{r})$) obtained from realistic electrostatic simulations can be represented by
\begin{align}
    V_{\rm{SL}}(\mathbf{r})&=\sum_\mathbf{Q} U(\mathbf{Q})e^{-i\mathbf{Q} \cdot \mathbf{r}}
\end{align}
where the Fourier component of the potential 
\begin{align}
    U(\mathbf{Q})=\frac{1}{N_s} \sum_{\mathbf{R},\alpha} V_{\rm{SL}} (\mathbf{R}+\bm{\tau}_{\alpha}) 
      e^{i\Q \cdot (\mathbf{R}+\bm{\tau}_{\alpha})}
\end{align}
where $\mathbf{R}$ are superlattice vectors and $\{\bm{\tau}_{\alpha}, \alpha=A, B, C\}$ are three sublattices of  kagome lattice and $\mathbf{Q}$ denotes reciprocal vector. To accurately describe the kagome superlattice potential, we include up to the fifth order of the Fourier components $U(\mathbf{Q})$.  
In Fig.~\ref{fig1}(b) we show the simulated superlattice potential distribution in real space for the case of $L_s=50$\,nm and $\delta V=-20$\,V, where etched holes on the substrate correspond to potential minima. 
 
Then, we consider the situation that low-energy electrons in rhombohedral multilayer graphene are coupled with the kagome superlattice potential, which can be properly described by the following non-interacting single-particle Hamiltonian 
\begin{equation}
    H^{0,\mu}=H^{0,\mu}_{n}+V_{\rm{SL}}\;.
\end{equation}
Here, $H^{0,\mu}_n$ is the non-interacting Hamiltonian of $n$-layer rhombohedral graphene of valley  index $\mu$, with $\mu=\mp$ denoting $K/K'$ valley of graphene's Brillouin zone, which is  derived from a widely used atomistic tight-biding model of graphene reported in Ref.~\cite{moon-tbg-prb13}. More details on the model Hamiltonian $H^{0,\mu}_n$ including the interaction renormalizations of the low-energy parameters of the model, are given in Supplementary Information \cite{supp_info}.

 To benchmark the accuracy of our theoretical modelling, we have experimentally synthesized patterned SiO$_2$ substrate with arrays of etched holes arranged into kagome lattice  with the lattice constant $L_s=120$\,nm, then place hBN-encapsulated BLG on top of it. Both bottom and top gate voltages are applied to tune the carrier density and superlattice potential strength, as  schematically shown in Fig.~\ref{fig1}(a). We have  measured resistance of such system as a function of carrier density (filling factor) under a weak $\delta V$ (see Supplementary Figure 1 \cite{supp_info}). There are a number of resistance peaks at relatively large filling factors as marked by dashed black lines in Fig.~\ref{fig1}(c), which  precisely correspond to the dips in the calculated  single-particle density of state as shown by the red solid line in Fig.~\ref{fig1}(c). This indicates that the above theoretical model can accurately capture the single-particle physics at large filling factors, which justifies the further exploration of  potential correlated topological physics  at lower fillings.

\begin{figure*}[bth!]
    \includegraphics[width=6in]{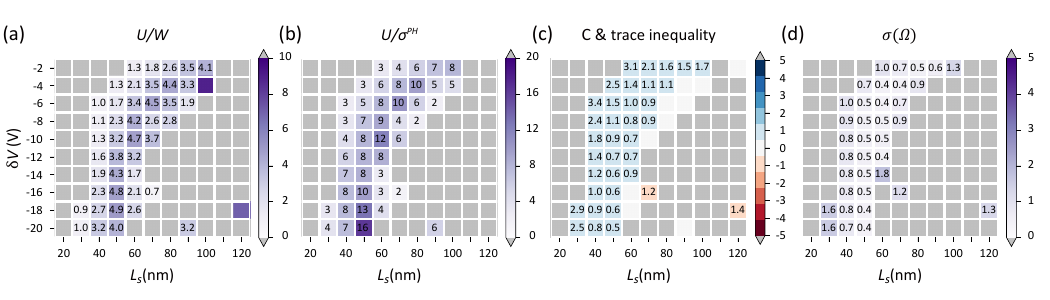}
\caption{Single particle phase diagrams of the HVB: (a) $U/W$, where $U=e^2/(4\pi\epsilon\epsilon_0 L_s)$ and $W$ is bandwidth, (b) $U/\sigma_{\rm{PH}}$ where $\sigma_{\rm{PH}}$ characterizes the PH symmetry violation of HVB \cite{supp_info}, $U/\sigma_{\rm{PH}} \rightarrow \infty$ if the PH symmetry is exact, (c) Chern number, and (d) normalized berry curvature standard deviation $\sigma(\Omega)$, in parameter space $ \left ( L_s, \delta V \right )$. The black numbers in (a), (b), (c) and (d) represent the values of $U/W$, $U/\sigma_{\rm{PH}}$, trace inequality and  $\sigma(\Omega)$, respectively. }
\label{fig2}
\end{figure*}

Low-energy flat bands with nontrivial topological properties can emerge in such BLG coupled with kagome superlattice potential. For example, when $L_s=50$\,nm and $\delta V=-20$\,V, the highest valence band (HVB) of the system has a Chern number $C=1$ while the lowest conduction band (LCB) has zero Chern number. Both of them are flat, with the dispersions $\lessapprox 5$\,meV, and are well separated from other bands as shown in Fig.~\ref{fig1}(d). In Fig.~\ref{fig1}(f) and (g) we further present the local charge density distributions of the HVB and LCB, respectively. Interestingly, electrons in the HVB with $C=1$ occupy the interstial regions between etched kagome-patterned holes, forming an emergent honeycomb lattice; while electrons in the LCB tend to occupy the central region surrounded by the kagome-patterned holes, forming an emergent triangular lattice. Most saliently, the topological HVB has a quite uniform Berry-curvature distribution in Brillouin zone as shown in Fig.~\ref{fig1}(e), with a normalized standard deviation as small as $\sigma(\Omega)=0.38$.
This demonstrates that isolated topological flat bands with desirable quantum geometric properties are likely to emerge in such system, which may favor fractional topological states at fractional fillings. Hence, we continue to explore the band structure characteristics, topological properties and quantum geometric properties of the low-energy states in the system in the parameter space of  $L_s$ and  $\delta V$.

Fig.~\ref{fig2} displays the ratio between characteristic $e$-$e$ interaction $U$ energy and bandwidth $W$ (Fig.~\ref{fig2}(a), denoted by $\eta=U/W$), Chern number (Fig.~\ref{fig2}(c)), Berry-curvature standard deviation (Fig.~\ref{fig2}(d)) and the ratio between $U$ and particle-hole (PH) symmetry violation $\sigma_{\rm{PH}}$ \cite{supp_info} (Fig.~\ref{fig2}(b)) of the HVB in the parameter space of ($L_s$, $\delta V$). When  $Ls$ is between $40\,nm$ and $100\,nm$ and $\delta V$ is between $-20\,V$ and $-2\,V$, there is a substantial region in which the HVB behaves as a nearly ``ideal" topological flat band, in the sense that its bandwidth is small  compared to $e$-$e$ interactions, having Chern number $C=1$ with a small Berry-curvature standard deviation ($\lessapprox 1$).  In the meanwhile the difference between the integral of trace of Fubini-Study metric of the Bloch states and  Chern number $|C|$ is small (see numbers marked in Fig.~\ref{fig2}(c)),
in close analogy with lowest-Landau-level (LLL) wavefunctions \cite{roy-prb14,claassen-prl15,wangjie-prl21}. Moreover, the HVB is nearly particle-hole symmetric as characterized by large $U/\sigma_{\rm{PH}}$ \cite{supp_info} shown in Fig.~\ref{fig2}(b). In comparison, the LLL possesses exact particle-hole symmetry with $U/\sigma_{\rm{PH}}\to \infty$, and $U/\sigma_{\rm{PH}}\approx 5$ for the HVB of $3.89^{\circ}$ twisted MoTe$_2$.
The ubiquitous topological flat bands in this system motivates us to further study the interacting ground states at fractional fillings, exploring possible fractional topological states.


We further consider $e$-$e$ interaction Hamiltonian projected to either HVB or LCB wavefunction from one valley and one spin, expressed as
\begin{align}
    &\hat{H}_{\rm{int}} = \nonumber \\ 
    &\frac{1}{2N_S}
    \sum_{\k,\k',\q}
    \sum_{ ll^{\prime}}
    \sum_\Q{
    V_{ll^{\prime}}
    (\Q+\q) 
    \Omega^{l,l^{\prime}}
    (\k,\k^{\prime},\q, \Q)} \nonumber\\
    &\times \hat{c}_{\k+\q}^{\dagger}
    \hat{c}_{\k^{\prime}-\q}^{\dagger}
    \hat{c}_{\k^{\prime}}
    \hat{c}_{\k}
\end{align}
where $\Omega^{l,l^{\prime}}(\k,\k^{\prime},\q, \Q)$ is the form factor \cite{supp_info},  $\k$, $\k^{\prime}$, $\q$ are the wave vectors within the mini Brillouin zone, and $\Q$ denotes reciprocal vector. $N_{S}$ is the number of primitive cells in the system, and $l$ and $\alpha$ are layer and sublattice indices, respectively. 
A layer-dependent,  screened Coulomb interaction $V_{ll'}(\q)$  with inverse screening length $\kappa = 0.0025\,\text{\AA}^{-1}$ is adopted with its detailed form given in Supplemental Materials \cite{supp_info}.

\begin{figure}[bth!]
    \includegraphics[width=3.5in]{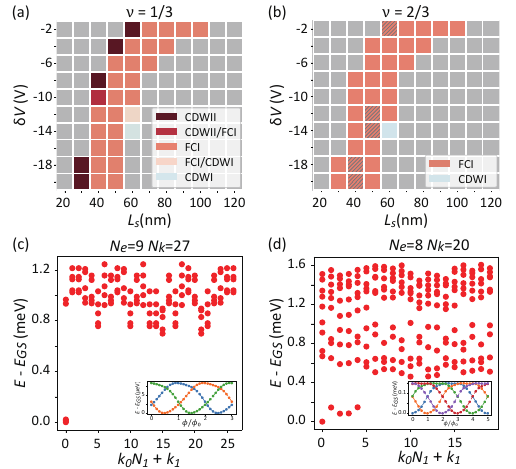}
\caption{Phase diagrams at (a) 1/3  filling, and (b) 2/3 filling. Energy spectra and spectral-flow characteristics at (c) 1/3 filling and (d)  2/5 filling, for the case of $L_s=50$\,nm, $\delta V=-20$\,V.}
\label{fig3}
\end{figure}

Assuming a full spin-valley polarization, which is well justified in experiments at low carrier densities ($\lessapprox 10^{11}\,\textrm{cm}^{-2}$) in multilayer graphene \cite{ju-chern-natnano2023,zhou-trilayer-nature21},  we continue to explore the  interacting  ground states at  filling factors $1/3$ and $2/3$ of the HVB (from one valley/spin) in the  parameter space of ($L_s,\delta V)$. The phase diagrams at 1/3 and 2/3 fillings of the HVB with Chern-number 1 are presented in Fig.~\ref{fig3}(a) and Fig.~\ref{fig3}(b), respectively. We find that the system stays in the FCI phase at 1/3 and 2/3 fillings in a substantial region of the parameter space \cite{comment-fci}. Taking the case of $L_s=50$\,nm and $\delta V=-20$\,V as an example, the  energy spectrum is characterized by a three-fold quasi-degenerate ground state manifold with a sizable gap $\sim 0.8$\,meV separated from excited states, as shown in Fig.~\ref{fig3}(c). The  three quasi-degenerate ground states all emerge at zero crystalline momentum sector for the case of 27 lattice sites, consistent with  the ``generalized Pauli principle" \cite{fci-prx11}. Furthermore,  as clearly shown in the inset of Fig.~\ref{fig3}(a), the adiabatic insertion of a magnetic flux $\phi$ causes the three quasi-degenerate ground states of the FCI at 1/3 filling to cyclically transition into one another, eventually returning to their original configuration when $\phi=3\phi_0$ ($\phi_0=h/e$ is the flux quantum). This process exhibits a characteristic spectral-flow behavior. There are also other correlated states such as two types of charge-density-wave (CDW\I\ and CDW\II) and the crossover states between CDWs and FCI \cite{supp_info}. The system also exhibits the Jain-sequence FCI states at 2/5 and 3/5 fillings of the HVB. As shown in Fig.~\ref{fig3}(d), at 2/5 filling with 20 lattice sites for $L_s=500\,\text{\AA}$ and $\delta V=-20$\,V, the energy spectrum consists of a five-fold quasi-degenerate ground-state manifold separated by a gap $\sim 0.5$\,meV from the excited states, which also exhibit the spectral-flow behavior as demonstrated in the inset of Fig.~\ref{fig3}(d). Similar result is obtained at 3/5 filling \cite{supp_info} thanks to the emergent PH symmetry.


\begin{figure*}[bth!]
    \includegraphics[width=7in]{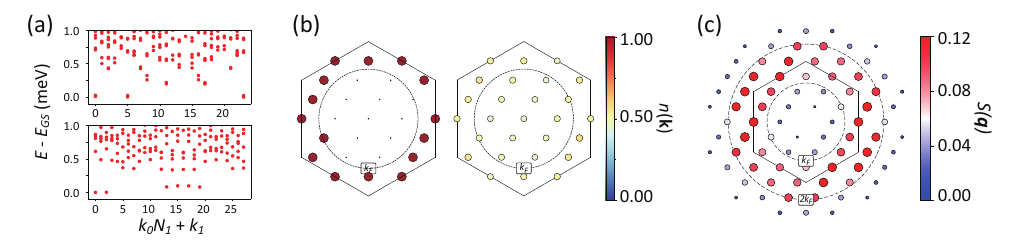}
\caption{CFL state at half filling of HVB for the case of $L_s=50$\,nm, $\delta V=-20$\,V: (a) energy spectrum of 24 sites (upper panel) and 28 sites (lower panel), (b) occupation number in $\k$ space (right panel) compared to that of FL state (left panel), and (c) structure factor. 
}
\label{fig4}
\end{figure*}
The FCI states discussed above are incompressible fractional topological states, which may be interpreted as the lattice analogue of FQH states realized under zero magnetic field \cite{qi-fqah-prl11,bernevig-fci-prb12,bernevig-fci-prl13,liu-fci-prb13}. At 1/2 filling of the lowest Landau level, the ground state becomes compressible as characterized by finite longitudinal conductivity and absence of quantized Hall plateau, which corresponds to a Fermi-liquid-like state consisting of composite fermions feeling vanishing effective magnetic field \cite{Jain-prl1989, Lopez-prb1991, Halperin-Patrick-Nicholas-prb1993, Son-prx2015}. Recently, the composite Fermi liquid (CFL) state has been generalized to the case of half-filled flat Chern band in twisted MoTe$_2$  under zero magnetic field \cite{ashvin-cfl-prl23,fu-cfl-prl23}. Here, we numerically achieved such peculiar compressible state when the HVB of BLG coupled with kagome superlattice potential is half filled. 
For $L_s=500\,\text{\AA}$ and $\delta V=-20$\,V with system size of 24, our ED calculations suggest at half filling of HVB there are  six-fold quasi-degenerate ground states located  as shown in the upper panel of Fig.~\ref{fig4}(a).  They  correspond to the three center-of-mass  momenta of compact Fermi sea configurations, consistent with those of half-filled LLL \cite{ashvin-cfl-prl23}. Increasing system size to 28, the ground-state energy spectrum at 1/2 filling is still consistent with that of half-filled LLL as shown in the lower panel of Fig.~\ref{fig4}(b) \cite{ashvin-cfl-prl23}, further confirming its CFL nature. Furthermore, the momentum-space electronic occupation number $n(\k)$ of the CFL state is relatively homogeneous with $\vert n(\tilde{\k})-1/2 \vert \lessapprox 0.15$, as shown in the right panel of Fig.~\ref{fig4}(b); in contrast, $n(\k)$ exhibits a sharp jump at Fermi wavevector $k_F=\sqrt{2\pi/\Omega_s}$ ($\Omega_s$ is the area of the primitive cell) in the regular Fermi-liquid state characterizing the presence of an electronic Fermi surface, as shown in left panel of Fig.~\ref{fig4}(b).
The band-projected structure factor $S(\q)$ of the zero-field CFL state shows notable peak around a circle in momentum space, and almost vanishes when $\vert\q\vert>2 k_F$ as shown in Fig.~\ref{fig4}(c). This implies the presence of a Fermi surface consisting of composite fermions with Fermi wavevector $k_F$ \cite{ashvin-cfl-prl23}. Combining all of these features, we conclude that the ground state at 1/2 filling of HVB is a CFL. 
The detailed phase diagram at 1/2 filling is given in Supplemental Materials \cite{supp_info}.


To summarize, we have theoretically studied the single-particle band structures, topological properties and quantum geometric properties of BLG coupled with a kagome superlattice potential. The single-particle properties of the system calculated using a realistic continuum model shows excellent agreement with  transport measurements in the weak-potential, large-filling-factor regime, showing the reliability of our  modelling. We further explore the interacting ground states at fractional fillings of the sub-bands, and find a variety of fractional topological states including both FCIs and CFLs. Besides BLG, we have studied the single-particle properties of  trilayer and tetralayer rhombohedral graphene coupled with kagome superlattice potential, which also exhibit well isolated topological flat bands with tunable Chern numbers \cite{supp_info}. Thus, we conclude that RMG coupled with kagome superlattice potential may be an ideal platform to explore zero-magnetic-field fractional topological phases.

\acknowledgements
J. L. thanks Zhao Liu for valuable discussions. This work is supported by the National Key Research and Development Program of China (grant no. 2024YFA1410400, no. 2020YFA0309601 and no. 2022YFA1604400/03), the National Natural Science Foundation of China (grant no. 12174257, no. 92365204 and no. 12274298) and Shanghai Science and Technology Innovation Action Plan (grant no. 24LZ1401100). The Flatiron Institute is a division of the Simons Foundation.

\widetext
\clearpage

\makeatletter
\def\@fnsymbol#1{\ensuremath{\ifcase#1\or \dagger\or \ddagger\or
		\mathsection\or \mathparagraph\or \|\or **\or \dagger\dagger
		\or \ddagger\ddagger \else\@ctrerr\fi}}
\makeatother

\begin{center}
\textbf{\large Supplemental Materials for “Fractional topological states in rhombohedral multilayer graphene modulated by kagome superlattice"} \\
\end{center}

\setcounter{equation}{0}
\setcounter{figure}{0}
\setcounter{table}{0}
\setcounter{section}{0}
\makeatletter
\renewcommand{\theequation}{S\arabic{equation}}
\renewcommand{\thesection}{S\arabic{section}}
\renewcommand{\figurename}{Supplementary Figure}
\renewcommand{\tablename}{Supplementary Table}

\def\bibsection{\section*{References}} 

\section{Device fabrication and transport data}
\label{sec:Experimental Result}
\begin{figure}[bth!]
    \includegraphics[width=3.5in]{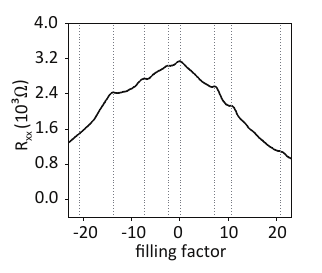}
\caption{The experimentally obtained longitudinal resistance $R_{\rm{xx}}$ as a function of filling factor.
}
\label{SIA}
\end{figure}

We experimentally implement the superlattice potential using a dielectric layer patterning technique. To achieve a high-quality periodic potential field, we employ Electron Beam Lithography (EBL) to define a kagome array with hole diameters of 40\,nm and a periodicity of 120\,nm on a \ce{SiO$_2$}/\ce{Si} substrate. The holes are then plasma-etched using a Reactive Ion Etching (RIE) system with \ce{CF4} gas at a flow rate of 20\,sccm, etching to a depth of 20-30\,nm. Next, we transfer Bernal-stacked bilayer graphene (BLG) encapsulated in hexagonal boron nitride (hBN), ensuring the bottom hBN layer thickness remains below 5\,nm to enhance the modulation amplitude of electrostatic potential. Graphite and the silicon substrate are used as the top gate and bottom gate, respectively. The heterostructure is then patterned into a Hall-bar geometry using standard nanofabrication techniques, and \ce{Cr}/\ce{Au} is evaporated as the contact electrode material to ensure robust one-dimensional edge contacts. \figurename~\ref{SIA} illustrates the experimental data of longitudinal resistance obtained in the measurement of our device. The measurements were taken under zero external magnetic field, with an excitation current of 100\,n\text{\AA}, at a temperature of 1.6\,K. The $V_{\rm{tg}}$ was scanned from -0.4\,V to 0.4\,V, while the $V_{\rm{bg}}$ was scanned from -4.68\,V to 4.68\,V, maintaining the nominal electric displacement field $D_{\rm{eff}}=(C_{\rm{tg}}V_{\rm{tg}}-C_{\rm{bg}}V_{\rm{bg}})/(2\epsilon_0)$ at zero;but there can be weak potential drop $\delta V=V_{\rm{tg}}-V_{\rm{bg}}$  on the order of a few V/nm applied to BLG due to the different values of $C_{\rm{tg}}$ and $C_{\rm{bg}}$. The filling factor was calculated as $\nu=n/n_0$, where $n=(C_{\rm{tg}} V_{\rm{tg}}+C_{\rm{bg}} V_{\rm{bg}})/e$ and $n_0=2/(\sqrt{3} L_s^2 )$. Here,  $C_{\rm{tg}}=1.9676 \times 10^{-7}$\,F/cm$^2$, $C_{\rm{bg}}=0.1682 \times 10^{-7}$\,F/cm$^2$, and $L_s=120$\,nm. $C_{\rm{tg}}$ is determined by parallel-plate capacitance formula $C_{\rm{tg}}= \epsilon/d$, with thickness $d$of the top hBN measured by atomic force microscopy. Given $C_{\rm{tg}}$, $C_{\rm{bg}}$ can be determined by the slope of charge neutrality line in the ($V_{\rm{bg}}$, $V_{\rm{tg}}$) parameter space \cite{prepare}.

\section{Superlattice potential simulations}
\label{sec:Superlattice potential simulation details}

For the purpose of numerically simulating artificially designed superlattice potential in a real device, we set up an electrostatic model that placed hBN-BLG-hBN on  top of etched \ce{SiO2}/\ce{Si} substrate. Top and bottom gate voltages $V_{\rm{tg}}$ and $V_{\rm{bg}}$ are acted on the graphite and doped silicon gates. The potential drop from top gate to bottom gate controls the strength of effective potential exerted to BLG. Firstly, we establish the geometry of this three-dimensional electrostatic capacitance model. The thickness of silica substrate is 290\,nm. BLG is placed at 5\,nm above the surface of silica, where etched holes array are arranged as Kagome lattice with variable lattice constant. The distance between graphite gate and BLG is 20\,nm. Then we set the top gate to zero potential point and let the potential of bottom gate change from -20\,V to 20\,V with step of 2\,V. The lattice constant of kagome-type holes array varies from 20\,nm to 120\,nm with the step of 10\,nm. The dielectric constant of \ce{SiO2}/\ce{Si} and hBN are set to 3.9 and 4, respectively. Finally, we obtain electric potential distribution in real space by  numerically solving Poisson's equation with boundary conditions defined above, which is performed using the Partial Differential Equation Toolbox of MATLAB. Due to the periodic arrangement of holes on the surface of the substrate, the potential drops unevenly, resulting in an effective  superlattice potential with the same geometry as the arrays of holes.

\section{Continuum  model and parameter renormalizations}
\subsection{Non-interacting continuum model}

Due to the existence of kagome superlattice potential, the energy bands of rhombohedral multilayer graphene (RMG) are folded into small mini Brillouin zone, which are likely to form flat bands with nontrivial topological properties thanks to the non-vanishing Berry curvatures possessed by the parent Bloch states of RMG. Focusing on the low energy physics in the vicinity of Dirac points, we derive a general effective continuum model of RMG directly from the atomistic tight-bonding Hamiltonian \cite{Moon-Koshino-prb2014, moon-tbg-prb13}. The non-interacting Hamiltonian contains two parts: one is the $\k \cdot \p$ model of RMG, the other is the periodic superlattice potential given by Eq.~(1)-(2) of main text. 
We define the primitive lattice vectors of graphene unit cell as $\a_1=a(1,0)$ and $\a_2=a(1/2, \sqrt{3}/2)$ with $a=2.46\,\text{\AA}$. The corresponding reciprocal lattices are constructed as $\b_1=2\pi/a (1,-1/\sqrt{3})$ and $\b_2=2\pi/a (0,2/\sqrt{3})$ from $\a_i \b_j=2\pi\delta_{ij}$. The two sublattices forming honeycomb lattice are seated at $\bm{\tau}_a=a(0,-1/\sqrt{3})$ and $\bm{\tau}_b=(0,0)$, respectively. We consider the situation that the next layer is shifted in the in-plane direction $a(0,-1/\sqrt{3})$ with respect to the previous layer, which defines the stacking chirality. The Dirac points of monolayer graphene are located at $\mathbf{K}_+= -4\pi/3a(1,0)$ and $\mathbf{K}_-=4\pi/3a(1,0)$. Then we derive the $\k \cdot \p$ model  from the atomistic Slater-Koster tight-bonding model based on carbon's $p_z$-like Wannier orbitals:
\begin{equation}
    H_{\rm{BLG}}=\sum_{i l\alpha,j l^{\prime}\alpha^{\prime}}
    {-t\left ( 
    \R_i+\bm{\tau}_{\alpha}+l d_0 \mathbf{e}_z
    -\R_j-\bm{\tau}_{\alpha^{\prime}}-l^{\prime} d_0 \mathbf{e}_z 
    \right )} 
    \hat{c}^{\dagger}_{i l \alpha}\hat{c}_{j l^{\prime}\alpha^{\prime}}\;,
\end{equation}
where $i$, $j$ represents lattice sites and $\R_i$, $\R_j$ represents lattice vectors in graphene. $l$ and $l^{\prime}$ are layer indexes while $\alpha$ and $\alpha^{\prime}$ are sublattice indexes. $d_0$ is the interlayer distance and $\mathbf{e}_z$ is an unit vector along out-of-plane direction. $t(\mathbf{d})$ is hopping amplitude between two $p_z$ orbitals displaced by vector $\mathbf{d}$, which is expressed in  the Slater-Kolster form:
\begin{align}
    -t(\bd) &= 
    V_{pp\pi}\left [ 1-\left ( \frac{\bd \cdot \mathbf{e}_z}{d} \right )^2  \right ] +V_{pp\sigma}\left ( \frac{\bd \cdot \mathbf{e}_z}{d} \right )^2 \\
    \begin{split}
        V_{pp\pi}&=V_{pp\pi}^0\exp\left ( -\frac{\left | \bd \right |-a/\sqrt{3}}{r_0}  \right ) \\
        V_{pp\sigma}&=V_{pp\sigma}^0\exp\left ( -\frac{\left | \bd \right |-d_0}{r_0}  \right ) 
    \end{split}
\end{align}
where $V_{pp\pi}^0=-2.7\, \eV$, $V_{pp\sigma}^0=0.48\, \eV$ and $r_0=0.184a$ \cite{moon-tbg-prb13,Moon-Koshino-prb2014}.

After Fourier transformation to $\k$ space, and expand the (Fourier transformed) tight-binding model in the vicinity of Dirac points $\mathbf{K}^{\mu}=-\mu4\pi/3a(1,0)$ ($\mu=\pm 1$ denoting valley index), we obtain the $\k\cdot\p$ model of RMG, with the intralayer and interlayer parts of Hamiltonian expressed as

\begin{align}
    h_{\rm{intra}}^{0,\mu}
    &= -\hbar v_F^0 \k \cdot \bm{\sigma} ^{\mu} \\
    h_{\rm{inter}}^{0,\mu}
    &=
    \begin{pmatrix}
        \hbar v_{\perp}(\mu k_x+ik_y) & t_{\perp} \\
        \hbar v_{\perp}(\mu k_x-ik_y) & \hbar v_{\perp}(\mu k_x+ik_y)
    \end{pmatrix}   
\end{align}
$\mu=\pm$ refers to valley index. $\k$ is the wave vector expanded around $\mathbf{K}_+/\mathbf{K}_-$ point. $\bm{\sigma}^\mu=(\mu \sigma _x,\sigma _y )$ are Pauli matrices defined in sublattice space. In term of Slater-Koster transfer integral form, $\hbar v_F^0=5.253\,\eVA$ is the non-interacting band velocity of Dirac fermions in monolayer graphene. $\hbar v_\perp=0.335\,\eVA$  and $t_\perp=0.34\,\eVA$ are Slater-Koster hopping parameters. The low-energy Hamiltonian of $n$-layer RMG of valley $\mu$, denoted as $H_n^{0,\mu}$, is just consisted of the intralayer term $h_{\rm{intra}}^{0,\mu}$ appearing in the diagonal block (in layer space) and the interlayer term $h_{\rm{inter}}^{0,\mu}$ coupling adjacent layers.

Low energy electrons of RMG are further coupled with the kagome superlattice potential $V_{\rm{SL}}$ (see Eq.~(1) of main text), leading to the following single-particle Hamiltonian for $n$-layer RMG of valley $\mu$
\begin{equation}
    H^{\mu}_{n}=H_{n}^{0, \mu}+V_{\rm{SL}}\;.
\end{equation}
Again, the superlattice $V_{\rm{SL}}$ is obtained from a realistic simulation of the electrostatic potential distribution of the device, as explained above.



\begin{figure}[bth!]
    \includegraphics[width=7in]{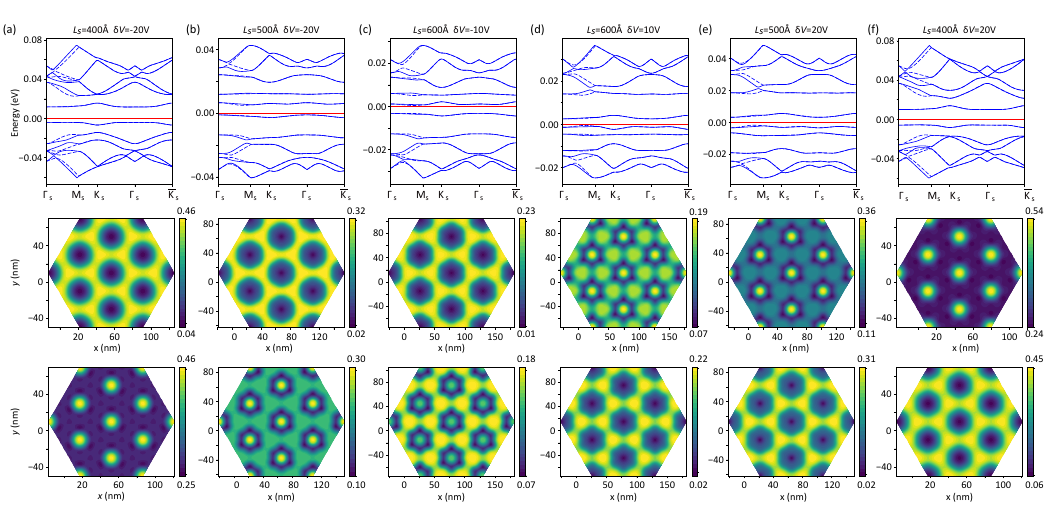}
\caption{Single particle energy band structure(top), local charge density distribution of the HVB(middle) and the LCB(bottom) for various parameters ($L_s=400\,\text{\AA}$, $\delta V=-20$\,V), ($L_s=500\,\text{\AA}$, $\delta V=-20$\,V), ($L_s=600\,\text{\AA}$, $\delta V=-10$\,V), ($L_s=600\,\text{\AA}$, $\delta V=10$\,V), ($L_s=500\,\text{\AA}$, $\delta V=20$\,V) and ($L_s=400\,\text{\AA}$, $\delta V=20$\,V), organized into columns (a) to (f).
}
\label{SIC1}
\end{figure}

We first take BLG as the simplest example. Based on the above Hamiltonian, we calculate the single-particle  band structure and local charge density distribution in real space for different superlattice constant $L_s$ and potential drop $\delta V$. As shown in \figurename ~\ref{SIC1}, we select six typical cases to present, with the corresponding $L_s$ and $\delta V$ values specified in the caption of \figurename ~\ref{SIC1}. When the potential drop $\delta V$ is negative, the HVB has Chern number $C=1$ while the LCB has Chern number $C=0$. However, the situation reserves when the potential drop $\delta V$ becomes positive, where the HVB has zero Chern number and the LCB has Chern number $C=1$. Both the HVB and the LCB are flat, with dispersions $\le 5$\,meV, and are well separated from other bands. The electron for the topological nontrivial Chern bands occupy the interstial regions between etched kagome-patterned holes, forming an emergent honeycomb lattice. While the electron for the topological trivial bands occupy the central region surrounded by the kagome-patterned holes, forming an emergent triangular lattice. 


\begin{figure}[bth!]
    \includegraphics[width=7in]{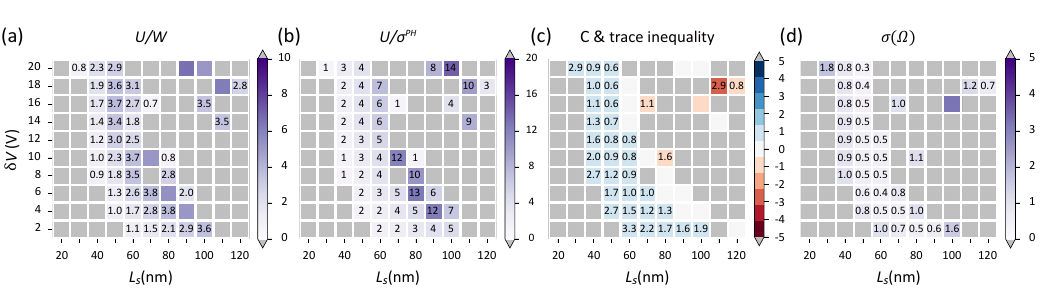}
\caption{Single particle phase diagrams of the LCB: (a) ratio between Coulomb interaction energy $U$ and bandwidth $W$, (b) ratio between $U$ and particle-hole symmetry violation $\sigma_{\rm{PH}}$, $U/\sigma_{\rm{PH}} \rightarrow \infty$ if the particle-hole symmetry is exact, (c)  Chern number, and (d) normalized berry curvature standard deviation $\sigma(\Omega)$ of the LCB in parameter space $ \left ( L_s, \delta V\right )$. The black numbers in (a), (b), (c) and (d) represent the values of $U/W$, $U/\sigma_{\rm{PH}}$, trace inequality and  $\sigma(\Omega)$, respectively.
}
\label{SIC2}
\end{figure}

We also present the phase diagrams of single-particle properties for the LCB of BLG in \figurename ~\ref{SIC2}. In particular, in \figurename~\ref{SIC2}(a), (b), (c) and (d), we present the ratio between characteristic interaction energy $U=e^2/(4\pi\epsilon L_s)$ and bandwidth $W$,  the ratio between $U$ and particle-hole symmetry violation $\sigma_{\rm{PH}}$ (see Sec.~\ref{sec:interaction}(C) for detailed explanation), the Chern number, and the Berry curvature standard deviation of the LCB in the parameter space of $\delta V$ and $L_s$. For the LCB, topological flat bands appear only for positive $\delta V$.
Similar to the phase diagrams for the HVB presented in the main text, there exists a substantial region where the LCB behaves as nearly ``ideal" topological flat band. The superlattice constant $L_s$ of this region ranges from 40\,nm to 100\,nm and the potential drop $\delta V$ of varies from 2\,V to 20\,V. This region hosts topological flat bands with Chern number $1$, as well as good quantum geometric properties with small Berry curvature standard deviations. However, the bandwidth is in general larger than that of HVB as can be seen from $U/W$ in \figurename~\ref{SIC2}(a), compared with that of the HVB shown in Fig.~2(a) of main text. Moreover, the particle-hole symmetry violation of LCB is also stronger compared to that of HVB, as indicated by the relatively small $U/\sigma_{\rm{PH}}$ values shown in \figurename~\ref{SIC2} ($U/\sigma_{\rm{PH}}$ of HVB is shown in Fig.~2(b) of main text). As a result, when potential drop is positive and at 1/3 and 2/3 fillings of  LCB, it turns out that the many-body phase diagrams are dominated by two types of competing states:  charge density wave and Fermi liquid.  FCIs emerge only in a small region of the parameter space. Detailed results are presented in Sec.~\ref{sec:interaction}(D) and \figurename~\ref{SID2}. 

\subsection{Renormalization of the model parameters due to interactions with remote-band electrons}
In our work, we mostly focus on the low-energy physics around charge neutrality point. We set up a low-energy window marked by $E_C^{*}$ which roughly includes three conduction subbands and three valence subbands per spin per valley. Electron-electron interactions  become important and non-perturbative within this low-energy window within $E_C^*$, while they can be approximately treated as perturbations to single-particle energy outside $E_C^*$. We call the electrons within the low-energy window as low-energy electrons, while those out-side $E_C^*$ as remote-band electrons.
Although we focus on the low-energy physics, the occupied remote-band electrons indeed play an important role considering $e$-$e$ interactions. The electrons in the filled remote bands will act through long-ranged Coulomb potential upon the properties of low-energy electrons. As a result, an effective low-energy single-particle Hamiltonian  (within $E_C^*$) would have parameters in general larger in amplitudes than the non-interacting ones. For example, it is well known that the Fermi velocity around the Dirac point in graphene would be amplified by the filled Dirac Fermi sea \cite{elias_natphys2011}. We take into account this effect using perturbative renormalization group approach \cite{vafek_prl2020,guo-prb24}.  Without going into detailed derivations, we directly write down the expressions of the renormalized model parameters as already reported in Ref.~\onlinecite{guo-prb24}
\begin{subequations}
\begin{align}
 v_F(E_C^*)&=v_F^0 \left(1+\frac{\alpha_0}{4\epsilon_r}\log{\frac{E_C}{E_C^*}} \right)\; \label{eq:H-RG-a}\\
 t_{\perp}(E_C^*)&=t_{\perp}\,\left(1+\frac{\alpha_0}{4\epsilon_r}\log{\frac{E_C}{E_C^*}}\right)\; \label{eq:H-RG-c}\\
 v_{\perp}(E_C^*)&=v_{\perp}\;.
 \label{eq:H-RG-f}
\end{align}
\label{eq:H-RG}
\end{subequations}
Here $E_C$ is the largest energy cutoff of the continuum model above which the Dirac-fermion description to graphene would no longer be valid. $\alpha_0=e^2/4\pi\epsilon_0\hbar v_F$ is the effective fine structure constant of grahene. We refer the readers to Ref.~\onlinecite{guo-prb24} for detailed derivations of the above equations.

\section{Interaction Hamiltonian, structure factor, particle-hole symmetry, and more results}
\label{sec:interaction}

\subsection{Interaction Hamiltonian}

We consider the $e$-$e$ Coulomb interactions
\begin{equation}
\hat{V}_\text{ee}=\frac{1}{2}\int d^2 r  d^2 r' \sum _{\sigma, \sigma '} \hat{\psi}_\sigma ^{\dagger}(\br)\hat{\psi}_{\sigma '}^{\dagger}(\br ') V_c (|\br -\br '|) \hat{\psi}_{\sigma '}(\br ') \hat{\psi}_{\sigma}(\br)
\label{eq:coulomb}
\end{equation}
where $\hat{\psi}_{\sigma}(\br)$ is real-space electron annihilation operator at $\br$ with spin $\sigma$. Such interaction can be expressed in Wannier basis as 
\begin{equation}
\hat{V}_\text{ee}=\frac{1}{2}\sum _{i i' j j'}\sum_{l l' m m'}\sum _{\alpha \alpha '\beta \beta ' }\sum _{\sigma \sigma '} \hat{c}^{\dagger}_{i, \sigma l \alpha}\hat{c}^{\dagger}_{i', \sigma ' l' \alpha '} V^{\alpha \beta l m \sigma , \alpha ' \beta ' l' m' \sigma '} _{ij,i'j'}\hat{c}_{j', \sigma ' m' \beta '} \hat{c}_{j, \sigma m \beta}\;,
\end{equation}
where
\begin{align}
&V^{\alpha \beta l m \sigma , \alpha ' \beta ' l' m' \sigma '} _{ij,i'j'} \nonumber \\
&=\int d^2 r d^2 r'  V_c (|\br -\br '|) \,\phi ^*_{l,\alpha} (\mathbf{r}-\mathbf{R}_i-\bm{\tau}_{l,\alpha})\,\phi_{m,\beta} (\mathbf{r}-\mathbf{R}_j- \bm{\tau}_{m,\beta}) \phi^*_{l', \alpha  '}(\br-\mathbf{R}_i'-\bm{\tau}_{l' , \alpha '})\phi _{m', \beta  '}(\br-\mathbf{R}_j'-\bm{\tau} _{m', \beta '}) \nonumber \\
&\quad \times \chi ^{\dagger}_\sigma \chi ^{\dagger}_{\sigma '}\chi _{\sigma '}\chi _{\sigma} .
\end{align}
Here $i$, $\alpha$, and $\sigma$ refer to atomic lattice vectors, layer/sublattice index, and spin index, respectively. $\phi$ is Wannier function and $\chi$ is the two-component spinor wave function. We further assume that the ``density-density" like interaction is dominant in the system, i.e., $V^{\alpha \beta l m \sigma , \alpha ' \beta ' l' m' \sigma '} _{ij,i'j'}\approx V^{\alpha \alpha l l \sigma , \alpha ' \alpha ' l' l' \sigma '} _{ii,i'i'}\equiv V_{i \sigma l \alpha  ,i' \sigma ' l' \alpha '}$, then the Coulomb interaction is simplified 
\begin{align}
\hat{V}_\text{ee}=&\frac{1}{2}\sum _{i i'}\sum _{\alpha \alpha ', l l'}\sum _{\sigma \sigma '}\hat{c}^{\dagger}_{i, \sigma l \alpha}\hat{c}^{\dagger}_{i', \sigma' l' \alpha} V_{i\sigma l \alpha, i' \sigma ' l' \alpha '}\hat{c}_{i', \sigma ' l' \alpha '}\hat{c}_{i, \sigma l \alpha} \nonumber \\
\approx&\frac{1}{2}\sum _{i l \alpha \neq i' l' \alpha '}\sum _{\sigma \sigma '}\hat{c}^{\dagger}_{i, \sigma l \alpha} \hat{c}^{\dagger}_{i', \sigma ' l' \alpha '} V_{i l \alpha,i' l' \alpha '}\hat{c}_{i', \sigma ' l' \alpha '}\hat{c}_{i, \sigma l \alpha} \nonumber 
\end{align}
Here we neglect on-site Coulomb interactions which is at least one order of magnitude weaker than long-range inter-site Coulomb interactions in the context of moir\'e superlattice and other long-period superlattices \cite{zhang_prl2022}. Given that the electron density is low ($10^{11}$~cm$^{-2}$ in our problem),  the chance that two electrons meet at the same atomic site is very low. Therefore, the Coulomb interactions between two electrons are mostly contributed by the inter-site ones. 

In order to model the screening effects  and capture the layer dependence of Coulomb interactions multilayer graphene, we introduce the following Coulomb potential in momentum space
\begin{align}
    &V_{ll} (\mathbf{q})=\frac{e^2}{2 \Omega_0 \epsilon_r \epsilon_0 \sqrt{q^2+\kappa^2}}\;\nn
    &V_{ll'}(\mathbf{q})=\frac{e^2}{2 \Omega_0 \epsilon_r \epsilon_0 q} e^{-q|l-l'|d_0},\,\hspace{6pt}\, \text{if } l\neq l'
    \label{eq:Vq}
\end{align}
where $\Omega _0$ is the area of the superlattice's primitive cell and $\kappa = 1/400$\,\AA$^{-1}$ is the inverse screening length. 

Since we are interested in the low-energy bands, the intersite Coulomb interactions can be divided into the intra-valley term and the inter-valley term. The intervalley term is at least two orders of magnitudes weaker than the intravalley one in our system because of the small Brillouin zone of the superlattice, thus is neglected in our present study. The intra-valley term $\hat{V}^{\text{intra}}$ is expressed as
\begin{equation}
\hat{V}^{\rm{intra}}=\frac{1}{2N_s}\sum_{\alpha\alpha ', l l'}\sum_{\mu\mu ',\sigma\sigma '}\sum_{\bk \bk ' \bq} V_{l l'}(\bq)\,
\hat{c}^{\dagger}_{\sigma \mu l \alpha}(\bk+\bq) \hat{c}^{\dagger}_{\sigma' \mu ' l' \alpha '}(\bk ' - \bq) \hat{c}_{\sigma ' \mu ' l' \alpha '}(\bk ')\hat{c}_{\sigma \mu l \alpha}(\bk)\;,
\label{eq:h-intra}
\end{equation}
where $N_s$ is the total number of the superlattice's sites. 

The electron annihilation operator can be transformed from the original basis to the band basis:
\begin{equation}
\hat{c}_{\sigma\mu l \alpha}(\bk)\equiv \hat{c}_{\sigma\mu l \alpha \G}(\btk) =\sum_n C_{ \mu l \alpha \mathbf{G},n}(\btk)\,\hat{c}_{\sigma \mu,n}(\btk)\;,
\label{eq:transform}
\end{equation}
where $C_{\mu l \alpha \mathbf{G},n}(\btk)$ is the expansion coefficient in the $n$-th Bloch eigenstate at $\btk$ of valley $\mu$: 
\begin{equation}
\ket{\sigma \mu, n; \btk}=\sum_{l \alpha \mathbf{G}}C_{\mu l \alpha \mathbf{G},n}(\btk)\,\ket{ \sigma, \mu, l, \alpha, \mathbf{G}; \btk }\;.
\end{equation}
We note that the non-interacting Bloch functions are spin degenerate due to the separate spin rotational symmetry ($SU(2)\otimes SU(2)$ symmetry) of each valley. Using the transformation given in Eq.~(\ref{eq:transform}), the intra-valley Coulomb interaction can be written in the band basis
\begin{align}
\hat{V}^{\rm{intra}}&=\frac{1}{2N_s}\sum _{\btk \btk'\btq}\sum_{\substack{\mu\mu' \\ \sigma\sigma'\\l l'}}\sum_{\substack{nm\\ n'm'}} \left(\sum _{\mathbf{Q}}\,V_{ll'} (\mathbf{Q}+\btq)\,\Omega^{\mu l ,\mu' l'}_{nm,n'm'}(\btk,\btk',\btq,\mathbf{Q})\right) \nonumber \\
&\times \hat{c}^{\dagger}_{\sigma\mu,n}(\btk+\btq) \hat{c}^{\dagger}_{\sigma'\mu',n'}(\btk'-\btq) \hat{c}_{\sigma'\mu',m'}(\btk') \hat{c}_{\sigma\mu,m}(\btk)
\label{eq:Hintra-band}
\end{align}
where the form factor $\Omega ^{\mu  l,\mu' l'}_{nm,n'm'}$ are written respectively as
\begin{equation}
\Omega ^{\mu l ,\mu' l'}_{nm,n'm'}(\btk,\btk',\btq,\mathbf{Q})
=\sum _{\alpha\alpha'\mathbf{G}\mathbf{G}'}C^*_{\mu l \alpha\mathbf{G}+\mathbf{Q},n}(\btk+\btq) C^*_{\mu'l'\alpha'\mathbf{G}'-\mathbf{Q},n'}(\btk'-\btq)C_{\mu'l'\alpha'\mathbf{G}',m'}(\btk')C_{\mu l \alpha\mathbf{G},m}(\btk).
\end{equation}

\subsection{Structure factor}
In this sector, we derive the expression for structure factor. The structure factor is actually the Fourier transformation of density-density correlation function. We express the density-density operator as  

\begin{equation}
\begin{aligned}
    \chi_0\left(\vr_i, \vr_j \right) 
    &= \left \langle \rho\left(\vr_i\right)\rho\left(\vr_j\right) \right \rangle \\
    &= \left \langle 
    \hat{\Psi}^{\dagger} \left ( \vr_i \right ) \Psi \left ( \vr_i \right ) \hat{\Psi}^{\dagger} \left ( \vr_j \right ) \Psi \left ( \vr_j \right ) 
    \right \rangle 
\end{aligned}
\end{equation}

$\hat{\Psi}\left(\vr_i\right)$ is electron annihilation operator at $\vr_i$ in real space. 

And the structure factor can be expressed as 

\begin{equation}
    S(\mathbf{q})=\rho(\q)\rho(-\q)\;,
\end{equation}
where $\rho(\mathbf{q}) = \sum_{\mathbf{k}}{\hat{c}^\dagger_{\mathbf{k}-\mathbf{q}}\hat{c}_{\mathbf{k}}}$ is  momentum-space density operator. The wave vector $\k$, $\q$ are expanded around the Dirac points graphene, and we fold them in to the mini Brillouin zone by $\k = \btk + \G$ and $\q = \btq + \Q$, with $\G$ and $\Q$ denoting reciprocal vectors of the superlattice. Here we only consider the momentum transfer $\q = \btq + \Q$ with $|\Q| \le 4\pi/\sqrt{3}L_s$.

We transform the expression for structure factor from original basis to Bloch band basis and project it onto the non-interacting spin-degenerate flat highest valence band, then we get the following expression of structure factor that have been used to calculate Fig~4(d) in main text:
\begin{equation}
\begin{aligned}
    S(\q) &= S(\btq, \Q) \\
    &= \sum_{l\alpha,l^{\prime}\beta}\sum_{\k, \k^{\prime}}
    \left \langle
    {\hat{c}^{\dagger}_{l\alpha, \k-\q}\hat{c}_{l\alpha, \k}
    \hat{c}^{\dagger}_{l^{\prime}\beta, \k^{\prime}+\q}
    \hat{c}_{l^{\prime}\beta, \k^{\prime}} }
    \right \rangle\\
    &= \sum_{l\alpha,l^{\prime}\beta}
    \sum_{\G, \G^{\prime}}\sum_{\btk, \btk^{\prime}}
    C^{*}_{l\alpha \G-\Q, \btk-\btq}
    C_{l\alpha \G, \btk}
    C^{*}_{l^{\prime}\beta \G^{\prime}+\Q, \btk^{\prime}+\btq}
    C_{l^{\prime}\beta \G^{\prime}, \btk^{\prime}} \\
    &\times
    \left\langle
    \hat{c}^{\dagger}_{\btk-\btq}
    \hat{c}_{\btk}
    \hat{c}^{\dagger}_{\btk^{\prime}+\btq}
    \hat{c}_{\btk^{\prime}} 
    \right\rangle\\
    &=\sum_{\btk, \btk^{\prime}}{
    \lambda^{*} \left ( \btk-\btq, \btq, \Q \right)
    \lambda \left ( \btk^{\prime}, \btq, \Q \right )
    \left \langle 
    \hat{c}^{\dagger}_{\btk-\btq}
    \hat{c}_{\btk}
    \hat{c}^{\dagger}_{\btk^{\prime}+\btq}
    \hat{c}_{\btk^{\prime}}
    \right \rangle} \\
    &=\sum_{\btk} \left | \lambda \left ( \btk, \btq, \Q \right) \right |^2 \hat{n} \left ( \btk \right )
    +\sum_{\btk,\btk^{\prime}}
    {
    \lambda^{*} \left ( \btk-\btq, \btq, \Q \right)
    \lambda \left ( \btk^{\prime}, \btq, \Q \right)
    \left\langle
    \hat{c}^{\dagger}_{\btk-\btq}
    \hat{c}^{\dagger}_{\btk^{\prime}+\btq}
    \hat{c}_{\btk^{\prime}}
    \hat{c}_{\btk}
    \right\rangle
    }
\end{aligned}
\end{equation}
where $\lambda\left ( \btk, \btq, \Q \right )$ is defined as:
\begin{equation}
    \begin{aligned}
        \lambda \left ( \btk, \btq, \Q \right )
        =\sum_{l \alpha, \G}
        {
        C^*_{l\alpha \G+\Q,\btk+\btq}C_{l\alpha\G,\btk}
        }
    \end{aligned}
\end{equation}

\subsection{Particle-hole symmetry breaking terms}
Here we evaluate to which extent  particle-hole symmetry is broken for the Hamiltonian projected to a single band \cite{moessner-fci-prl13}.  A general band-projected two-body interaction Hamiltonian can be expressed as 
\begin{equation}
    \hat{V}_{\text{proj}}= \frac{1}{2N_s}\sum_{\kt,\kt',\qt}
    {\Omega(\kt, \kt', \qt)\hc^\dagger_{\kt+\qt}\hc^\dagger_{\kt'-\qt}\hc_{\kt'}\hc_{\kt}}
\end{equation}
where $\Omega(\kt, \kt',\qt)$ is form factor from band projection. We define the particle-hole transformation $\hc^\dagger_{\kt} \to \hd_{\kt}$. Then, after some straightfoward algebra, the projected Coulomb interaction transforms as:
\begin{align}
    &\hat{V}^{\rm{PH}}_{\rm{proj}} \nonumber\\
    &= \frac{1}{2N_s}\sum_{\kt,\kt',\qt}{\Omega^*(\kt,\kt'\qt)\hd_{\kt+\qt}\hd_{\kt'-\qt}\hd^\dagger_{\kt'}\hd^\dagger_{\kt}} \nonumber \\
    &= -\frac{1}{2N_s}\sum_{\kt,\kt'}\left\{ \Omega^*(\kt,\kt',\mathbf{0})+\Omega(\kt,\kt',\mathbf{0}) \right\} \hd^\dagger_{\kt}\hd_{\kt}+\frac{1}{2N_s}\sum_{\kt,\kt'}\left\{ \Omega^*(\kt,\kt',\kt'-\kt) + \Omega(\kt,\kt',\kt'-\kt)\right\} \hd^\dagger_{\kt}\hd_{\kt} \nonumber \\ 
    & \quad +\frac{1}{2N_s}\sum_{\kt,\kt',\qt}\Omega(\kt,\kt',\qt) \hd^\dagger_{\kt+\qt}\hd^\dagger_{\kt'-\qt}\hd_{\kt'}\hd_{\kt} + \rm{constant}
\end{align}
We note that after particle-hole transformation, there are two single-particle terms arising from the band-projected Coulomb interactions as given in the third line of the above equation.

Kinetic energy term transforms under particle-hole operation as 
\begin{align}
    \hat{T}^{\text{PH}}  &= -\sum_{\kt} \epsilon_{\kt}\hd^\dagger_{\kt}\hd_{\kt} + \rm{constant}
\end{align}
Therefore, after particle-hole transforamtion, the effective single-particle Hamiltonian for the holes can be expressed as
\begin{align}
&\hat{H}^{h}_{\rm{single}}\;\nonumber\\
=& -\sum_{\kt} \epsilon_{\kt}\hd^\dagger_{\kt}\hd_{\kt} -\frac{1}{2N_s}\sum_{\kt,\kt'}\left\{ \Omega^*(\kt,\kt',\mathbf{0})+\Omega(\kt,\kt',\mathbf{0}) \right\} \hd^\dagger_{\kt}\hd_{\kt}\nonumber\\
&+\frac{1}{2N_s}\sum_{\kt,\kt'}\left\{ \Omega^*(\kt,\kt',\kt'-\kt) + \Omega(\kt,\kt',\kt'-\kt)\right\} \hd^\dagger_{\kt}\hd_{\kt} 
\end{align}
The difference between the single-particle energy of electrons  and that of holes  characterizes the degrees of particle-hole symmetry breaking of the target band. Note that this term is proportional to the interaction strength $U$. Therefore, we consider the quantity
\begin{align}
&\delta E\nonumber\\
=&\sum_{\kt}\delta\varepsilon_{\kt}\nonumber\\
=&\sum_{\kt}\,\Big(-2\epsilon_{\kt} + \frac{1}{N_s}\sum_{\kt'}\left\{ -\rm{Re}(\Omega(\kt,\kt',\mathbf{0}))+\rm{Re}(\Omega(\kt,\kt',\kt'-\kt))  \right\}\,\Big)\;,
\end{align}
and define $\sigma^{\rm{PH}}$ as the standard deviation of $\left\{ \delta\varepsilon_{\kt} \right\}$.  Then, the ratio between characteristic Coulomb interaction energy $U=e^2/4\pi\epsilon L_s$ and $\sigma^{\rm{PH}}$ characterizes the degrees of particle-hole symmetry breaking for the target band. If $U/\sigma^{\rm{PH}}\to\infty$, it would corresponds to the situation of lowest Landau level, where the particle-hole symmetry is exact.

\subsection{More results on exact-diagonalization calculations}

\begin{figure}[bth!]
    \includegraphics[width=3in]{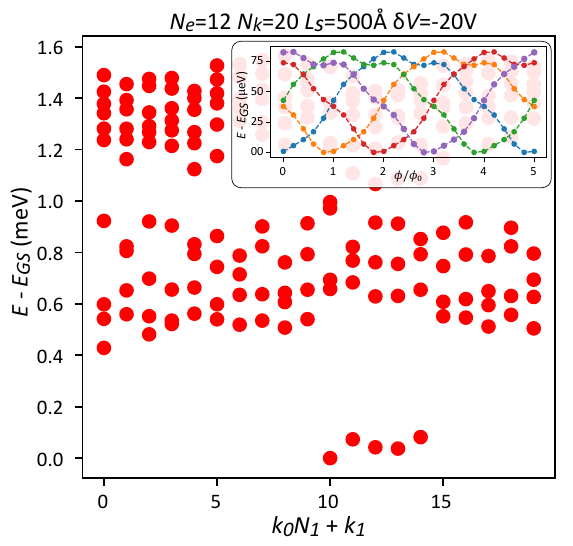}
\caption{Total energy $vs.$ total momentum at 3/5  filling of the HVB with $L_s = 50\,$nm and $\delta V =-20\,$V. The inset indicates the corresponding spectra flow behavior. Note the similarities with Fig.~3(d) of main text (for filling factor 2/5) due to the approximate particle-hole symmetry of HVB (up to a global momentum shift of $(\pi, \pi)$). }
\label{SID1}
\end{figure}

In this section, we provide more results on exact-diagonalization calculations of bilayer graphene system modulated by kagome-patterned superlattice potential.  Firstly, we present the energy spectrum (total energy $vs.$ total crystalline momentum) when 12 electrons occupying 20 Bloch orbitals of the HVB (3/5 filling), with $L_s=50$\,nm and $\delta V = -20$\,V. As shown in \figurename ~\ref{SID1}, the energy spectrum consists of a five-fold quasi-degenerate ground-state manifold separated by a gap $\sim 0.4$\,meV from the excited states. Moreover, the inset displays the corresponding spectra flow behavior: upon the adiabatic flux insertion, these five nearly degenerate ground states will interchange with each other and return to the original configurations when $\phi=5\phi_0$. This confirms the topological nature of such a degenerate many-body state as composite-Fermion type FCI, similar to that at 2/5 filling of HVB.

\begin{figure}[bth!]
    \includegraphics[width=7in]{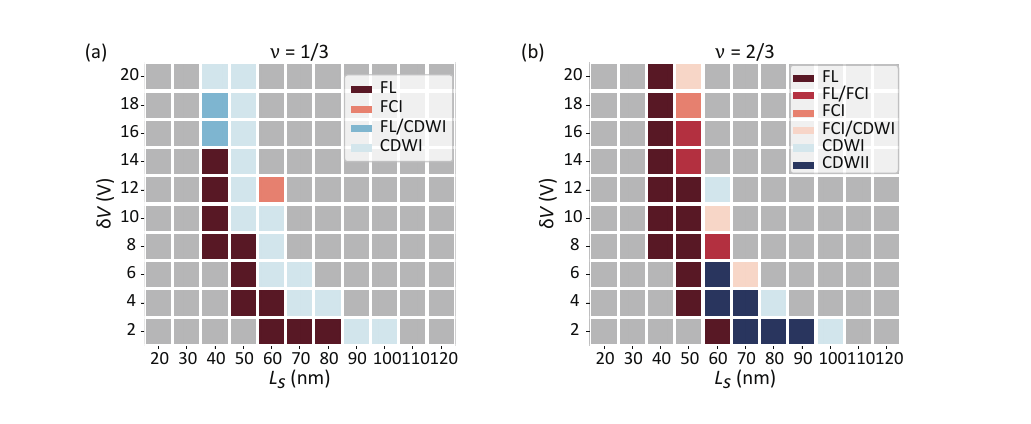}
\caption{Phase diagram of the LCB at 1/3 and 2/3 electron fillings obtained from exact-diagonalization calculations. (a) Phase diagram at 1/3 electron filling. (b)Phase diagram at 2/3 electron filling. ``FL" stands for Fermi liquid state, ``CDW" stands for charge density wave, ``FL/CDW" stands for cross-over state between FL and CDW. ``FCI" represents fractional chern insulator, ``FL/FCL" denotes cross-over state between FL and FCL and ``FCL/CDW" denotes cross-over state between FCL and CDW. 
}
\label{SID2}
\end{figure}

\begin{figure}[bth!]
    \includegraphics[width=7in]{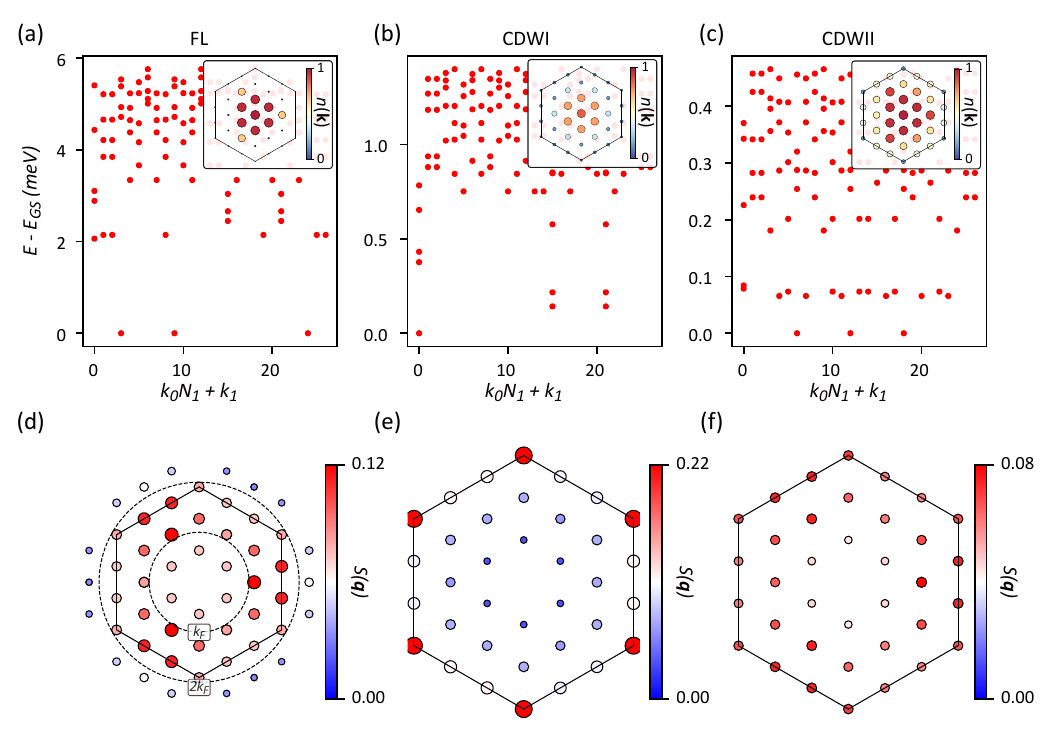}
\caption{Many-body energy spectrum of (a) Fermi liquid state, (b) CDW\I\ state and (c) CDW\II\ state. Structure factor of (d) Fermi liquid state, (e) CDW\I\ state and CDW\II\ state.}
\label{SID3}
\end{figure}

We further provide the many-body phase diagrams  at 1/3 and 2/3 fillings of LCB in \figurename~\ref{SID2}(a) and (b), respectively, where we only consider positive $\delta V$ as otherwise LCB has mostly has zero Chern number. 
We find that there are various many-body  states competing with each other. 
At 1/3  filling, there are two states, Fermi liquid (FL) and CDW states (marked as CDW\I) that are competing with each other.  The FL state is characterized by a sharp Fermi surface with abrupt jump of occupation number from (nearly) $1$ to (nearly) 0 as shown in the  inset of \figurename~\ref{SID3}(a). There are three $\k$ points at which the occupation is around 1/3, this is because with the system size of 27 and 9 occupied electrons, one cannot construct a compact Fermi sea preserving $C_{3}$ symmetry, and  one of  the three sites at the corners of the Fermi sea has to be empty. This also leads to the three-degenerate ground state as shown in \figurename~\ref{SID3}(a). This is certainly an artifact due to finite-size effect, in the  thermodynamic limit, one would expect a compact Fermi sea respecting $C_{3}$ symmetry of the system. Therefore, to restore $C_{3}$ symmetry, in the inset of \figurename~\ref{SID3}(a) we present the $\k$-space occupation number averaged over the three ground states, leading to the 1/3 filling at the corners which is completely an artifact due  to finite size effects. We have also calculated the structure factor of the FL state as shown in \figurename~\ref{SID3}(d), which drastically decays when $\vert\q\vert>2k_F$, consistent with the FL behavior. Most importantly, we checked this state is qualitatively the same as the non-interacting ground state of 9 electrons occupying 27 sites in the sense that they have the same three-fold degenerate ground at the same crystalline momenta and have the similar sharp Fermi surface configurations in $n(\k)$. The seemingly ``gapped" behavior of the FL as shown in \figurename~\ref{SID3}(a) is attributed to the interaction-enhanced Fermi velocities, such that it costs more energy to create a particle-hole excitations $\sim \hbar v_F\delta q$. In the thermodynamic limit with $\delta q\to 0$, the gapless behavior of the FL state would be recovered. The interaction-enhanced kinetic energy is also seen in monolayer graphene \cite{elias-np11} and twisted bilayer graphene \cite{kang-prl21}. 
The energy spectrum of CDW\I\ state is given in    \figurename~\ref{SID3}(b), where there are three low-lying states at wavevectors of $\Gamma_s$, $K_s$ and $K_s'$, separated by a tiny gap from excited states. The ground state at zero momentum is characterized by structure factor strongly peaked at $K_s/K_s'$ point as shown in \figurename~\ref{SID3}(e), suggesting that it is a CDW state with $\sqrt{3}\times\sqrt{3}$ cell enlargement. There are also crossover states between FL and CDW\I, which are marked by ``FL/CDW\I" in the phase diagram. 
Corresponding to the bandwidth of LCB shown in \figurename ~\ref{SIC2}(a), the global many-body ground state at 1/3 filling of LCB evolves from FL state to CDW\I\ as the bandwidth becomes smaller. At $L_s=60\,$nm and $\delta V=12\,$V, the ground state becomes FCI.

At 2/3  filling of LCB, a new type of charge density wave state emerges and is denoted by CDW\II. We see that there is a substantial region where FL is the global many-body ground state at 2/3 filling due to the large non-interacting bandwidth as shown in \figurename~\ref{SIC2}(a). When tuning $L_s$ and $\delta V$ such that the bandwidth becomes smaller, FCI state and two types of CDW states (marked as CDW\I\ and CDW\II\ in \figurename~\ref{SID2}(b)) emerge and compete with each other. In \figurename~\ref{SID3}(c), we present the many-body energy spectra of the CDW\II\ state, characterized by three degenerate ground states at crystalline momenta corresponding 1/3 of the three reciprocal vectors. The orbital occupation number seems to be characterized by a Fermi surface that breaks $C_3$ symmetry, suggesting that this state is a metallic CDW state. The structure factor of CDW\II\ state is given in \figurename~\ref{SID3}(f), where the largest amplitude occurs at (1/3,0) wavevector.
 
There are also many crossover states between different many-body interacting ground states such as FL/CDW\I, FL/FCI, FCI/CDW\I\ etc. The crossover states may be an artifact due to finite-size effects. In the thermodynamic limit, the crossover region may become a sharp phase boundary.

\begin{figure}[bth!]
    \includegraphics[width=7in]{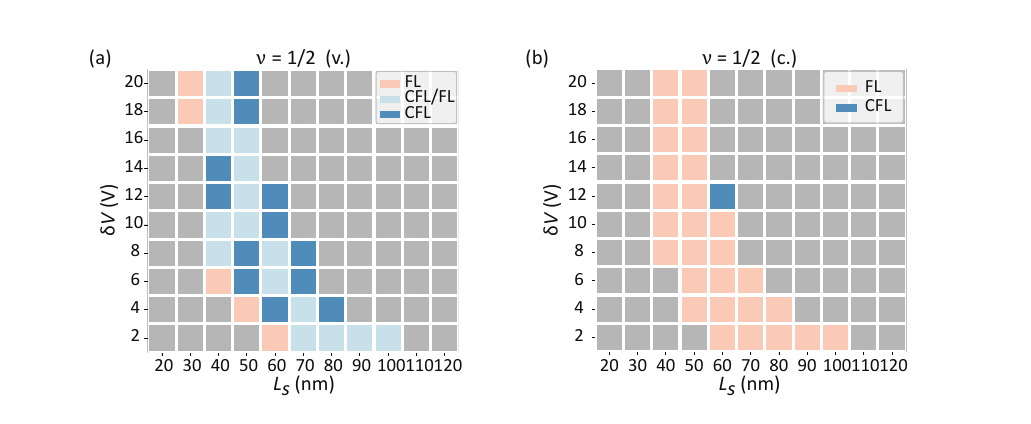}
\caption{Phase diagram of the HVB(a) and the LCB(b) at 1/2 electron fillings obtained from exact-diagonalization calculations. "FL" stands for Fermi liquid state, "CFL" denotes composite fermi liquid state and "FL/CFL" represents cross-over state between FL and CFL.
}
\label{SID4}
\end{figure}

Then we also provide the phase diagrams for the HVB and the LCB at 1/2 electron filling in \figurename ~\ref{SID4}. For the ground state of the HVB, as shown in \figurename ~\ref{SID4}(a), CFL and FL compete with each other in the selected region. The crossover CFL/FL state may be an artifact due to finite-size effect. In the thermodynamic limit, the crossover region may become a sharp phase boundary between CFL and FL. While the ground state of the LCB are mainly fermi liquid state, according to \figurename ~\ref{SID4}(b).

\section{Results on trilayer and tetralayer graphene}
\begin{figure}[bth!]
    \includegraphics[width=5in]{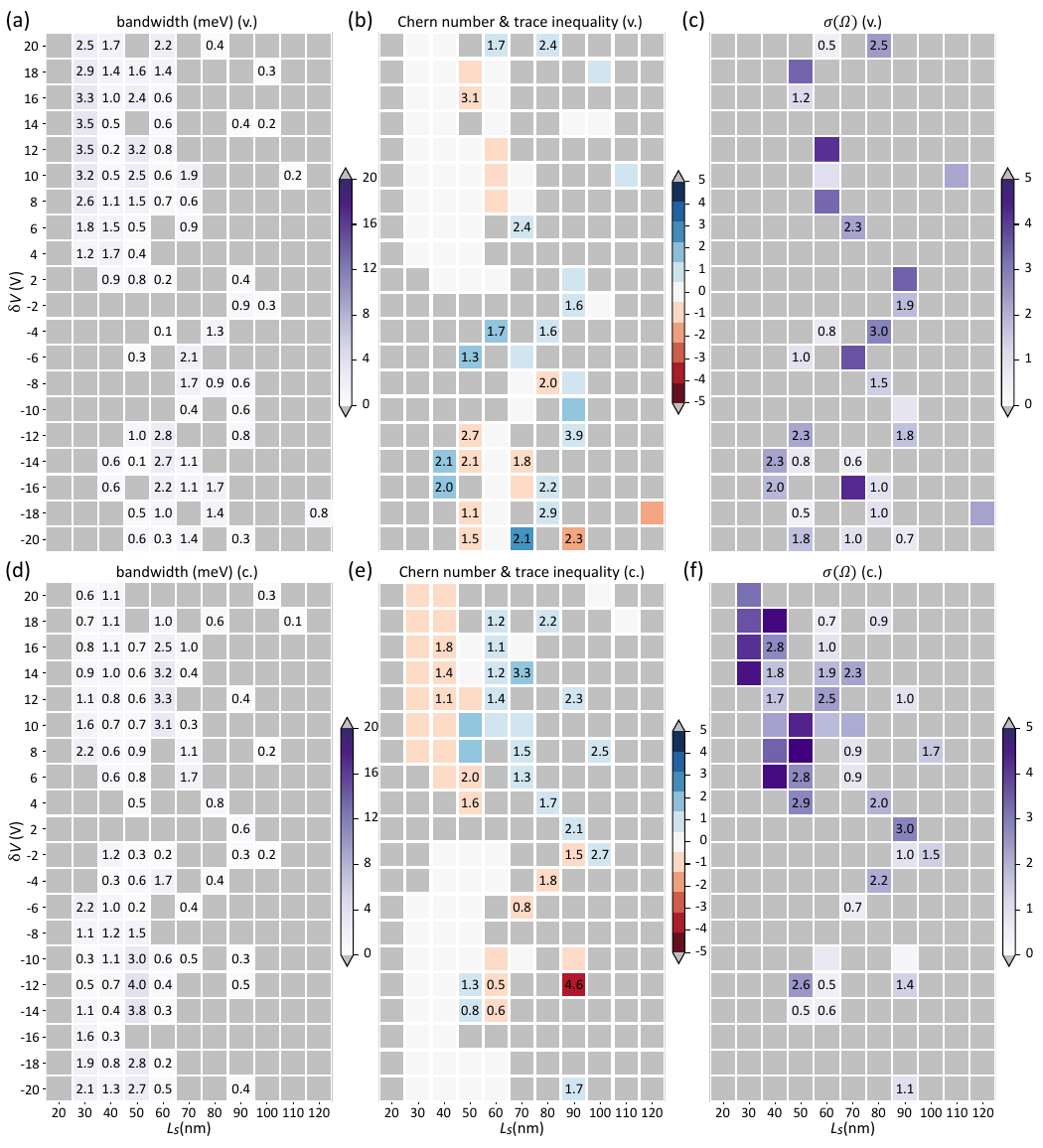}
\caption{Single particle phase diagrams of the HVB and the LCB for trilayer graphene system modulated by kagome potential. (a) Bandwidth, (b) Chern number and (c) normalized berry curvature standard deviation $\sigma(\Omega)$ of the HVB in parameter space. The numbers in (a), (b) and (c) denotes the values of bandwidth, trace inequality and $\sigma(\Omega)$, respectively. (d)-(f) are phase diagrams for the LCB.
}
\label{SIE1}
\end{figure}

\begin{figure}[bth!]
    \includegraphics[width=5in]{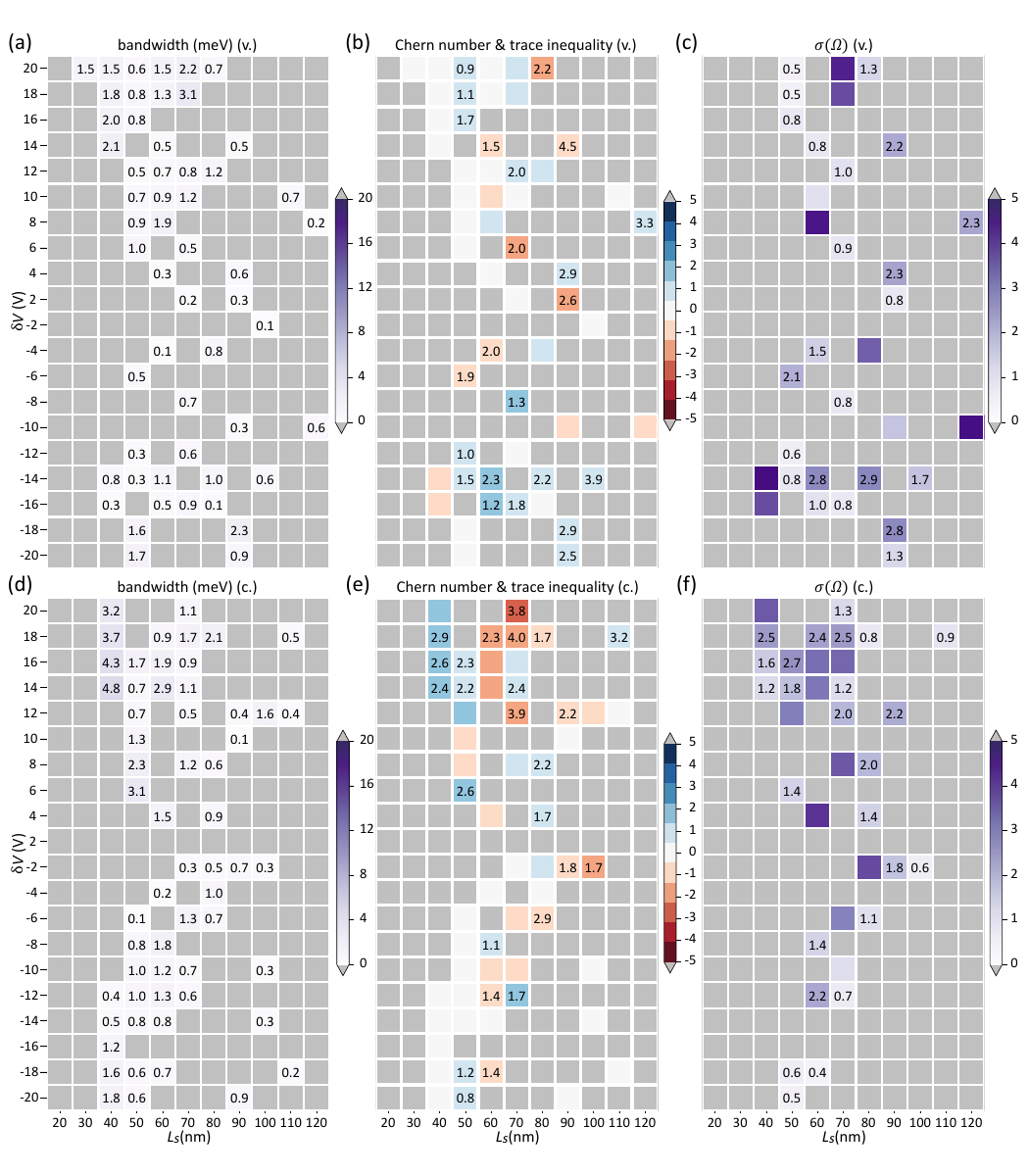}
\caption{Single particle phase diagrams of the second HVB and the second LCB for trilayer graphene system modulated by kagome potential. (a) Bandwidth, (b) Chern number and (c) normalized berry curvature standard deviation $\sigma(\Omega)$ of the second HVB in parameter space. The numbers in (a), (b) and (c) denotes the values of bandwidth, trace inequality and $\sigma(\Omega)$, respectively. (d)-(f) are phase diagrams for the second LCB.
}
\label{SIE2}
\end{figure}

\begin{figure}[bth!]
    \includegraphics[width=5in]{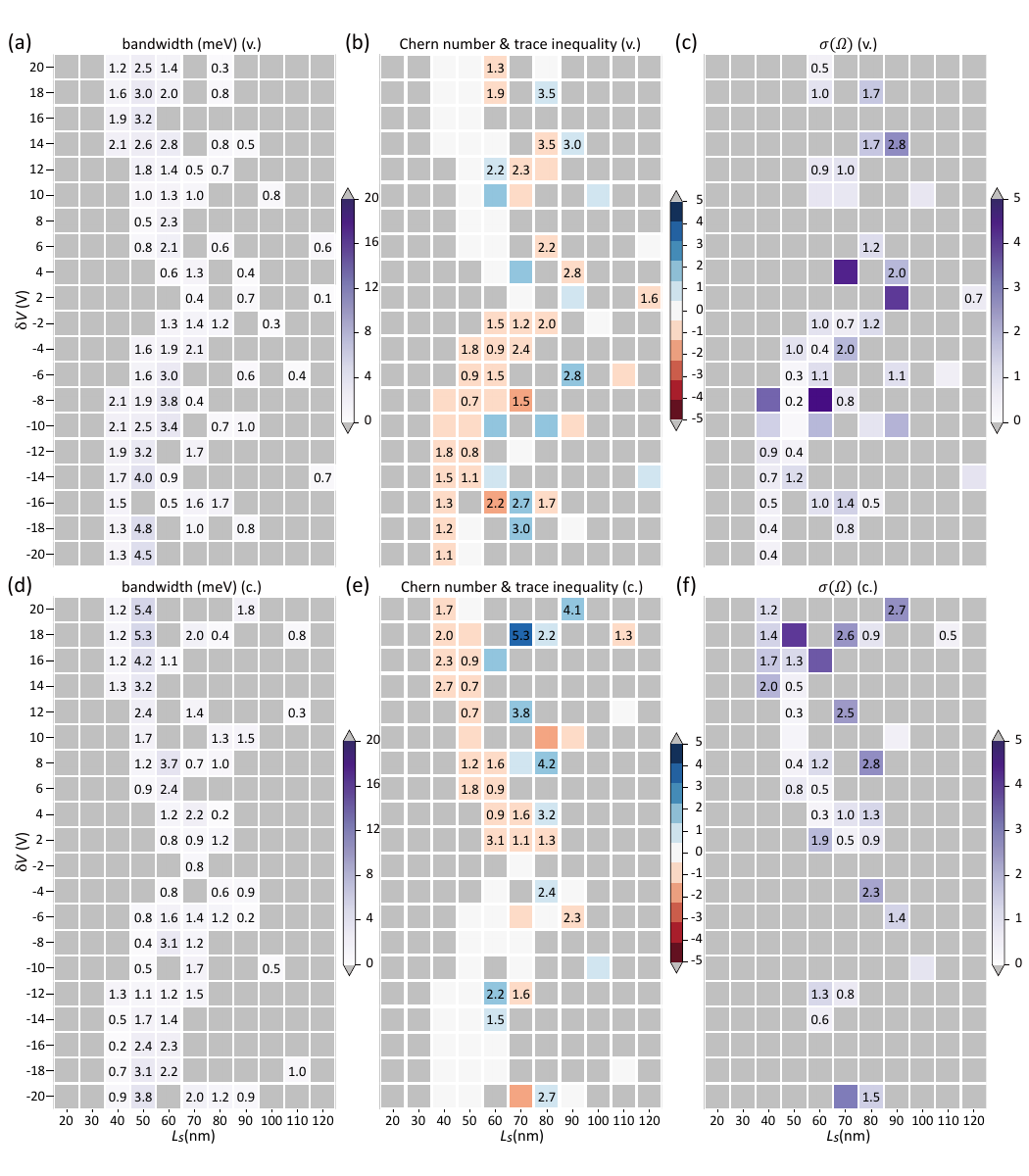}
\caption{Single particle phase diagrams of the third HVB and the third LCB for trilayer graphene system modulated by kagome potential. (a) Bandwidth, (b) Chern number and (c) normalized berry curvature standard deviation $\sigma(\Omega)$ of the third HVB in parameter space. The numbers in (a), (b) and (c) denotes the values of bandwidth, trace inequality and $\sigma(\Omega)$, respectively. (d)-(f) are phase diagrams for the third LCB.
}
\label{SIE3}
\end{figure}

\begin{figure}[bth!]
    \includegraphics[width=5in]{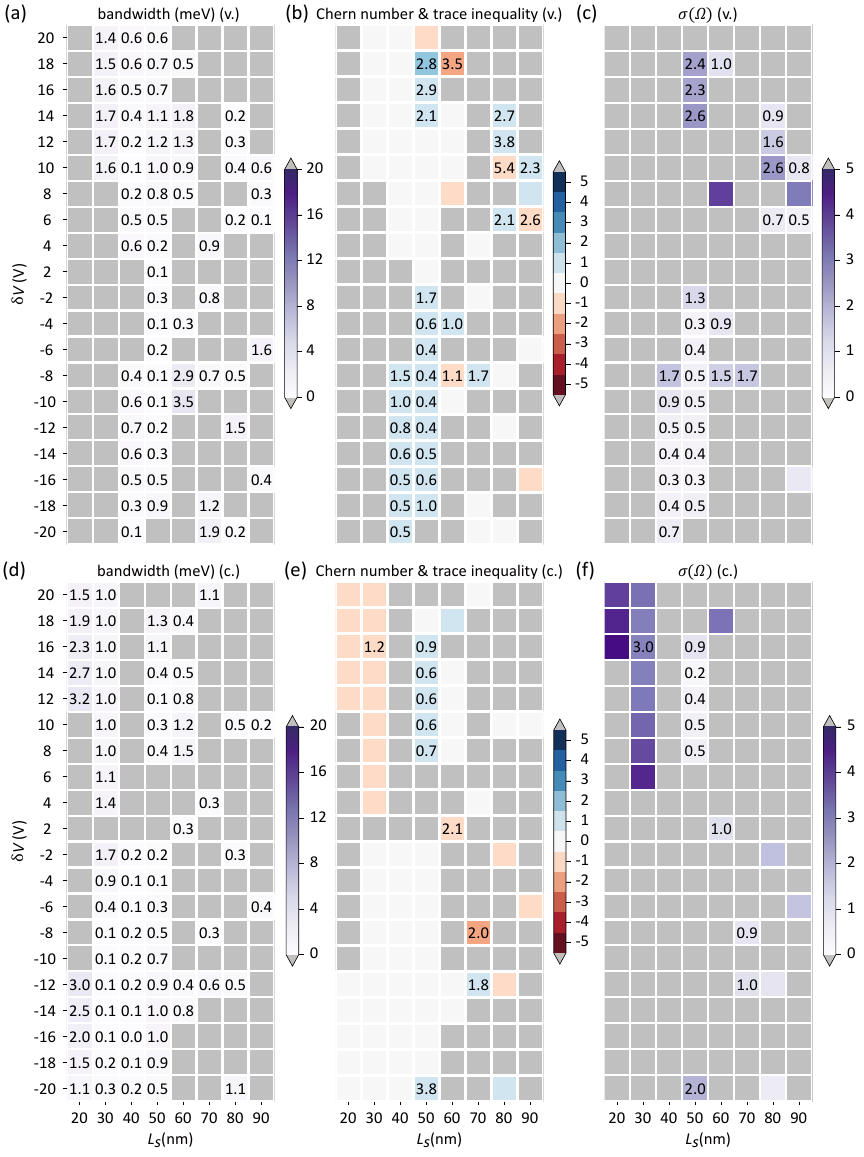}
\caption{Single particle phase diagrams of the HVB and the LCB for  tetralayer graphene system modulated by kagome potential ignoring screening effect. (a) Bandwidth, (b) Chern number and (c) normalized berry curvature standard deviation $\sigma(\Omega)$ of the HVB in parameter space. The numbers in (a), (b) and (c) denotes the values of bandwidth, trace inequality and $\sigma(\Omega)$, respectively. (d)-(f) are phase diagrams for the LCB.
}
\label{SIF1}
\end{figure}

\begin{figure}[bth!]
    \includegraphics[width=5in]{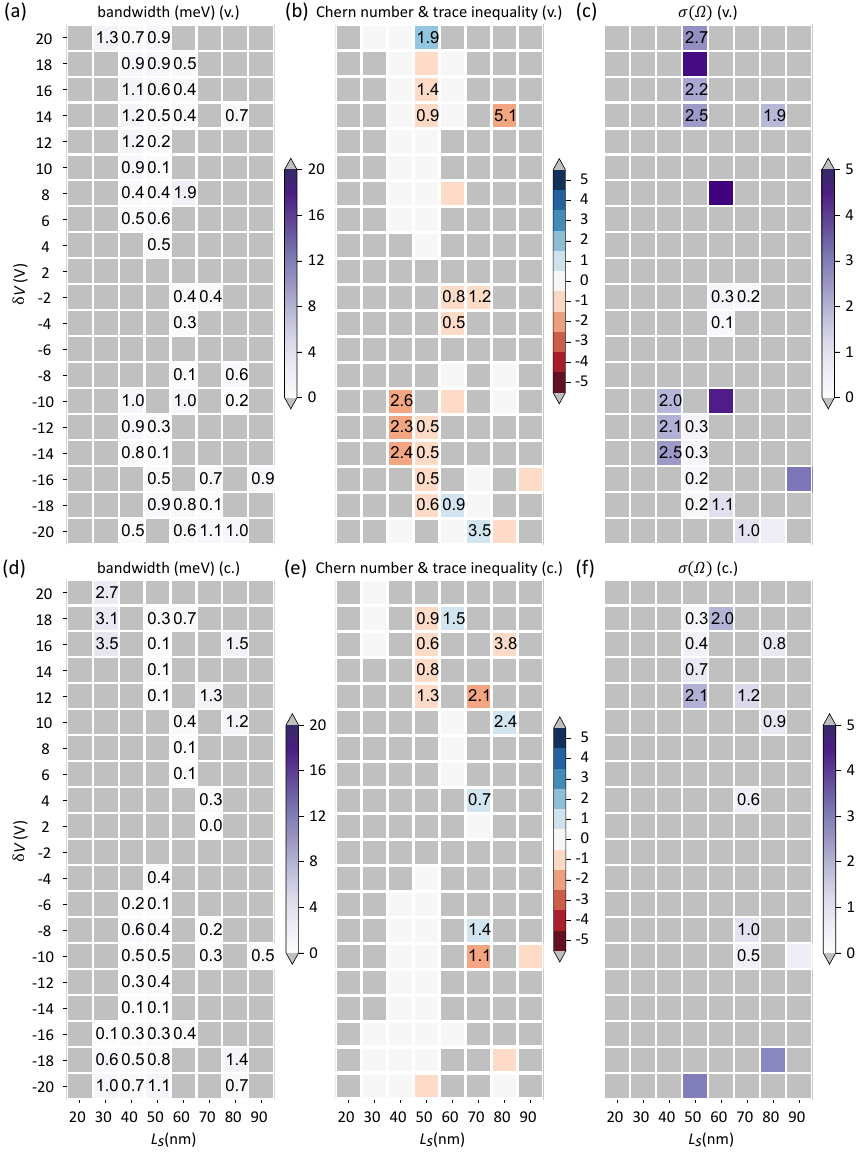}
\caption{Single particle phase diagrams of the second HVB and the second LCB for tetralayer graphene system modulated by kagome potential. (a) Bandwidth, (b) Chern number and (c) normalized berry curvature standard deviation $\sigma(\Omega)$ of the second HVB in parameter space. The numbers in (a), (b) and (c) denotes the values of bandwidth, trace inequality and $\sigma(\Omega)$, respectively. (d)-(f) are phase diagrams for the second LCB.
}
\label{SIF2}
\end{figure}

In this section, we provide results of single particle properties for rhombohedral trilayer graphene  in \figurename \ref{SIE1}-\ref{SIE3} and rhombohedral tetralayer graphene  in \figurename \ref{SIF1}-\ref{SIF2}, both are coupled with a  kagome-patterned superlattice potential. Similar to bilayer graphene system discussed in the main text and above, there exist regions that possess nearly ideal topological flat bands in the trilayer and tetralayer graphene systems. However, the regions are smaller than that in the bilayer graphene system. This is because more subbands are generated during the process of folding original bands into moir\'e Brillouin Zone and energy band structure evolves more drastically under the influence of the superlattice potential.   More interestingly, in trilayer and tetralayer systems, we can obtain diverse topological flat bands with high Chern numbers.   

\clearpage

\bibliography{reference}

\begin{thebibliography}{87}%
\makeatletter
\providecommand \@ifxundefined [1]{%
 \@ifx{#1\undefined}
}%
\providecommand \@ifnum [1]{%
 \ifnum #1\expandafter \@firstoftwo
 \else \expandafter \@secondoftwo
 \fi
}%
\providecommand \@ifx [1]{%
 \ifx #1\expandafter \@firstoftwo
 \else \expandafter \@secondoftwo
 \fi
}%
\providecommand \natexlab [1]{#1}%
\providecommand \enquote  [1]{``#1''}%
\providecommand \bibnamefont  [1]{#1}%
\providecommand \bibfnamefont [1]{#1}%
\providecommand \citenamefont [1]{#1}%
\providecommand \href@noop [0]{\@secondoftwo}%
\providecommand \href [0]{\begingroup \@sanitize@url \@href}%
\providecommand \@href[1]{\@@startlink{#1}\@@href}%
\providecommand \@@href[1]{\endgroup#1\@@endlink}%
\providecommand \@sanitize@url [0]{\catcode `\\12\catcode `\$12\catcode
  `\&12\catcode `\#12\catcode `\^12\catcode `\_12\catcode `\%12\relax}%
\providecommand \@@startlink[1]{}%
\providecommand \@@endlink[0]{}%
\providecommand \url  [0]{\begingroup\@sanitize@url \@url }%
\providecommand \@url [1]{\endgroup\@href {#1}{\urlprefix }}%
\providecommand \urlprefix  [0]{URL }%
\providecommand \Eprint [0]{\href }%
\providecommand \doibase [0]{https://doi.org/}%
\providecommand \selectlanguage [0]{\@gobble}%
\providecommand \bibinfo  [0]{\@secondoftwo}%
\providecommand \bibfield  [0]{\@secondoftwo}%
\providecommand \translation [1]{[#1]}%
\providecommand \BibitemOpen [0]{}%
\providecommand \bibitemStop [0]{}%
\providecommand \bibitemNoStop [0]{.\EOS\space}%
\providecommand \EOS [0]{\spacefactor3000\relax}%
\providecommand \BibitemShut  [1]{\csname bibitem#1\endcsname}%
\let\auto@bib@innerbib\@empty
\bibitem [{\citenamefont {Park}\ \emph {et~al.}(2023)\citenamefont {Park},
  \citenamefont {Cai}, \citenamefont {Anderson}, \citenamefont {Zhang},
  \citenamefont {Zhu}, \citenamefont {Liu}, \citenamefont {Wang}, \citenamefont
  {Holtzmann}, \citenamefont {Hu}, \citenamefont {Liu}, \citenamefont
  {Taniguchi}, \citenamefont {Watanabe}, \citenamefont {Chu}, \citenamefont
  {Cao}, \citenamefont {Fu}, \citenamefont {Yao}, \citenamefont {Chang},
  \citenamefont {Cobden}, \citenamefont {Xiao},\ and\ \citenamefont
  {Xu}}]{fqah-nature23}%
  \BibitemOpen
  \bibfield  {author} {\bibinfo {author} {\bibfnamefont {H.}~\bibnamefont
  {Park}}, \bibinfo {author} {\bibfnamefont {J.}~\bibnamefont {Cai}}, \bibinfo
  {author} {\bibfnamefont {E.}~\bibnamefont {Anderson}}, \bibinfo {author}
  {\bibfnamefont {Y.}~\bibnamefont {Zhang}}, \bibinfo {author} {\bibfnamefont
  {J.}~\bibnamefont {Zhu}}, \bibinfo {author} {\bibfnamefont {X.}~\bibnamefont
  {Liu}}, \bibinfo {author} {\bibfnamefont {C.}~\bibnamefont {Wang}}, \bibinfo
  {author} {\bibfnamefont {W.}~\bibnamefont {Holtzmann}}, \bibinfo {author}
  {\bibfnamefont {C.}~\bibnamefont {Hu}}, \bibinfo {author} {\bibfnamefont
  {Z.}~\bibnamefont {Liu}}, \bibinfo {author} {\bibfnamefont {T.}~\bibnamefont
  {Taniguchi}}, \bibinfo {author} {\bibfnamefont {K.}~\bibnamefont {Watanabe}},
  \bibinfo {author} {\bibfnamefont {J.-H.}\ \bibnamefont {Chu}}, \bibinfo
  {author} {\bibfnamefont {T.}~\bibnamefont {Cao}}, \bibinfo {author}
  {\bibfnamefont {L.}~\bibnamefont {Fu}}, \bibinfo {author} {\bibfnamefont
  {W.}~\bibnamefont {Yao}}, \bibinfo {author} {\bibfnamefont {C.-Z.}\
  \bibnamefont {Chang}}, \bibinfo {author} {\bibfnamefont {D.}~\bibnamefont
  {Cobden}}, \bibinfo {author} {\bibfnamefont {D.}~\bibnamefont {Xiao}},\ and\
  \bibinfo {author} {\bibfnamefont {X.}~\bibnamefont {Xu}},\ }\href
  {https://doi.org/10.1038/s41586-023-06536-0} {\bibfield  {journal} {\bibinfo
  {journal} {Nature}\ }\textbf {\bibinfo {volume} {622}},\ \bibinfo {pages}
  {74} (\bibinfo {year} {2023})}\BibitemShut {NoStop}%
\bibitem [{\citenamefont {Xu}\ \emph {et~al.}(2023{\natexlab{a}})\citenamefont
  {Xu}, \citenamefont {Sun}, \citenamefont {Jia}, \citenamefont {Liu},
  \citenamefont {Xu}, \citenamefont {Li}, \citenamefont {Gu}, \citenamefont
  {Watanabe}, \citenamefont {Taniguchi}, \citenamefont {Tong}, \citenamefont
  {Jia}, \citenamefont {Shi}, \citenamefont {Jiang}, \citenamefont {Zhang},
  \citenamefont {Liu},\ and\ \citenamefont {Li}}]{fqah-prx23}%
  \BibitemOpen
  \bibfield  {author} {\bibinfo {author} {\bibfnamefont {F.}~\bibnamefont
  {Xu}}, \bibinfo {author} {\bibfnamefont {Z.}~\bibnamefont {Sun}}, \bibinfo
  {author} {\bibfnamefont {T.}~\bibnamefont {Jia}}, \bibinfo {author}
  {\bibfnamefont {C.}~\bibnamefont {Liu}}, \bibinfo {author} {\bibfnamefont
  {C.}~\bibnamefont {Xu}}, \bibinfo {author} {\bibfnamefont {C.}~\bibnamefont
  {Li}}, \bibinfo {author} {\bibfnamefont {Y.}~\bibnamefont {Gu}}, \bibinfo
  {author} {\bibfnamefont {K.}~\bibnamefont {Watanabe}}, \bibinfo {author}
  {\bibfnamefont {T.}~\bibnamefont {Taniguchi}}, \bibinfo {author}
  {\bibfnamefont {B.}~\bibnamefont {Tong}}, \bibinfo {author} {\bibfnamefont
  {J.}~\bibnamefont {Jia}}, \bibinfo {author} {\bibfnamefont {Z.}~\bibnamefont
  {Shi}}, \bibinfo {author} {\bibfnamefont {S.}~\bibnamefont {Jiang}}, \bibinfo
  {author} {\bibfnamefont {Y.}~\bibnamefont {Zhang}}, \bibinfo {author}
  {\bibfnamefont {X.}~\bibnamefont {Liu}},\ and\ \bibinfo {author}
  {\bibfnamefont {T.}~\bibnamefont {Li}},\ }\href
  {https://doi.org/10.1103/PhysRevX.13.031037} {\bibfield  {journal} {\bibinfo
  {journal} {Phys. Rev. X}\ }\textbf {\bibinfo {volume} {13}},\ \bibinfo
  {pages} {031037} (\bibinfo {year} {2023}{\natexlab{a}})}\BibitemShut
  {NoStop}%
\bibitem [{\citenamefont {Cai}\ \emph {et~al.}(2023)\citenamefont {Cai},
  \citenamefont {Anderson}, \citenamefont {Wang}, \citenamefont {Zhang},
  \citenamefont {Liu}, \citenamefont {Holtzmann}, \citenamefont {Zhang},
  \citenamefont {Fan}, \citenamefont {Taniguchi}, \citenamefont {Watanabe},
  \citenamefont {Ran}, \citenamefont {Cao}, \citenamefont {Fu}, \citenamefont
  {Xiao}, \citenamefont {Yao},\ and\ \citenamefont
  {Xu}}]{fqah-optics-xu-nature23}%
  \BibitemOpen
  \bibfield  {author} {\bibinfo {author} {\bibfnamefont {J.}~\bibnamefont
  {Cai}}, \bibinfo {author} {\bibfnamefont {E.}~\bibnamefont {Anderson}},
  \bibinfo {author} {\bibfnamefont {C.}~\bibnamefont {Wang}}, \bibinfo {author}
  {\bibfnamefont {X.}~\bibnamefont {Zhang}}, \bibinfo {author} {\bibfnamefont
  {X.}~\bibnamefont {Liu}}, \bibinfo {author} {\bibfnamefont {W.}~\bibnamefont
  {Holtzmann}}, \bibinfo {author} {\bibfnamefont {Y.}~\bibnamefont {Zhang}},
  \bibinfo {author} {\bibfnamefont {F.}~\bibnamefont {Fan}}, \bibinfo {author}
  {\bibfnamefont {T.}~\bibnamefont {Taniguchi}}, \bibinfo {author}
  {\bibfnamefont {K.}~\bibnamefont {Watanabe}}, \bibinfo {author}
  {\bibfnamefont {Y.}~\bibnamefont {Ran}}, \bibinfo {author} {\bibfnamefont
  {T.}~\bibnamefont {Cao}}, \bibinfo {author} {\bibfnamefont {L.}~\bibnamefont
  {Fu}}, \bibinfo {author} {\bibfnamefont {D.}~\bibnamefont {Xiao}}, \bibinfo
  {author} {\bibfnamefont {W.}~\bibnamefont {Yao}},\ and\ \bibinfo {author}
  {\bibfnamefont {X.}~\bibnamefont {Xu}},\ }\href
  {https://doi.org/10.1038/s41586-023-06289-w} {\bibfield  {journal} {\bibinfo
  {journal} {Nature}\ }\textbf {\bibinfo {volume} {622}},\ \bibinfo {pages}
  {63} (\bibinfo {year} {2023})}\BibitemShut {NoStop}%
\bibitem [{\citenamefont {Zeng}\ \emph {et~al.}(2023)\citenamefont {Zeng},
  \citenamefont {Xia}, \citenamefont {Kang}, \citenamefont {Zhu}, \citenamefont
  {Kn{\"u}ppel}, \citenamefont {Vaswani}, \citenamefont {Watanabe},
  \citenamefont {Taniguchi}, \citenamefont {Mak},\ and\ \citenamefont
  {Shan}}]{fqah-mak-nature23}%
  \BibitemOpen
  \bibfield  {author} {\bibinfo {author} {\bibfnamefont {Y.}~\bibnamefont
  {Zeng}}, \bibinfo {author} {\bibfnamefont {Z.}~\bibnamefont {Xia}}, \bibinfo
  {author} {\bibfnamefont {K.}~\bibnamefont {Kang}}, \bibinfo {author}
  {\bibfnamefont {J.}~\bibnamefont {Zhu}}, \bibinfo {author} {\bibfnamefont
  {P.}~\bibnamefont {Kn{\"u}ppel}}, \bibinfo {author} {\bibfnamefont
  {C.}~\bibnamefont {Vaswani}}, \bibinfo {author} {\bibfnamefont
  {K.}~\bibnamefont {Watanabe}}, \bibinfo {author} {\bibfnamefont
  {T.}~\bibnamefont {Taniguchi}}, \bibinfo {author} {\bibfnamefont {K.~F.}\
  \bibnamefont {Mak}},\ and\ \bibinfo {author} {\bibfnamefont {J.}~\bibnamefont
  {Shan}},\ }\href {https://doi.org/10.1038/s41586-023-06452-3} {\bibfield
  {journal} {\bibinfo  {journal} {Nature}\ }\textbf {\bibinfo {volume} {622}},\
  \bibinfo {pages} {69} (\bibinfo {year} {2023})}\BibitemShut {NoStop}%
\bibitem [{\citenamefont {Lu}\ \emph {et~al.}(2024)\citenamefont {Lu},
  \citenamefont {Han}, \citenamefont {Yao}, \citenamefont {Reddy},
  \citenamefont {Yang}, \citenamefont {Seo}, \citenamefont {Watanabe},
  \citenamefont {Taniguchi}, \citenamefont {Fu},\ and\ \citenamefont
  {Ju}}]{fqah-ju-nature24}%
  \BibitemOpen
  \bibfield  {author} {\bibinfo {author} {\bibfnamefont {Z.}~\bibnamefont
  {Lu}}, \bibinfo {author} {\bibfnamefont {T.}~\bibnamefont {Han}}, \bibinfo
  {author} {\bibfnamefont {Y.}~\bibnamefont {Yao}}, \bibinfo {author}
  {\bibfnamefont {A.~P.}\ \bibnamefont {Reddy}}, \bibinfo {author}
  {\bibfnamefont {J.}~\bibnamefont {Yang}}, \bibinfo {author} {\bibfnamefont
  {J.}~\bibnamefont {Seo}}, \bibinfo {author} {\bibfnamefont {K.}~\bibnamefont
  {Watanabe}}, \bibinfo {author} {\bibfnamefont {T.}~\bibnamefont {Taniguchi}},
  \bibinfo {author} {\bibfnamefont {L.}~\bibnamefont {Fu}},\ and\ \bibinfo
  {author} {\bibfnamefont {L.}~\bibnamefont {Ju}},\ }\href
  {https://doi.org/10.1038/s41586-023-07010-7} {\bibfield  {journal} {\bibinfo
  {journal} {Nature}\ }\textbf {\bibinfo {volume} {626}},\ \bibinfo {pages}
  {759} (\bibinfo {year} {2024})}\BibitemShut {NoStop}%
\bibitem [{\citenamefont {Xie}\ \emph {et~al.}(2024)\citenamefont {Xie},
  \citenamefont {Huo}, \citenamefont {Lu}, \citenamefont {Feng}, \citenamefont
  {Zhang}, \citenamefont {Wang}, \citenamefont {Yang}, \citenamefont
  {Watanabe}, \citenamefont {Taniguchi}, \citenamefont {Liu}, \citenamefont
  {Song}, \citenamefont {Xie}, \citenamefont {Liu},\ and\ \citenamefont
  {Lu}}]{lu-hexlayer-arxiv24}%
  \BibitemOpen
  \bibfield  {author} {\bibinfo {author} {\bibfnamefont {J.}~\bibnamefont
  {Xie}}, \bibinfo {author} {\bibfnamefont {Z.}~\bibnamefont {Huo}}, \bibinfo
  {author} {\bibfnamefont {X.}~\bibnamefont {Lu}}, \bibinfo {author}
  {\bibfnamefont {Z.}~\bibnamefont {Feng}}, \bibinfo {author} {\bibfnamefont
  {Z.}~\bibnamefont {Zhang}}, \bibinfo {author} {\bibfnamefont
  {W.}~\bibnamefont {Wang}}, \bibinfo {author} {\bibfnamefont {Q.}~\bibnamefont
  {Yang}}, \bibinfo {author} {\bibfnamefont {K.}~\bibnamefont {Watanabe}},
  \bibinfo {author} {\bibfnamefont {T.}~\bibnamefont {Taniguchi}}, \bibinfo
  {author} {\bibfnamefont {K.}~\bibnamefont {Liu}}, \bibinfo {author}
  {\bibfnamefont {Z.}~\bibnamefont {Song}}, \bibinfo {author} {\bibfnamefont
  {X.~C.}\ \bibnamefont {Xie}}, \bibinfo {author} {\bibfnamefont
  {J.}~\bibnamefont {Liu}},\ and\ \bibinfo {author} {\bibfnamefont
  {X.}~\bibnamefont {Lu}},\ }\href@noop {} {\bibinfo {title} {Even- and
  odd-denominator fractional quantum anomalous hall effect in graphene moire
  superlattices}} (\bibinfo {year} {2024}),\ \Eprint
  {https://arxiv.org/abs/2405.16944} {arXiv:2405.16944 [cond-mat.mes-hall]}
  \BibitemShut {NoStop}%
\bibitem [{\citenamefont {Lu}\ \emph {et~al.}(2025)\citenamefont {Lu},
  \citenamefont {Han}, \citenamefont {Yao}, \citenamefont {Hadjri},
  \citenamefont {Yang}, \citenamefont {Seo}, \citenamefont {Shi}, \citenamefont
  {Ye}, \citenamefont {Watanabe}, \citenamefont {Taniguchi},\ and\
  \citenamefont {Ju}}]{ju-eqah-nature25}%
  \BibitemOpen
  \bibfield  {author} {\bibinfo {author} {\bibfnamefont {Z.}~\bibnamefont
  {Lu}}, \bibinfo {author} {\bibfnamefont {T.}~\bibnamefont {Han}}, \bibinfo
  {author} {\bibfnamefont {Y.}~\bibnamefont {Yao}}, \bibinfo {author}
  {\bibfnamefont {Z.}~\bibnamefont {Hadjri}}, \bibinfo {author} {\bibfnamefont
  {J.}~\bibnamefont {Yang}}, \bibinfo {author} {\bibfnamefont {J.}~\bibnamefont
  {Seo}}, \bibinfo {author} {\bibfnamefont {L.}~\bibnamefont {Shi}}, \bibinfo
  {author} {\bibfnamefont {S.}~\bibnamefont {Ye}}, \bibinfo {author}
  {\bibfnamefont {K.}~\bibnamefont {Watanabe}}, \bibinfo {author}
  {\bibfnamefont {T.}~\bibnamefont {Taniguchi}},\ and\ \bibinfo {author}
  {\bibfnamefont {L.}~\bibnamefont {Ju}},\ }\bibfield  {journal} {\bibinfo
  {journal} {Nature}\ }\href {https://doi.org/10.1038/s41586-024-08470-1}
  {10.1038/s41586-024-08470-1} (\bibinfo {year} {2025})\BibitemShut {NoStop}%
\bibitem [{\citenamefont {Regnault}\ and\ \citenamefont
  {Bernevig}(2011)}]{fci-prx11}%
  \BibitemOpen
  \bibfield  {author} {\bibinfo {author} {\bibfnamefont {N.}~\bibnamefont
  {Regnault}}\ and\ \bibinfo {author} {\bibfnamefont {B.~A.}\ \bibnamefont
  {Bernevig}},\ }\href {https://doi.org/10.1103/PhysRevX.1.021014} {\bibfield
  {journal} {\bibinfo  {journal} {Phys. Rev. X}\ }\textbf {\bibinfo {volume}
  {1}},\ \bibinfo {pages} {021014} (\bibinfo {year} {2011})}\BibitemShut
  {NoStop}%
\bibitem [{\citenamefont {Sheng}\ \emph {et~al.}(2011)\citenamefont {Sheng},
  \citenamefont {Gu}, \citenamefont {Sun},\ and\ \citenamefont
  {Sheng}}]{sheng-fci-nc11}%
  \BibitemOpen
  \bibfield  {author} {\bibinfo {author} {\bibfnamefont {D.~N.}\ \bibnamefont
  {Sheng}}, \bibinfo {author} {\bibfnamefont {Z.-C.}\ \bibnamefont {Gu}},
  \bibinfo {author} {\bibfnamefont {K.}~\bibnamefont {Sun}},\ and\ \bibinfo
  {author} {\bibfnamefont {L.}~\bibnamefont {Sheng}},\ }\href
  {https://doi.org/10.1038/ncomms1380} {\bibfield  {journal} {\bibinfo
  {journal} {Nature Communications}\ }\textbf {\bibinfo {volume} {2}},\
  \bibinfo {pages} {389} (\bibinfo {year} {2011})}\BibitemShut {NoStop}%
\bibitem [{\citenamefont {Neupert}\ \emph {et~al.}(2011)\citenamefont
  {Neupert}, \citenamefont {Santos}, \citenamefont {Chamon},\ and\
  \citenamefont {Mudry}}]{murdy-fci-prl11}%
  \BibitemOpen
  \bibfield  {author} {\bibinfo {author} {\bibfnamefont {T.}~\bibnamefont
  {Neupert}}, \bibinfo {author} {\bibfnamefont {L.}~\bibnamefont {Santos}},
  \bibinfo {author} {\bibfnamefont {C.}~\bibnamefont {Chamon}},\ and\ \bibinfo
  {author} {\bibfnamefont {C.}~\bibnamefont {Mudry}},\ }\href
  {https://doi.org/10.1103/PhysRevLett.106.236804} {\bibfield  {journal}
  {\bibinfo  {journal} {Phys. Rev. Lett.}\ }\textbf {\bibinfo {volume} {106}},\
  \bibinfo {pages} {236804} (\bibinfo {year} {2011})}\BibitemShut {NoStop}%
\bibitem [{\citenamefont {Tang}\ \emph {et~al.}(2011)\citenamefont {Tang},
  \citenamefont {Mei},\ and\ \citenamefont {Wen}}]{wen-kagome-prl11}%
  \BibitemOpen
  \bibfield  {author} {\bibinfo {author} {\bibfnamefont {E.}~\bibnamefont
  {Tang}}, \bibinfo {author} {\bibfnamefont {J.-W.}\ \bibnamefont {Mei}},\ and\
  \bibinfo {author} {\bibfnamefont {X.-G.}\ \bibnamefont {Wen}},\ }\href
  {https://doi.org/10.1103/PhysRevLett.106.236802} {\bibfield  {journal}
  {\bibinfo  {journal} {Phys. Rev. Lett.}\ }\textbf {\bibinfo {volume} {106}},\
  \bibinfo {pages} {236802} (\bibinfo {year} {2011})}\BibitemShut {NoStop}%
\bibitem [{\citenamefont {Sun}\ \emph {et~al.}(2011)\citenamefont {Sun},
  \citenamefont {Gu}, \citenamefont {Katsura},\ and\ \citenamefont
  {Das~Sarma}}]{sarma-flatchern-prl11}%
  \BibitemOpen
  \bibfield  {author} {\bibinfo {author} {\bibfnamefont {K.}~\bibnamefont
  {Sun}}, \bibinfo {author} {\bibfnamefont {Z.}~\bibnamefont {Gu}}, \bibinfo
  {author} {\bibfnamefont {H.}~\bibnamefont {Katsura}},\ and\ \bibinfo {author}
  {\bibfnamefont {S.}~\bibnamefont {Das~Sarma}},\ }\href
  {https://doi.org/10.1103/PhysRevLett.106.236803} {\bibfield  {journal}
  {\bibinfo  {journal} {Phys. Rev. Lett.}\ }\textbf {\bibinfo {volume} {106}},\
  \bibinfo {pages} {236803} (\bibinfo {year} {2011})}\BibitemShut {NoStop}%
\bibitem [{\citenamefont {M\"oller}\ and\ \citenamefont
  {Cooper}(2009)}]{cooper-fci-prl09}%
  \BibitemOpen
  \bibfield  {author} {\bibinfo {author} {\bibfnamefont {G.}~\bibnamefont
  {M\"oller}}\ and\ \bibinfo {author} {\bibfnamefont {N.~R.}\ \bibnamefont
  {Cooper}},\ }\href {https://doi.org/10.1103/PhysRevLett.103.105303}
  {\bibfield  {journal} {\bibinfo  {journal} {Phys. Rev. Lett.}\ }\textbf
  {\bibinfo {volume} {103}},\ \bibinfo {pages} {105303} (\bibinfo {year}
  {2009})}\BibitemShut {NoStop}%
\bibitem [{\citenamefont {Qi}(2011)}]{qi-fqah-prl11}%
  \BibitemOpen
  \bibfield  {author} {\bibinfo {author} {\bibfnamefont {X.-L.}\ \bibnamefont
  {Qi}},\ }\href {https://doi.org/10.1103/PhysRevLett.107.126803} {\bibfield
  {journal} {\bibinfo  {journal} {Phys. Rev. Lett.}\ }\textbf {\bibinfo
  {volume} {107}},\ \bibinfo {pages} {126803} (\bibinfo {year}
  {2011})}\BibitemShut {NoStop}%
\bibitem [{\citenamefont {Tsui}\ \emph {et~al.}(1982)\citenamefont {Tsui},
  \citenamefont {Stormer},\ and\ \citenamefont {Gossard}}]{fqhe-prl82}%
  \BibitemOpen
  \bibfield  {author} {\bibinfo {author} {\bibfnamefont {D.~C.}\ \bibnamefont
  {Tsui}}, \bibinfo {author} {\bibfnamefont {H.~L.}\ \bibnamefont {Stormer}},\
  and\ \bibinfo {author} {\bibfnamefont {A.~C.}\ \bibnamefont {Gossard}},\
  }\href {https://doi.org/10.1103/PhysRevLett.48.1559} {\bibfield  {journal}
  {\bibinfo  {journal} {Phys. Rev. Lett.}\ }\textbf {\bibinfo {volume} {48}},\
  \bibinfo {pages} {1559} (\bibinfo {year} {1982})}\BibitemShut {NoStop}%
\bibitem [{\citenamefont {Laughlin}(1983)}]{laughlin-prl83}%
  \BibitemOpen
  \bibfield  {author} {\bibinfo {author} {\bibfnamefont {R.~B.}\ \bibnamefont
  {Laughlin}},\ }\href@noop {} {\bibfield  {journal} {\bibinfo  {journal}
  {Physical Review Letters}\ }\textbf {\bibinfo {volume} {50}},\ \bibinfo
  {pages} {1395} (\bibinfo {year} {1983})}\BibitemShut {NoStop}%
\bibitem [{\citenamefont {Jain}(1989{\natexlab{a}})}]{jain-prl89}%
  \BibitemOpen
  \bibfield  {author} {\bibinfo {author} {\bibfnamefont {J.~K.}\ \bibnamefont
  {Jain}},\ }\href@noop {} {\bibfield  {journal} {\bibinfo  {journal} {Physical
  review letters}\ }\textbf {\bibinfo {volume} {63}},\ \bibinfo {pages} {199}
  (\bibinfo {year} {1989}{\natexlab{a}})}\BibitemShut {NoStop}%
\bibitem [{\citenamefont {Moore}\ and\ \citenamefont
  {Read}(1991)}]{moore-read}%
  \BibitemOpen
  \bibfield  {author} {\bibinfo {author} {\bibfnamefont {G.}~\bibnamefont
  {Moore}}\ and\ \bibinfo {author} {\bibfnamefont {N.}~\bibnamefont {Read}},\
  }\href@noop {} {\bibfield  {journal} {\bibinfo  {journal} {Nuclear Physics
  B}\ }\textbf {\bibinfo {volume} {360}},\ \bibinfo {pages} {362} (\bibinfo
  {year} {1991})}\BibitemShut {NoStop}%
\bibitem [{\citenamefont {Stormer}\ \emph {et~al.}(1999)\citenamefont
  {Stormer}, \citenamefont {Tsui},\ and\ \citenamefont {Gossard}}]{fqhe-rmp99}%
  \BibitemOpen
  \bibfield  {author} {\bibinfo {author} {\bibfnamefont {H.~L.}\ \bibnamefont
  {Stormer}}, \bibinfo {author} {\bibfnamefont {D.~C.}\ \bibnamefont {Tsui}},\
  and\ \bibinfo {author} {\bibfnamefont {A.~C.}\ \bibnamefont {Gossard}},\
  }\href@noop {} {\bibfield  {journal} {\bibinfo  {journal} {Reviews of Modern
  Physics}\ }\textbf {\bibinfo {volume} {71}},\ \bibinfo {pages} {S298}
  (\bibinfo {year} {1999})}\BibitemShut {NoStop}%
\bibitem [{\citenamefont {Cage}\ \emph {et~al.}(2012)\citenamefont {Cage},
  \citenamefont {Klitzing}, \citenamefont {Chang}, \citenamefont {Duncan},
  \citenamefont {Haldane}, \citenamefont {Laughlin}, \citenamefont {Pruisken},\
  and\ \citenamefont {Thouless}}]{qhe-book-2012}%
  \BibitemOpen
  \bibfield  {author} {\bibinfo {author} {\bibfnamefont {M.~E.}\ \bibnamefont
  {Cage}}, \bibinfo {author} {\bibfnamefont {K.}~\bibnamefont {Klitzing}},
  \bibinfo {author} {\bibfnamefont {A.}~\bibnamefont {Chang}}, \bibinfo
  {author} {\bibfnamefont {F.}~\bibnamefont {Duncan}}, \bibinfo {author}
  {\bibfnamefont {M.}~\bibnamefont {Haldane}}, \bibinfo {author} {\bibfnamefont
  {R.~B.}\ \bibnamefont {Laughlin}}, \bibinfo {author} {\bibfnamefont
  {A.}~\bibnamefont {Pruisken}},\ and\ \bibinfo {author} {\bibfnamefont
  {D.}~\bibnamefont {Thouless}},\ }\href@noop {} {\emph {\bibinfo {title} {The
  quantum Hall effect}}}\ (\bibinfo  {publisher} {Springer Science \& Business
  Media},\ \bibinfo {year} {2012})\BibitemShut {NoStop}%
\bibitem [{\citenamefont {Wang}\ \emph {et~al.}(2011)\citenamefont {Wang},
  \citenamefont {Gu}, \citenamefont {Gong},\ and\ \citenamefont
  {Sheng}}]{fqh-boson-prl11}%
  \BibitemOpen
  \bibfield  {author} {\bibinfo {author} {\bibfnamefont {Y.-F.}\ \bibnamefont
  {Wang}}, \bibinfo {author} {\bibfnamefont {Z.-C.}\ \bibnamefont {Gu}},
  \bibinfo {author} {\bibfnamefont {C.-D.}\ \bibnamefont {Gong}},\ and\
  \bibinfo {author} {\bibfnamefont {D.~N.}\ \bibnamefont {Sheng}},\ }\href
  {https://doi.org/10.1103/PhysRevLett.107.146803} {\bibfield  {journal}
  {\bibinfo  {journal} {Phys. Rev. Lett.}\ }\textbf {\bibinfo {volume} {107}},\
  \bibinfo {pages} {146803} (\bibinfo {year} {2011})}\BibitemShut {NoStop}%
\bibitem [{\citenamefont {Wu}\ \emph {et~al.}(2012{\natexlab{a}})\citenamefont
  {Wu}, \citenamefont {Bernevig},\ and\ \citenamefont
  {Regnault}}]{fci-zoo-prb12}%
  \BibitemOpen
  \bibfield  {author} {\bibinfo {author} {\bibfnamefont {Y.-L.}\ \bibnamefont
  {Wu}}, \bibinfo {author} {\bibfnamefont {B.~A.}\ \bibnamefont {Bernevig}},\
  and\ \bibinfo {author} {\bibfnamefont {N.}~\bibnamefont {Regnault}},\ }\href
  {https://doi.org/10.1103/PhysRevB.85.075116} {\bibfield  {journal} {\bibinfo
  {journal} {Phys. Rev. B}\ }\textbf {\bibinfo {volume} {85}},\ \bibinfo
  {pages} {075116} (\bibinfo {year} {2012}{\natexlab{a}})}\BibitemShut
  {NoStop}%
\bibitem [{\citenamefont {Liu}\ \emph {et~al.}(2012)\citenamefont {Liu},
  \citenamefont {Bergholtz}, \citenamefont {Fan},\ and\ \citenamefont
  {L\"auchli}}]{liuzhao-fci-prl12}%
  \BibitemOpen
  \bibfield  {author} {\bibinfo {author} {\bibfnamefont {Z.}~\bibnamefont
  {Liu}}, \bibinfo {author} {\bibfnamefont {E.~J.}\ \bibnamefont {Bergholtz}},
  \bibinfo {author} {\bibfnamefont {H.}~\bibnamefont {Fan}},\ and\ \bibinfo
  {author} {\bibfnamefont {A.~M.}\ \bibnamefont {L\"auchli}},\ }\href
  {https://doi.org/10.1103/PhysRevLett.109.186805} {\bibfield  {journal}
  {\bibinfo  {journal} {Phys. Rev. Lett.}\ }\textbf {\bibinfo {volume} {109}},\
  \bibinfo {pages} {186805} (\bibinfo {year} {2012})}\BibitemShut {NoStop}%
\bibitem [{\citenamefont {Venderbos}\ \emph {et~al.}(2012)\citenamefont
  {Venderbos}, \citenamefont {Kourtis}, \citenamefont {van~den Brink},\ and\
  \citenamefont {Daghofer}}]{vanderbos-fci-prl12}%
  \BibitemOpen
  \bibfield  {author} {\bibinfo {author} {\bibfnamefont {J.~W.~F.}\
  \bibnamefont {Venderbos}}, \bibinfo {author} {\bibfnamefont {S.}~\bibnamefont
  {Kourtis}}, \bibinfo {author} {\bibfnamefont {J.}~\bibnamefont {van~den
  Brink}},\ and\ \bibinfo {author} {\bibfnamefont {M.}~\bibnamefont
  {Daghofer}},\ }\href {https://doi.org/10.1103/PhysRevLett.108.126405}
  {\bibfield  {journal} {\bibinfo  {journal} {Phys. Rev. Lett.}\ }\textbf
  {\bibinfo {volume} {108}},\ \bibinfo {pages} {126405} (\bibinfo {year}
  {2012})}\BibitemShut {NoStop}%
\bibitem [{\citenamefont {Liu}\ \emph {et~al.}(2013)\citenamefont {Liu},
  \citenamefont {Repellin}, \citenamefont {Bernevig},\ and\ \citenamefont
  {Regnault}}]{liu-fci-prb13}%
  \BibitemOpen
  \bibfield  {author} {\bibinfo {author} {\bibfnamefont {T.}~\bibnamefont
  {Liu}}, \bibinfo {author} {\bibfnamefont {C.}~\bibnamefont {Repellin}},
  \bibinfo {author} {\bibfnamefont {B.~A.}\ \bibnamefont {Bernevig}},\ and\
  \bibinfo {author} {\bibfnamefont {N.}~\bibnamefont {Regnault}},\ }\href
  {https://doi.org/10.1103/PhysRevB.87.205136} {\bibfield  {journal} {\bibinfo
  {journal} {Phys. Rev. B}\ }\textbf {\bibinfo {volume} {87}},\ \bibinfo
  {pages} {205136} (\bibinfo {year} {2013})}\BibitemShut {NoStop}%
\bibitem [{\citenamefont {Hu}\ \emph {et~al.}(2011)\citenamefont {Hu},
  \citenamefont {Kargarian},\ and\ \citenamefont {Fiete}}]{fiete-ruby-prb11}%
  \BibitemOpen
  \bibfield  {author} {\bibinfo {author} {\bibfnamefont {X.}~\bibnamefont
  {Hu}}, \bibinfo {author} {\bibfnamefont {M.}~\bibnamefont {Kargarian}},\ and\
  \bibinfo {author} {\bibfnamefont {G.~A.}\ \bibnamefont {Fiete}},\ }\href
  {https://doi.org/10.1103/PhysRevB.84.155116} {\bibfield  {journal} {\bibinfo
  {journal} {Phys. Rev. B}\ }\textbf {\bibinfo {volume} {84}},\ \bibinfo
  {pages} {155116} (\bibinfo {year} {2011})}\BibitemShut {NoStop}%
\bibitem [{\citenamefont {Reddy}\ \emph {et~al.}(2023)\citenamefont {Reddy},
  \citenamefont {Alsallom}, \citenamefont {Zhang}, \citenamefont {Devakul},\
  and\ \citenamefont {Fu}}]{liangfu-fqah-tmd-prb23}%
  \BibitemOpen
  \bibfield  {author} {\bibinfo {author} {\bibfnamefont {A.~P.}\ \bibnamefont
  {Reddy}}, \bibinfo {author} {\bibfnamefont {F.}~\bibnamefont {Alsallom}},
  \bibinfo {author} {\bibfnamefont {Y.}~\bibnamefont {Zhang}}, \bibinfo
  {author} {\bibfnamefont {T.}~\bibnamefont {Devakul}},\ and\ \bibinfo {author}
  {\bibfnamefont {L.}~\bibnamefont {Fu}},\ }\href
  {https://doi.org/10.1103/PhysRevB.108.085117} {\bibfield  {journal} {\bibinfo
   {journal} {Phys. Rev. B}\ }\textbf {\bibinfo {volume} {108}},\ \bibinfo
  {pages} {085117} (\bibinfo {year} {2023})}\BibitemShut {NoStop}%
\bibitem [{\citenamefont {Wang}\ \emph {et~al.}(2023)\citenamefont {Wang},
  \citenamefont {Zhang}, \citenamefont {Liu}, \citenamefont {He}, \citenamefont
  {Xu}, \citenamefont {Ran}, \citenamefont {Cao},\ and\ \citenamefont
  {Xiao}}]{xiao-fqah-arxiv23}%
  \BibitemOpen
  \bibfield  {author} {\bibinfo {author} {\bibfnamefont {C.}~\bibnamefont
  {Wang}}, \bibinfo {author} {\bibfnamefont {X.-W.}\ \bibnamefont {Zhang}},
  \bibinfo {author} {\bibfnamefont {X.}~\bibnamefont {Liu}}, \bibinfo {author}
  {\bibfnamefont {Y.}~\bibnamefont {He}}, \bibinfo {author} {\bibfnamefont
  {X.}~\bibnamefont {Xu}}, \bibinfo {author} {\bibfnamefont {Y.}~\bibnamefont
  {Ran}}, \bibinfo {author} {\bibfnamefont {T.}~\bibnamefont {Cao}},\ and\
  \bibinfo {author} {\bibfnamefont {D.}~\bibnamefont {Xiao}},\ }\href@noop {}
  {\bibinfo {title} {Fractional chern insulator in twisted bilayer mote$_2$}}
  (\bibinfo {year} {2023}),\ \Eprint {https://arxiv.org/abs/2304.11864}
  {arXiv:2304.11864 [cond-mat.str-el]} \BibitemShut {NoStop}%
\bibitem [{\citenamefont {Xu}\ \emph {et~al.}(2023{\natexlab{b}})\citenamefont
  {Xu}, \citenamefont {Li}, \citenamefont {Xu}, \citenamefont {Bi},\ and\
  \citenamefont {Zhang}}]{zhangyang-fqah-tmd-arxiv23}%
  \BibitemOpen
  \bibfield  {author} {\bibinfo {author} {\bibfnamefont {C.}~\bibnamefont
  {Xu}}, \bibinfo {author} {\bibfnamefont {J.}~\bibnamefont {Li}}, \bibinfo
  {author} {\bibfnamefont {Y.}~\bibnamefont {Xu}}, \bibinfo {author}
  {\bibfnamefont {Z.}~\bibnamefont {Bi}},\ and\ \bibinfo {author}
  {\bibfnamefont {Y.}~\bibnamefont {Zhang}},\ }\href@noop {} {\bibinfo {title}
  {Maximally localized wannier orbitals, interaction models and fractional
  quantum anomalous hall effect in twisted bilayer mote2}} (\bibinfo {year}
  {2023}{\natexlab{b}}),\ \Eprint {https://arxiv.org/abs/2308.09697}
  {arXiv:2308.09697 [cond-mat.str-el]} \BibitemShut {NoStop}%
\bibitem [{\citenamefont {Yu}\ \emph {et~al.}(2024{\natexlab{a}})\citenamefont
  {Yu}, \citenamefont {Herzog-Arbeitman}, \citenamefont {Wang}, \citenamefont
  {Vafek}, \citenamefont {Bernevig},\ and\ \citenamefont
  {Regnault}}]{regnault-fci-mote2-prb24}%
  \BibitemOpen
  \bibfield  {author} {\bibinfo {author} {\bibfnamefont {J.}~\bibnamefont
  {Yu}}, \bibinfo {author} {\bibfnamefont {J.}~\bibnamefont
  {Herzog-Arbeitman}}, \bibinfo {author} {\bibfnamefont {M.}~\bibnamefont
  {Wang}}, \bibinfo {author} {\bibfnamefont {O.}~\bibnamefont {Vafek}},
  \bibinfo {author} {\bibfnamefont {B.~A.}\ \bibnamefont {Bernevig}},\ and\
  \bibinfo {author} {\bibfnamefont {N.}~\bibnamefont {Regnault}},\ }\href
  {https://doi.org/10.1103/PhysRevB.109.045147} {\bibfield  {journal} {\bibinfo
   {journal} {Phys. Rev. B}\ }\textbf {\bibinfo {volume} {109}},\ \bibinfo
  {pages} {045147} (\bibinfo {year} {2024}{\natexlab{a}})}\BibitemShut
  {NoStop}%
\bibitem [{\citenamefont {Dong}\ \emph
  {et~al.}(2024{\natexlab{a}})\citenamefont {Dong}, \citenamefont {Patri},\
  and\ \citenamefont {Senthil}}]{senthil-fci-prl24}%
  \BibitemOpen
  \bibfield  {author} {\bibinfo {author} {\bibfnamefont {Z.}~\bibnamefont
  {Dong}}, \bibinfo {author} {\bibfnamefont {A.~S.}\ \bibnamefont {Patri}},\
  and\ \bibinfo {author} {\bibfnamefont {T.}~\bibnamefont {Senthil}},\ }\href
  {https://doi.org/10.1103/PhysRevLett.133.206502} {\bibfield  {journal}
  {\bibinfo  {journal} {Phys. Rev. Lett.}\ }\textbf {\bibinfo {volume} {133}},\
  \bibinfo {pages} {206502} (\bibinfo {year} {2024}{\natexlab{a}})}\BibitemShut
  {NoStop}%
\bibitem [{\citenamefont {Zhou}\ \emph {et~al.}(2024)\citenamefont {Zhou},
  \citenamefont {Yang},\ and\ \citenamefont {Zhang}}]{zhangyh-fci-prl24}%
  \BibitemOpen
  \bibfield  {author} {\bibinfo {author} {\bibfnamefont {B.}~\bibnamefont
  {Zhou}}, \bibinfo {author} {\bibfnamefont {H.}~\bibnamefont {Yang}},\ and\
  \bibinfo {author} {\bibfnamefont {Y.-H.}\ \bibnamefont {Zhang}},\ }\href
  {https://doi.org/10.1103/PhysRevLett.133.206504} {\bibfield  {journal}
  {\bibinfo  {journal} {Phys. Rev. Lett.}\ }\textbf {\bibinfo {volume} {133}},\
  \bibinfo {pages} {206504} (\bibinfo {year} {2024})}\BibitemShut {NoStop}%
\bibitem [{\citenamefont {Dong}\ \emph
  {et~al.}(2024{\natexlab{b}})\citenamefont {Dong}, \citenamefont {Wang},
  \citenamefont {Wang}, \citenamefont {Soejima}, \citenamefont {Zaletel},
  \citenamefont {Vishwanath},\ and\ \citenamefont {Parker}}]{ashvin-fci-prl24}%
  \BibitemOpen
  \bibfield  {author} {\bibinfo {author} {\bibfnamefont {J.}~\bibnamefont
  {Dong}}, \bibinfo {author} {\bibfnamefont {T.}~\bibnamefont {Wang}}, \bibinfo
  {author} {\bibfnamefont {T.}~\bibnamefont {Wang}}, \bibinfo {author}
  {\bibfnamefont {T.}~\bibnamefont {Soejima}}, \bibinfo {author} {\bibfnamefont
  {M.~P.}\ \bibnamefont {Zaletel}}, \bibinfo {author} {\bibfnamefont
  {A.}~\bibnamefont {Vishwanath}},\ and\ \bibinfo {author} {\bibfnamefont
  {D.~E.}\ \bibnamefont {Parker}},\ }\href
  {https://doi.org/10.1103/PhysRevLett.133.206503} {\bibfield  {journal}
  {\bibinfo  {journal} {Phys. Rev. Lett.}\ }\textbf {\bibinfo {volume} {133}},\
  \bibinfo {pages} {206503} (\bibinfo {year} {2024}{\natexlab{b}})}\BibitemShut
  {NoStop}%
\bibitem [{\citenamefont {Guo}\ \emph {et~al.}(2024)\citenamefont {Guo},
  \citenamefont {Lu}, \citenamefont {Xie},\ and\ \citenamefont
  {Liu}}]{guo-prb24}%
  \BibitemOpen
  \bibfield  {author} {\bibinfo {author} {\bibfnamefont {Z.}~\bibnamefont
  {Guo}}, \bibinfo {author} {\bibfnamefont {X.}~\bibnamefont {Lu}}, \bibinfo
  {author} {\bibfnamefont {B.}~\bibnamefont {Xie}},\ and\ \bibinfo {author}
  {\bibfnamefont {J.}~\bibnamefont {Liu}},\ }\href
  {https://doi.org/10.1103/PhysRevB.110.075109} {\bibfield  {journal} {\bibinfo
   {journal} {Phys. Rev. B}\ }\textbf {\bibinfo {volume} {110}},\ \bibinfo
  {pages} {075109} (\bibinfo {year} {2024})}\BibitemShut {NoStop}%
\bibitem [{\citenamefont {Kwan}\ \emph {et~al.}(2023)\citenamefont {Kwan},
  \citenamefont {Yu}, \citenamefont {Herzog-Arbeitman}, \citenamefont {Efetov},
  \citenamefont {Regnault},\ and\ \citenamefont
  {Bernevig}}]{bernevig-fci-iii-arxiv23}%
  \BibitemOpen
  \bibfield  {author} {\bibinfo {author} {\bibfnamefont {Y.~H.}\ \bibnamefont
  {Kwan}}, \bibinfo {author} {\bibfnamefont {J.}~\bibnamefont {Yu}}, \bibinfo
  {author} {\bibfnamefont {J.}~\bibnamefont {Herzog-Arbeitman}}, \bibinfo
  {author} {\bibfnamefont {D.~K.}\ \bibnamefont {Efetov}}, \bibinfo {author}
  {\bibfnamefont {N.}~\bibnamefont {Regnault}},\ and\ \bibinfo {author}
  {\bibfnamefont {B.~A.}\ \bibnamefont {Bernevig}},\ }\href@noop {} {\bibinfo
  {title} {Moir\'e fractional chern insulators iii: Hartree-fock phase diagram,
  magic angle regime for chern insulator states, the role of the moir\'e
  potential and goldstone gaps in rhombohedral graphene superlattices}}
  (\bibinfo {year} {2023}),\ \Eprint {https://arxiv.org/abs/2312.11617}
  {arXiv:2312.11617 [cond-mat.str-el]} \BibitemShut {NoStop}%
\bibitem [{\citenamefont {Yu}\ \emph {et~al.}(2024{\natexlab{b}})\citenamefont
  {Yu}, \citenamefont {Herzog-Arbeitman}, \citenamefont {Kwan}, \citenamefont
  {Regnault},\ and\ \citenamefont {Bernevig}}]{bernevig-fci-iv-arxiv24}%
  \BibitemOpen
  \bibfield  {author} {\bibinfo {author} {\bibfnamefont {J.}~\bibnamefont
  {Yu}}, \bibinfo {author} {\bibfnamefont {J.}~\bibnamefont
  {Herzog-Arbeitman}}, \bibinfo {author} {\bibfnamefont {Y.~H.}\ \bibnamefont
  {Kwan}}, \bibinfo {author} {\bibfnamefont {N.}~\bibnamefont {Regnault}},\
  and\ \bibinfo {author} {\bibfnamefont {B.~A.}\ \bibnamefont {Bernevig}},\
  }\href@noop {} {\bibinfo {title} {Moir\'e fractional chern insulators iv:
  Fluctuation-driven collapse of fcis in multi-band exact diagonalization
  calculations on rhombohedral graphene}} (\bibinfo {year}
  {2024}{\natexlab{b}}),\ \Eprint {https://arxiv.org/abs/2407.13770}
  {arXiv:2407.13770 [cond-mat.str-el]} \BibitemShut {NoStop}%
\bibitem [{\citenamefont {Huang}\ \emph {et~al.}(2024)\citenamefont {Huang},
  \citenamefont {Li}, \citenamefont {Das~Sarma},\ and\ \citenamefont
  {Zhang}}]{lixiao-fci-prb24}%
  \BibitemOpen
  \bibfield  {author} {\bibinfo {author} {\bibfnamefont {K.}~\bibnamefont
  {Huang}}, \bibinfo {author} {\bibfnamefont {X.}~\bibnamefont {Li}}, \bibinfo
  {author} {\bibfnamefont {S.}~\bibnamefont {Das~Sarma}},\ and\ \bibinfo
  {author} {\bibfnamefont {F.}~\bibnamefont {Zhang}},\ }\href
  {https://doi.org/10.1103/PhysRevB.110.115146} {\bibfield  {journal} {\bibinfo
   {journal} {Phys. Rev. B}\ }\textbf {\bibinfo {volume} {110}},\ \bibinfo
  {pages} {115146} (\bibinfo {year} {2024})}\BibitemShut {NoStop}%
\bibitem [{\citenamefont {Ledwith}\ \emph {et~al.}(2020)\citenamefont
  {Ledwith}, \citenamefont {Tarnopolsky}, \citenamefont {Khalaf},\ and\
  \citenamefont {Vishwanath}}]{ashvin-fci-tbg-prr20}%
  \BibitemOpen
  \bibfield  {author} {\bibinfo {author} {\bibfnamefont {P.~J.}\ \bibnamefont
  {Ledwith}}, \bibinfo {author} {\bibfnamefont {G.}~\bibnamefont
  {Tarnopolsky}}, \bibinfo {author} {\bibfnamefont {E.}~\bibnamefont
  {Khalaf}},\ and\ \bibinfo {author} {\bibfnamefont {A.}~\bibnamefont
  {Vishwanath}},\ }\href {https://doi.org/10.1103/PhysRevResearch.2.023237}
  {\bibfield  {journal} {\bibinfo  {journal} {Phys. Rev. Res.}\ }\textbf
  {\bibinfo {volume} {2}},\ \bibinfo {pages} {023237} (\bibinfo {year}
  {2020})}\BibitemShut {NoStop}%
\bibitem [{\citenamefont {Wang}\ \emph
  {et~al.}(2021{\natexlab{a}})\citenamefont {Wang}, \citenamefont {Cano},
  \citenamefont {Millis}, \citenamefont {Liu},\ and\ \citenamefont
  {Yang}}]{wang-geometry-prl21}%
  \BibitemOpen
  \bibfield  {author} {\bibinfo {author} {\bibfnamefont {J.}~\bibnamefont
  {Wang}}, \bibinfo {author} {\bibfnamefont {J.}~\bibnamefont {Cano}}, \bibinfo
  {author} {\bibfnamefont {A.~J.}\ \bibnamefont {Millis}}, \bibinfo {author}
  {\bibfnamefont {Z.}~\bibnamefont {Liu}},\ and\ \bibinfo {author}
  {\bibfnamefont {B.}~\bibnamefont {Yang}},\ }\href
  {https://doi.org/10.1103/PhysRevLett.127.246403} {\bibfield  {journal}
  {\bibinfo  {journal} {Phys. Rev. Lett.}\ }\textbf {\bibinfo {volume} {127}},\
  \bibinfo {pages} {246403} (\bibinfo {year} {2021}{\natexlab{a}})}\BibitemShut
  {NoStop}%
\bibitem [{\citenamefont {Ledwith}\ \emph {et~al.}(2022)\citenamefont
  {Ledwith}, \citenamefont {Vishwanath},\ and\ \citenamefont
  {Khalaf}}]{eslam-tmg-prl22}%
  \BibitemOpen
  \bibfield  {author} {\bibinfo {author} {\bibfnamefont {P.~J.}\ \bibnamefont
  {Ledwith}}, \bibinfo {author} {\bibfnamefont {A.}~\bibnamefont
  {Vishwanath}},\ and\ \bibinfo {author} {\bibfnamefont {E.}~\bibnamefont
  {Khalaf}},\ }\href {https://doi.org/10.1103/PhysRevLett.128.176404}
  {\bibfield  {journal} {\bibinfo  {journal} {Phys. Rev. Lett.}\ }\textbf
  {\bibinfo {volume} {128}},\ \bibinfo {pages} {176404} (\bibinfo {year}
  {2022})}\BibitemShut {NoStop}%
\bibitem [{\citenamefont {Wang}\ and\ \citenamefont
  {Liu}(2022)}]{wang-tmg-prl22}%
  \BibitemOpen
  \bibfield  {author} {\bibinfo {author} {\bibfnamefont {J.}~\bibnamefont
  {Wang}}\ and\ \bibinfo {author} {\bibfnamefont {Z.}~\bibnamefont {Liu}},\
  }\href {https://doi.org/10.1103/PhysRevLett.128.176403} {\bibfield  {journal}
  {\bibinfo  {journal} {Phys. Rev. Lett.}\ }\textbf {\bibinfo {volume} {128}},\
  \bibinfo {pages} {176403} (\bibinfo {year} {2022})}\BibitemShut {NoStop}%
\bibitem [{\citenamefont {Tarnopolsky}\ \emph {et~al.}(2019)\citenamefont
  {Tarnopolsky}, \citenamefont {Kruchkov},\ and\ \citenamefont
  {Vishwanath}}]{origin-magic-angle-prl19}%
  \BibitemOpen
  \bibfield  {author} {\bibinfo {author} {\bibfnamefont {G.}~\bibnamefont
  {Tarnopolsky}}, \bibinfo {author} {\bibfnamefont {A.~J.}\ \bibnamefont
  {Kruchkov}},\ and\ \bibinfo {author} {\bibfnamefont {A.}~\bibnamefont
  {Vishwanath}},\ }\href {https://doi.org/10.1103/PhysRevLett.122.106405}
  {\bibfield  {journal} {\bibinfo  {journal} {Phys. Rev. Lett.}\ }\textbf
  {\bibinfo {volume} {122}},\ \bibinfo {pages} {106405} (\bibinfo {year}
  {2019})}\BibitemShut {NoStop}%
\bibitem [{\citenamefont {Liu}\ \emph {et~al.}(2019)\citenamefont {Liu},
  \citenamefont {Liu},\ and\ \citenamefont {Dai}}]{jpliu-prb19}%
  \BibitemOpen
  \bibfield  {author} {\bibinfo {author} {\bibfnamefont {J.}~\bibnamefont
  {Liu}}, \bibinfo {author} {\bibfnamefont {J.}~\bibnamefont {Liu}},\ and\
  \bibinfo {author} {\bibfnamefont {X.}~\bibnamefont {Dai}},\ }\href
  {https://doi.org/10.1103/PhysRevB.99.155415} {\bibfield  {journal} {\bibinfo
  {journal} {Phys. Rev. B}\ }\textbf {\bibinfo {volume} {99}},\ \bibinfo
  {pages} {155415} (\bibinfo {year} {2019})}\BibitemShut {NoStop}%
\bibitem [{\citenamefont {Bultinck}\ \emph {et~al.}(2020)\citenamefont
  {Bultinck}, \citenamefont {Chatterjee},\ and\ \citenamefont
  {Zaletel}}]{zaletel-tbg-2019}%
  \BibitemOpen
  \bibfield  {author} {\bibinfo {author} {\bibfnamefont {N.}~\bibnamefont
  {Bultinck}}, \bibinfo {author} {\bibfnamefont {S.}~\bibnamefont
  {Chatterjee}},\ and\ \bibinfo {author} {\bibfnamefont {M.~P.}\ \bibnamefont
  {Zaletel}},\ }\href {https://doi.org/10.1103/PhysRevLett.124.166601}
  {\bibfield  {journal} {\bibinfo  {journal} {Phys. Rev. Lett.}\ }\textbf
  {\bibinfo {volume} {124}},\ \bibinfo {pages} {166601} (\bibinfo {year}
  {2020})}\BibitemShut {NoStop}%
\bibitem [{\citenamefont {Morales-Dur\'an}\ \emph {et~al.}(2024)\citenamefont
  {Morales-Dur\'an}, \citenamefont {Wei}, \citenamefont {Shi},\ and\
  \citenamefont {MacDonald}}]{macdonald-tmd-ll-prl24}%
  \BibitemOpen
  \bibfield  {author} {\bibinfo {author} {\bibfnamefont {N.}~\bibnamefont
  {Morales-Dur\'an}}, \bibinfo {author} {\bibfnamefont {N.}~\bibnamefont
  {Wei}}, \bibinfo {author} {\bibfnamefont {J.}~\bibnamefont {Shi}},\ and\
  \bibinfo {author} {\bibfnamefont {A.~H.}\ \bibnamefont {MacDonald}},\ }\href
  {https://doi.org/10.1103/PhysRevLett.132.096602} {\bibfield  {journal}
  {\bibinfo  {journal} {Phys. Rev. Lett.}\ }\textbf {\bibinfo {volume} {132}},\
  \bibinfo {pages} {096602} (\bibinfo {year} {2024})}\BibitemShut {NoStop}%
\bibitem [{\citenamefont {Li}\ and\ \citenamefont
  {Wu}(2024)}]{wu-mote2-arxiv24}%
  \BibitemOpen
  \bibfield  {author} {\bibinfo {author} {\bibfnamefont {B.}~\bibnamefont
  {Li}}\ and\ \bibinfo {author} {\bibfnamefont {F.}~\bibnamefont {Wu}},\
  }\href@noop {} {\bibinfo {title} {Variational mapping of chern bands to
  landau levels: Application to fractional chern insulators in twisted
  mote$_2$}} (\bibinfo {year} {2024}),\ \Eprint
  {https://arxiv.org/abs/2405.20307} {arXiv:2405.20307 [cond-mat.mes-hall]}
  \BibitemShut {NoStop}%
\bibitem [{\citenamefont {Li}\ \emph {et~al.}(2024)\citenamefont {Li},
  \citenamefont {Su}, \citenamefont {Kim}, \citenamefont {Kee}, \citenamefont
  {Sun},\ and\ \citenamefont {Lin}}]{lin-gauge-prb24}%
  \BibitemOpen
  \bibfield  {author} {\bibinfo {author} {\bibfnamefont {H.}~\bibnamefont
  {Li}}, \bibinfo {author} {\bibfnamefont {Y.}~\bibnamefont {Su}}, \bibinfo
  {author} {\bibfnamefont {Y.~B.}\ \bibnamefont {Kim}}, \bibinfo {author}
  {\bibfnamefont {H.-Y.}\ \bibnamefont {Kee}}, \bibinfo {author} {\bibfnamefont
  {K.}~\bibnamefont {Sun}},\ and\ \bibinfo {author} {\bibfnamefont {S.-Z.}\
  \bibnamefont {Lin}},\ }\href {https://doi.org/10.1103/PhysRevB.109.245131}
  {\bibfield  {journal} {\bibinfo  {journal} {Phys. Rev. B}\ }\textbf {\bibinfo
  {volume} {109}},\ \bibinfo {pages} {245131} (\bibinfo {year}
  {2024})}\BibitemShut {NoStop}%
\bibitem [{\citenamefont {Uri}\ \emph {et~al.}(2020)\citenamefont {Uri},
  \citenamefont {Grover}, \citenamefont {Cao}, \citenamefont {Crosse},
  \citenamefont {Bagani}, \citenamefont {Rodan-Legrain}, \citenamefont
  {Myasoedov}, \citenamefont {Watanabe}, \citenamefont {Taniguchi},
  \citenamefont {Moon}, \citenamefont {Koshino}, \citenamefont
  {Jarillo-Herrero},\ and\ \citenamefont {Zeldov}}]{zeldov-disorder-np20}%
  \BibitemOpen
  \bibfield  {author} {\bibinfo {author} {\bibfnamefont {A.}~\bibnamefont
  {Uri}}, \bibinfo {author} {\bibfnamefont {S.}~\bibnamefont {Grover}},
  \bibinfo {author} {\bibfnamefont {Y.}~\bibnamefont {Cao}}, \bibinfo {author}
  {\bibfnamefont {J.~A.}\ \bibnamefont {Crosse}}, \bibinfo {author}
  {\bibfnamefont {K.}~\bibnamefont {Bagani}}, \bibinfo {author} {\bibfnamefont
  {D.}~\bibnamefont {Rodan-Legrain}}, \bibinfo {author} {\bibfnamefont
  {Y.}~\bibnamefont {Myasoedov}}, \bibinfo {author} {\bibfnamefont
  {K.}~\bibnamefont {Watanabe}}, \bibinfo {author} {\bibfnamefont
  {T.}~\bibnamefont {Taniguchi}}, \bibinfo {author} {\bibfnamefont
  {P.}~\bibnamefont {Moon}}, \bibinfo {author} {\bibfnamefont {M.}~\bibnamefont
  {Koshino}}, \bibinfo {author} {\bibfnamefont {P.}~\bibnamefont
  {Jarillo-Herrero}},\ and\ \bibinfo {author} {\bibfnamefont {E.}~\bibnamefont
  {Zeldov}},\ }\href {https://doi.org/10.1038/s41586-020-2255-3} {\bibfield
  {journal} {\bibinfo  {journal} {Nature}\ }\textbf {\bibinfo {volume} {581}},\
  \bibinfo {pages} {47} (\bibinfo {year} {2020})}\BibitemShut {NoStop}%
\bibitem [{\citenamefont {Kazmierczak}\ \emph {et~al.}(2021)\citenamefont
  {Kazmierczak}, \citenamefont {Van~Winkle}, \citenamefont {Ophus},
  \citenamefont {Bustillo}, \citenamefont {Carr}, \citenamefont {Brown},
  \citenamefont {Ciston}, \citenamefont {Taniguchi}, \citenamefont {Watanabe},\
  and\ \citenamefont {Bediako}}]{kazmierczak2021strain}%
  \BibitemOpen
  \bibfield  {author} {\bibinfo {author} {\bibfnamefont {N.~P.}\ \bibnamefont
  {Kazmierczak}}, \bibinfo {author} {\bibfnamefont {M.}~\bibnamefont
  {Van~Winkle}}, \bibinfo {author} {\bibfnamefont {C.}~\bibnamefont {Ophus}},
  \bibinfo {author} {\bibfnamefont {K.~C.}\ \bibnamefont {Bustillo}}, \bibinfo
  {author} {\bibfnamefont {S.}~\bibnamefont {Carr}}, \bibinfo {author}
  {\bibfnamefont {H.~G.}\ \bibnamefont {Brown}}, \bibinfo {author}
  {\bibfnamefont {J.}~\bibnamefont {Ciston}}, \bibinfo {author} {\bibfnamefont
  {T.}~\bibnamefont {Taniguchi}}, \bibinfo {author} {\bibfnamefont
  {K.}~\bibnamefont {Watanabe}},\ and\ \bibinfo {author} {\bibfnamefont
  {D.~K.}\ \bibnamefont {Bediako}},\ }\href@noop {} {\bibfield  {journal}
  {\bibinfo  {journal} {Nature materials}\ }\textbf {\bibinfo {volume} {20}},\
  \bibinfo {pages} {956} (\bibinfo {year} {2021})}\BibitemShut {NoStop}%
\bibitem [{\citenamefont {Forsythe}\ \emph {et~al.}(2018)\citenamefont
  {Forsythe}, \citenamefont {Zhou}, \citenamefont {Watanabe}, \citenamefont
  {Taniguchi}, \citenamefont {Pasupathy}, \citenamefont {Moon}, \citenamefont
  {Koshino}, \citenamefont {Kim},\ and\ \citenamefont {Dean}}]{dean-nn18}%
  \BibitemOpen
  \bibfield  {author} {\bibinfo {author} {\bibfnamefont {C.}~\bibnamefont
  {Forsythe}}, \bibinfo {author} {\bibfnamefont {X.}~\bibnamefont {Zhou}},
  \bibinfo {author} {\bibfnamefont {K.}~\bibnamefont {Watanabe}}, \bibinfo
  {author} {\bibfnamefont {T.}~\bibnamefont {Taniguchi}}, \bibinfo {author}
  {\bibfnamefont {A.}~\bibnamefont {Pasupathy}}, \bibinfo {author}
  {\bibfnamefont {P.}~\bibnamefont {Moon}}, \bibinfo {author} {\bibfnamefont
  {M.}~\bibnamefont {Koshino}}, \bibinfo {author} {\bibfnamefont
  {P.}~\bibnamefont {Kim}},\ and\ \bibinfo {author} {\bibfnamefont {C.~R.}\
  \bibnamefont {Dean}},\ }\href {https://doi.org/10.1038/s41565-018-0138-7}
  {\bibfield  {journal} {\bibinfo  {journal} {Nature Nanotechnology}\ }\textbf
  {\bibinfo {volume} {13}},\ \bibinfo {pages} {566} (\bibinfo {year}
  {2018})}\BibitemShut {NoStop}%
\bibitem [{\citenamefont {Chen}\ \emph {et~al.}(2020)\citenamefont {Chen},
  \citenamefont {Kraft}, \citenamefont {Danneau}, \citenamefont {Richter},\
  and\ \citenamefont {Liu}}]{chen-cp20}%
  \BibitemOpen
  \bibfield  {author} {\bibinfo {author} {\bibfnamefont {S.-C.}\ \bibnamefont
  {Chen}}, \bibinfo {author} {\bibfnamefont {R.}~\bibnamefont {Kraft}},
  \bibinfo {author} {\bibfnamefont {R.}~\bibnamefont {Danneau}}, \bibinfo
  {author} {\bibfnamefont {K.}~\bibnamefont {Richter}},\ and\ \bibinfo {author}
  {\bibfnamefont {M.-H.}\ \bibnamefont {Liu}},\ }\href
  {https://doi.org/10.1038/s42005-020-0335-1} {\bibfield  {journal} {\bibinfo
  {journal} {Communications Physics}\ }\textbf {\bibinfo {volume} {3}},\
  \bibinfo {pages} {71} (\bibinfo {year} {2020})}\BibitemShut {NoStop}%
\bibitem [{\citenamefont {Li}\ \emph {et~al.}(2021)\citenamefont {Li},
  \citenamefont {Dietrich}, \citenamefont {Forsythe}, \citenamefont
  {Taniguchi}, \citenamefont {Watanabe}, \citenamefont {Moon},\ and\
  \citenamefont {Dean}}]{dean-nn21}%
  \BibitemOpen
  \bibfield  {author} {\bibinfo {author} {\bibfnamefont {Y.}~\bibnamefont
  {Li}}, \bibinfo {author} {\bibfnamefont {S.}~\bibnamefont {Dietrich}},
  \bibinfo {author} {\bibfnamefont {C.}~\bibnamefont {Forsythe}}, \bibinfo
  {author} {\bibfnamefont {T.}~\bibnamefont {Taniguchi}}, \bibinfo {author}
  {\bibfnamefont {K.}~\bibnamefont {Watanabe}}, \bibinfo {author}
  {\bibfnamefont {P.}~\bibnamefont {Moon}},\ and\ \bibinfo {author}
  {\bibfnamefont {C.~R.}\ \bibnamefont {Dean}},\ }\href
  {https://doi.org/10.1038/s41565-021-00849-9} {\bibfield  {journal} {\bibinfo
  {journal} {Nature Nanotechnology}\ }\textbf {\bibinfo {volume} {16}},\
  \bibinfo {pages} {525} (\bibinfo {year} {2021})}\BibitemShut {NoStop}%
\bibitem [{\citenamefont {Barcons~Ruiz}\ \emph {et~al.}(2022)\citenamefont
  {Barcons~Ruiz}, \citenamefont {Herzig~Sheinfux}, \citenamefont {Hoffmann},
  \citenamefont {Torre}, \citenamefont {Agarwal}, \citenamefont {Kumar},
  \citenamefont {Vistoli}, \citenamefont {Taniguchi}, \citenamefont {Watanabe},
  \citenamefont {Bachtold},\ and\ \citenamefont {Koppens}}]{ruiz-nc22}%
  \BibitemOpen
  \bibfield  {author} {\bibinfo {author} {\bibfnamefont {D.}~\bibnamefont
  {Barcons~Ruiz}}, \bibinfo {author} {\bibfnamefont {H.}~\bibnamefont
  {Herzig~Sheinfux}}, \bibinfo {author} {\bibfnamefont {R.}~\bibnamefont
  {Hoffmann}}, \bibinfo {author} {\bibfnamefont {I.}~\bibnamefont {Torre}},
  \bibinfo {author} {\bibfnamefont {H.}~\bibnamefont {Agarwal}}, \bibinfo
  {author} {\bibfnamefont {R.~K.}\ \bibnamefont {Kumar}}, \bibinfo {author}
  {\bibfnamefont {L.}~\bibnamefont {Vistoli}}, \bibinfo {author} {\bibfnamefont
  {T.}~\bibnamefont {Taniguchi}}, \bibinfo {author} {\bibfnamefont
  {K.}~\bibnamefont {Watanabe}}, \bibinfo {author} {\bibfnamefont
  {A.}~\bibnamefont {Bachtold}},\ and\ \bibinfo {author} {\bibfnamefont
  {F.~H.~L.}\ \bibnamefont {Koppens}},\ }\href
  {https://doi.org/10.1038/s41467-022-34734-3} {\bibfield  {journal} {\bibinfo
  {journal} {Nature Communications}\ }\textbf {\bibinfo {volume} {13}},\
  \bibinfo {pages} {6926} (\bibinfo {year} {2022})}\BibitemShut {NoStop}%
\bibitem [{\citenamefont {Wang}\ \emph {et~al.}(2024)\citenamefont {Wang},
  \citenamefont {Zhan}, \citenamefont {Fan}, \citenamefont {Li}, \citenamefont
  {Pantale\'on}, \citenamefont {Ye}, \citenamefont {He}, \citenamefont {Wei},
  \citenamefont {Li}, \citenamefont {Guinea}, \citenamefont {Yuan},\ and\
  \citenamefont {Zeng}}]{zeng-prl24}%
  \BibitemOpen
  \bibfield  {author} {\bibinfo {author} {\bibfnamefont {S.}~\bibnamefont
  {Wang}}, \bibinfo {author} {\bibfnamefont {Z.}~\bibnamefont {Zhan}}, \bibinfo
  {author} {\bibfnamefont {X.}~\bibnamefont {Fan}}, \bibinfo {author}
  {\bibfnamefont {Y.}~\bibnamefont {Li}}, \bibinfo {author} {\bibfnamefont
  {P.~A.}\ \bibnamefont {Pantale\'on}}, \bibinfo {author} {\bibfnamefont
  {C.}~\bibnamefont {Ye}}, \bibinfo {author} {\bibfnamefont {Z.}~\bibnamefont
  {He}}, \bibinfo {author} {\bibfnamefont {L.}~\bibnamefont {Wei}}, \bibinfo
  {author} {\bibfnamefont {L.}~\bibnamefont {Li}}, \bibinfo {author}
  {\bibfnamefont {F.}~\bibnamefont {Guinea}}, \bibinfo {author} {\bibfnamefont
  {S.}~\bibnamefont {Yuan}},\ and\ \bibinfo {author} {\bibfnamefont
  {C.}~\bibnamefont {Zeng}},\ }\href
  {https://doi.org/10.1103/PhysRevLett.133.066302} {\bibfield  {journal}
  {\bibinfo  {journal} {Phys. Rev. Lett.}\ }\textbf {\bibinfo {volume} {133}},\
  \bibinfo {pages} {066302} (\bibinfo {year} {2024})}\BibitemShut {NoStop}%
\bibitem [{\citenamefont {Sun}\ \emph {et~al.}(2024)\citenamefont {Sun},
  \citenamefont {Akbar~Ghorashi}, \citenamefont {Watanabe}, \citenamefont
  {Taniguchi}, \citenamefont {Camino}, \citenamefont {Cano},\ and\
  \citenamefont {Du}}]{du-nanoletter24}%
  \BibitemOpen
  \bibfield  {author} {\bibinfo {author} {\bibfnamefont {J.}~\bibnamefont
  {Sun}}, \bibinfo {author} {\bibfnamefont {S.~A.}\ \bibnamefont
  {Akbar~Ghorashi}}, \bibinfo {author} {\bibfnamefont {K.}~\bibnamefont
  {Watanabe}}, \bibinfo {author} {\bibfnamefont {T.}~\bibnamefont {Taniguchi}},
  \bibinfo {author} {\bibfnamefont {F.}~\bibnamefont {Camino}}, \bibinfo
  {author} {\bibfnamefont {J.}~\bibnamefont {Cano}},\ and\ \bibinfo {author}
  {\bibfnamefont {X.}~\bibnamefont {Du}},\ }\href
  {https://doi.org/10.1021/acs.nanolett.4c03238} {\bibfield  {journal}
  {\bibinfo  {journal} {Nano Letters}\ }\textbf {\bibinfo {volume} {24}},\
  \bibinfo {pages} {13600} (\bibinfo {year} {2024})},\ \bibinfo {note} {pMID:
  39432385},\ \Eprint
  {https://arxiv.org/abs/https://doi.org/10.1021/acs.nanolett.4c03238}
  {https://doi.org/10.1021/acs.nanolett.4c03238} \BibitemShut {NoStop}%
\bibitem [{\citenamefont {Ghorashi}\ \emph {et~al.}(2023)\citenamefont
  {Ghorashi}, \citenamefont {Dunbrack}, \citenamefont {Abouelkomsan},
  \citenamefont {Sun}, \citenamefont {Du},\ and\ \citenamefont
  {Cano}}]{cano-bilayer-prl23}%
  \BibitemOpen
  \bibfield  {author} {\bibinfo {author} {\bibfnamefont {S.~A.~A.}\
  \bibnamefont {Ghorashi}}, \bibinfo {author} {\bibfnamefont {A.}~\bibnamefont
  {Dunbrack}}, \bibinfo {author} {\bibfnamefont {A.}~\bibnamefont
  {Abouelkomsan}}, \bibinfo {author} {\bibfnamefont {J.}~\bibnamefont {Sun}},
  \bibinfo {author} {\bibfnamefont {X.}~\bibnamefont {Du}},\ and\ \bibinfo
  {author} {\bibfnamefont {J.}~\bibnamefont {Cano}},\ }\href
  {https://doi.org/10.1103/PhysRevLett.130.196201} {\bibfield  {journal}
  {\bibinfo  {journal} {Phys. Rev. Lett.}\ }\textbf {\bibinfo {volume} {130}},\
  \bibinfo {pages} {196201} (\bibinfo {year} {2023})}\BibitemShut {NoStop}%
\bibitem [{\citenamefont {Lu}\ \emph {et~al.}(2023)\citenamefont {Lu},
  \citenamefont {Zhang}, \citenamefont {Wang}, \citenamefont {Gao},
  \citenamefont {Yang}, \citenamefont {Guo}, \citenamefont {Gao}, \citenamefont
  {Ye}, \citenamefont {Han},\ and\ \citenamefont {Liu}}]{lu-nc23}%
  \BibitemOpen
  \bibfield  {author} {\bibinfo {author} {\bibfnamefont {X.}~\bibnamefont
  {Lu}}, \bibinfo {author} {\bibfnamefont {S.}~\bibnamefont {Zhang}}, \bibinfo
  {author} {\bibfnamefont {Y.}~\bibnamefont {Wang}}, \bibinfo {author}
  {\bibfnamefont {X.}~\bibnamefont {Gao}}, \bibinfo {author} {\bibfnamefont
  {K.}~\bibnamefont {Yang}}, \bibinfo {author} {\bibfnamefont {Z.}~\bibnamefont
  {Guo}}, \bibinfo {author} {\bibfnamefont {Y.}~\bibnamefont {Gao}}, \bibinfo
  {author} {\bibfnamefont {Y.}~\bibnamefont {Ye}}, \bibinfo {author}
  {\bibfnamefont {Z.}~\bibnamefont {Han}},\ and\ \bibinfo {author}
  {\bibfnamefont {J.}~\bibnamefont {Liu}},\ }\href
  {https://doi.org/10.1038/s41467-023-41293-8} {\bibfield  {journal} {\bibinfo
  {journal} {Nature Communications}\ }\textbf {\bibinfo {volume} {14}},\
  \bibinfo {pages} {5550} (\bibinfo {year} {2023})}\BibitemShut {NoStop}%
\bibitem [{\citenamefont {Ghorashi}\ and\ \citenamefont
  {Cano}(2023)}]{cano-multilayer-prb23}%
  \BibitemOpen
  \bibfield  {author} {\bibinfo {author} {\bibfnamefont {S.~A.~A.}\
  \bibnamefont {Ghorashi}}\ and\ \bibinfo {author} {\bibfnamefont
  {J.}~\bibnamefont {Cano}},\ }\href
  {https://doi.org/10.1103/PhysRevB.107.195423} {\bibfield  {journal} {\bibinfo
   {journal} {Phys. Rev. B}\ }\textbf {\bibinfo {volume} {107}},\ \bibinfo
  {pages} {195423} (\bibinfo {year} {2023})}\BibitemShut {NoStop}%
\bibitem [{\citenamefont {Seleznev}\ \emph {et~al.}(2024)\citenamefont
  {Seleznev}, \citenamefont {Cano},\ and\ \citenamefont
  {Vanderbilt}}]{vanderbilt-bilayer-prb24}%
  \BibitemOpen
  \bibfield  {author} {\bibinfo {author} {\bibfnamefont {D.}~\bibnamefont
  {Seleznev}}, \bibinfo {author} {\bibfnamefont {J.}~\bibnamefont {Cano}},\
  and\ \bibinfo {author} {\bibfnamefont {D.}~\bibnamefont {Vanderbilt}},\
  }\href {https://doi.org/10.1103/PhysRevB.110.205115} {\bibfield  {journal}
  {\bibinfo  {journal} {Phys. Rev. B}\ }\textbf {\bibinfo {volume} {110}},\
  \bibinfo {pages} {205115} (\bibinfo {year} {2024})}\BibitemShut {NoStop}%
\bibitem [{\citenamefont {Zhan}\ \emph {et~al.}(2025)\citenamefont {Zhan},
  \citenamefont {Li},\ and\ \citenamefont
  {Pantale\'on}}]{zhan-patterned-prb25}%
  \BibitemOpen
  \bibfield  {author} {\bibinfo {author} {\bibfnamefont {Z.}~\bibnamefont
  {Zhan}}, \bibinfo {author} {\bibfnamefont {Y.}~\bibnamefont {Li}},\ and\
  \bibinfo {author} {\bibfnamefont {P.~A.}\ \bibnamefont {Pantale\'on}},\
  }\href {https://doi.org/10.1103/PhysRevB.111.045148} {\bibfield  {journal}
  {\bibinfo  {journal} {Phys. Rev. B}\ }\textbf {\bibinfo {volume} {111}},\
  \bibinfo {pages} {045148} (\bibinfo {year} {2025})}\BibitemShut {NoStop}%
\bibitem [{\citenamefont {Tan}\ \emph {et~al.}(2024)\citenamefont {Tan},
  \citenamefont {Reddy}, \citenamefont {Fu},\ and\ \citenamefont
  {Devakul}}]{trithep-semiconductor-prl24}%
  \BibitemOpen
  \bibfield  {author} {\bibinfo {author} {\bibfnamefont {T.}~\bibnamefont
  {Tan}}, \bibinfo {author} {\bibfnamefont {A.~P.}\ \bibnamefont {Reddy}},
  \bibinfo {author} {\bibfnamefont {L.}~\bibnamefont {Fu}},\ and\ \bibinfo
  {author} {\bibfnamefont {T.}~\bibnamefont {Devakul}},\ }\href
  {https://doi.org/10.1103/PhysRevLett.133.206601} {\bibfield  {journal}
  {\bibinfo  {journal} {Phys. Rev. Lett.}\ }\textbf {\bibinfo {volume} {133}},\
  \bibinfo {pages} {206601} (\bibinfo {year} {2024})}\BibitemShut {NoStop}%
\bibitem [{\citenamefont {Miao}\ \emph {et~al.}(2024)\citenamefont {Miao},
  \citenamefont {Rashidi},\ and\ \citenamefont {Dai}}]{dai-artificial-arxiv24}%
  \BibitemOpen
  \bibfield  {author} {\bibinfo {author} {\bibfnamefont {W.}~\bibnamefont
  {Miao}}, \bibinfo {author} {\bibfnamefont {A.}~\bibnamefont {Rashidi}},\ and\
  \bibinfo {author} {\bibfnamefont {X.}~\bibnamefont {Dai}},\ }\href@noop {}
  {\bibinfo {title} {Artificial moir\'{e} engineering for an ideal bhz model}}
  (\bibinfo {year} {2024}),\ \Eprint {https://arxiv.org/abs/2409.08540}
  {arXiv:2409.08540 [cond-mat.mes-hall]} \BibitemShut {NoStop}%
\bibitem [{\citenamefont {Bergman}\ \emph {et~al.}(2008)\citenamefont
  {Bergman}, \citenamefont {Wu},\ and\ \citenamefont
  {Balents}}]{balents-prb08}%
  \BibitemOpen
  \bibfield  {author} {\bibinfo {author} {\bibfnamefont {D.~L.}\ \bibnamefont
  {Bergman}}, \bibinfo {author} {\bibfnamefont {C.}~\bibnamefont {Wu}},\ and\
  \bibinfo {author} {\bibfnamefont {L.}~\bibnamefont {Balents}},\ }\href
  {https://doi.org/10.1103/PhysRevB.78.125104} {\bibfield  {journal} {\bibinfo
  {journal} {Phys. Rev. B}\ }\textbf {\bibinfo {volume} {78}},\ \bibinfo
  {pages} {125104} (\bibinfo {year} {2008})}\BibitemShut {NoStop}%
\bibitem [{sup()}]{supp_info}%
  \BibitemOpen
  \href@noop {} {}\bibinfo {note} {See Supplemental Materials for: (a) details
  of the device fabrication of bilayer graphene coupled with kagome-patterned
  dielectric substrate and its transport data under weak electrostatic
  potential, (b) details of numerical simulations of electrostatic potential
  distributions in the device, (c) details of the continuum model describing
  rhombohedral multilayer graphene coupled with kagome superlattice potential,
  (d) interacting Hamiltonian and more details on exact diagonalization
  calculations, and (e) band structures, quantum geometric properties, and
  topological properties for low-energy subbands in trilayer and tetralayer
  rhombohedral graphene coupled with kagome superlattice
  potential.}\BibitemShut {Stop}%
\bibitem [{\citenamefont {Moon}\ and\ \citenamefont
  {Koshino}(2013)}]{moon-tbg-prb13}%
  \BibitemOpen
  \bibfield  {author} {\bibinfo {author} {\bibfnamefont {P.}~\bibnamefont
  {Moon}}\ and\ \bibinfo {author} {\bibfnamefont {M.}~\bibnamefont {Koshino}},\
  }\href@noop {} {\bibfield  {journal} {\bibinfo  {journal} {Physical Review
  B}\ }\textbf {\bibinfo {volume} {87}},\ \bibinfo {pages} {205404} (\bibinfo
  {year} {2013})}\BibitemShut {NoStop}%
\bibitem [{\citenamefont {Roy}(2014)}]{roy-prb14}%
  \BibitemOpen
  \bibfield  {author} {\bibinfo {author} {\bibfnamefont {R.}~\bibnamefont
  {Roy}},\ }\href {https://doi.org/10.1103/PhysRevB.90.165139} {\bibfield
  {journal} {\bibinfo  {journal} {Phys. Rev. B}\ }\textbf {\bibinfo {volume}
  {90}},\ \bibinfo {pages} {165139} (\bibinfo {year} {2014})}\BibitemShut
  {NoStop}%
\bibitem [{\citenamefont {Claassen}\ \emph {et~al.}(2015)\citenamefont
  {Claassen}, \citenamefont {Lee}, \citenamefont {Thomale}, \citenamefont
  {Qi},\ and\ \citenamefont {Devereaux}}]{claassen-prl15}%
  \BibitemOpen
  \bibfield  {author} {\bibinfo {author} {\bibfnamefont {M.}~\bibnamefont
  {Claassen}}, \bibinfo {author} {\bibfnamefont {C.~H.}\ \bibnamefont {Lee}},
  \bibinfo {author} {\bibfnamefont {R.}~\bibnamefont {Thomale}}, \bibinfo
  {author} {\bibfnamefont {X.-L.}\ \bibnamefont {Qi}},\ and\ \bibinfo {author}
  {\bibfnamefont {T.~P.}\ \bibnamefont {Devereaux}},\ }\href
  {https://doi.org/10.1103/PhysRevLett.114.236802} {\bibfield  {journal}
  {\bibinfo  {journal} {Phys. Rev. Lett.}\ }\textbf {\bibinfo {volume} {114}},\
  \bibinfo {pages} {236802} (\bibinfo {year} {2015})}\BibitemShut {NoStop}%
\bibitem [{\citenamefont {Wang}\ \emph
  {et~al.}(2021{\natexlab{b}})\citenamefont {Wang}, \citenamefont {Cano},
  \citenamefont {Millis}, \citenamefont {Liu},\ and\ \citenamefont
  {Yang}}]{wangjie-prl21}%
  \BibitemOpen
  \bibfield  {author} {\bibinfo {author} {\bibfnamefont {J.}~\bibnamefont
  {Wang}}, \bibinfo {author} {\bibfnamefont {J.}~\bibnamefont {Cano}}, \bibinfo
  {author} {\bibfnamefont {A.~J.}\ \bibnamefont {Millis}}, \bibinfo {author}
  {\bibfnamefont {Z.}~\bibnamefont {Liu}},\ and\ \bibinfo {author}
  {\bibfnamefont {B.}~\bibnamefont {Yang}},\ }\href
  {https://doi.org/10.1103/PhysRevLett.127.246403} {\bibfield  {journal}
  {\bibinfo  {journal} {Phys. Rev. Lett.}\ }\textbf {\bibinfo {volume} {127}},\
  \bibinfo {pages} {246403} (\bibinfo {year} {2021}{\natexlab{b}})}\BibitemShut
  {NoStop}%
\bibitem [{\citenamefont {Han}\ \emph {et~al.}(2023)\citenamefont {Han},
  \citenamefont {Lu}, \citenamefont {Scuri}, \citenamefont {Sung},
  \citenamefont {Wang}, \citenamefont {Han}, \citenamefont {Watanabe},
  \citenamefont {Taniguchi}, \citenamefont {Park},\ and\ \citenamefont
  {Ju}}]{ju-chern-natnano2023}%
  \BibitemOpen
  \bibfield  {author} {\bibinfo {author} {\bibfnamefont {T.}~\bibnamefont
  {Han}}, \bibinfo {author} {\bibfnamefont {Z.}~\bibnamefont {Lu}}, \bibinfo
  {author} {\bibfnamefont {G.}~\bibnamefont {Scuri}}, \bibinfo {author}
  {\bibfnamefont {J.}~\bibnamefont {Sung}}, \bibinfo {author} {\bibfnamefont
  {J.}~\bibnamefont {Wang}}, \bibinfo {author} {\bibfnamefont {T.}~\bibnamefont
  {Han}}, \bibinfo {author} {\bibfnamefont {K.}~\bibnamefont {Watanabe}},
  \bibinfo {author} {\bibfnamefont {T.}~\bibnamefont {Taniguchi}}, \bibinfo
  {author} {\bibfnamefont {H.}~\bibnamefont {Park}},\ and\ \bibinfo {author}
  {\bibfnamefont {L.}~\bibnamefont {Ju}},\ }\href@noop {} {\bibfield  {journal}
  {\bibinfo  {journal} {Nature Nanotechnology}\ ,\ \bibinfo {pages} {1}}
  (\bibinfo {year} {2023})}\BibitemShut {NoStop}%
\bibitem [{\citenamefont {Zhou}\ \emph {et~al.}(2021)\citenamefont {Zhou},
  \citenamefont {Xie}, \citenamefont {Ghazaryan}, \citenamefont {Holder},
  \citenamefont {Ehrets}, \citenamefont {Spanton}, \citenamefont {Taniguchi},
  \citenamefont {Watanabe}, \citenamefont {Berg}, \citenamefont {Serbyn} \emph
  {et~al.}}]{zhou-trilayer-nature21}%
  \BibitemOpen
  \bibfield  {author} {\bibinfo {author} {\bibfnamefont {H.}~\bibnamefont
  {Zhou}}, \bibinfo {author} {\bibfnamefont {T.}~\bibnamefont {Xie}}, \bibinfo
  {author} {\bibfnamefont {A.}~\bibnamefont {Ghazaryan}}, \bibinfo {author}
  {\bibfnamefont {T.}~\bibnamefont {Holder}}, \bibinfo {author} {\bibfnamefont
  {J.~R.}\ \bibnamefont {Ehrets}}, \bibinfo {author} {\bibfnamefont {E.~M.}\
  \bibnamefont {Spanton}}, \bibinfo {author} {\bibfnamefont {T.}~\bibnamefont
  {Taniguchi}}, \bibinfo {author} {\bibfnamefont {K.}~\bibnamefont {Watanabe}},
  \bibinfo {author} {\bibfnamefont {E.}~\bibnamefont {Berg}}, \bibinfo {author}
  {\bibfnamefont {M.}~\bibnamefont {Serbyn}}, \emph {et~al.},\ }\href@noop {}
  {\bibfield  {journal} {\bibinfo  {journal} {Nature}\ }\textbf {\bibinfo
  {volume} {598}},\ \bibinfo {pages} {429} (\bibinfo {year}
  {2021})}\BibitemShut {NoStop}%
\bibitem [{com()}]{comment-fci}%
  \BibitemOpen
  \href@noop {} {}\bibinfo {note} {The orange blocks with slight shadow in
  Fig.~3(a)-(b) indicate states with a low-energy spectrum similar to FCI, but
  the gap between the 4th and 3rd states is smaller than the energy spread
  within the first three states}\BibitemShut {NoStop}%
\bibitem [{\citenamefont {Wu}\ \emph {et~al.}(2012{\natexlab{b}})\citenamefont
  {Wu}, \citenamefont {Regnault},\ and\ \citenamefont
  {Bernevig}}]{bernevig-fci-prb12}%
  \BibitemOpen
  \bibfield  {author} {\bibinfo {author} {\bibfnamefont {Y.-L.}\ \bibnamefont
  {Wu}}, \bibinfo {author} {\bibfnamefont {N.}~\bibnamefont {Regnault}},\ and\
  \bibinfo {author} {\bibfnamefont {B.~A.}\ \bibnamefont {Bernevig}},\ }\href
  {https://doi.org/10.1103/PhysRevB.86.085129} {\bibfield  {journal} {\bibinfo
  {journal} {Phys. Rev. B}\ }\textbf {\bibinfo {volume} {86}},\ \bibinfo
  {pages} {085129} (\bibinfo {year} {2012}{\natexlab{b}})}\BibitemShut
  {NoStop}%
\bibitem [{\citenamefont {Wu}\ \emph {et~al.}(2013)\citenamefont {Wu},
  \citenamefont {Regnault},\ and\ \citenamefont
  {Bernevig}}]{bernevig-fci-prl13}%
  \BibitemOpen
  \bibfield  {author} {\bibinfo {author} {\bibfnamefont {Y.-L.}\ \bibnamefont
  {Wu}}, \bibinfo {author} {\bibfnamefont {N.}~\bibnamefont {Regnault}},\ and\
  \bibinfo {author} {\bibfnamefont {B.~A.}\ \bibnamefont {Bernevig}},\ }\href
  {https://doi.org/10.1103/PhysRevLett.110.106802} {\bibfield  {journal}
  {\bibinfo  {journal} {Phys. Rev. Lett.}\ }\textbf {\bibinfo {volume} {110}},\
  \bibinfo {pages} {106802} (\bibinfo {year} {2013})}\BibitemShut {NoStop}%
\bibitem [{\citenamefont {Jain}(1989{\natexlab{b}})}]{Jain-prl1989}%
  \BibitemOpen
  \bibfield  {author} {\bibinfo {author} {\bibfnamefont {J.~K.}\ \bibnamefont
  {Jain}},\ }\href {https://doi.org/10.1103/PhysRevLett.63.199} {\bibfield
  {journal} {\bibinfo  {journal} {Phys. Rev. Lett.}\ }\textbf {\bibinfo
  {volume} {63}},\ \bibinfo {pages} {199} (\bibinfo {year}
  {1989}{\natexlab{b}})}\BibitemShut {NoStop}%
\bibitem [{\citenamefont {Lopez}\ and\ \citenamefont
  {Fradkin}(1991)}]{Lopez-prb1991}%
  \BibitemOpen
  \bibfield  {author} {\bibinfo {author} {\bibfnamefont {A.}~\bibnamefont
  {Lopez}}\ and\ \bibinfo {author} {\bibfnamefont {E.}~\bibnamefont
  {Fradkin}},\ }\href {https://doi.org/10.1103/PhysRevB.44.5246} {\bibfield
  {journal} {\bibinfo  {journal} {Phys. Rev. B}\ }\textbf {\bibinfo {volume}
  {44}},\ \bibinfo {pages} {5246} (\bibinfo {year} {1991})}\BibitemShut
  {NoStop}%
\bibitem [{\citenamefont {Halperin}\ \emph {et~al.}(1993)\citenamefont
  {Halperin}, \citenamefont {Lee},\ and\ \citenamefont
  {Read}}]{Halperin-Patrick-Nicholas-prb1993}%
  \BibitemOpen
  \bibfield  {author} {\bibinfo {author} {\bibfnamefont {B.~I.}\ \bibnamefont
  {Halperin}}, \bibinfo {author} {\bibfnamefont {P.~A.}\ \bibnamefont {Lee}},\
  and\ \bibinfo {author} {\bibfnamefont {N.}~\bibnamefont {Read}},\ }\href
  {https://doi.org/10.1103/PhysRevB.47.7312} {\bibfield  {journal} {\bibinfo
  {journal} {Phys. Rev. B}\ }\textbf {\bibinfo {volume} {47}},\ \bibinfo
  {pages} {7312} (\bibinfo {year} {1993})}\BibitemShut {NoStop}%
\bibitem [{\citenamefont {Son}(2015)}]{Son-prx2015}%
  \BibitemOpen
  \bibfield  {author} {\bibinfo {author} {\bibfnamefont {D.~T.}\ \bibnamefont
  {Son}},\ }\href {https://doi.org/10.1103/PhysRevX.5.031027} {\bibfield
  {journal} {\bibinfo  {journal} {Phys. Rev. X}\ }\textbf {\bibinfo {volume}
  {5}},\ \bibinfo {pages} {031027} (\bibinfo {year} {2015})}\BibitemShut
  {NoStop}%
\bibitem [{\citenamefont {Dong}\ \emph {et~al.}(2023)\citenamefont {Dong},
  \citenamefont {Wang}, \citenamefont {Ledwith}, \citenamefont {Vishwanath},\
  and\ \citenamefont {Parker}}]{ashvin-cfl-prl23}%
  \BibitemOpen
  \bibfield  {author} {\bibinfo {author} {\bibfnamefont {J.}~\bibnamefont
  {Dong}}, \bibinfo {author} {\bibfnamefont {J.}~\bibnamefont {Wang}}, \bibinfo
  {author} {\bibfnamefont {P.~J.}\ \bibnamefont {Ledwith}}, \bibinfo {author}
  {\bibfnamefont {A.}~\bibnamefont {Vishwanath}},\ and\ \bibinfo {author}
  {\bibfnamefont {D.~E.}\ \bibnamefont {Parker}},\ }\href
  {https://doi.org/10.1103/PhysRevLett.131.136502} {\bibfield  {journal}
  {\bibinfo  {journal} {Phys. Rev. Lett.}\ }\textbf {\bibinfo {volume} {131}},\
  \bibinfo {pages} {136502} (\bibinfo {year} {2023})}\BibitemShut {NoStop}%
\bibitem [{\citenamefont {Goldman}\ \emph {et~al.}(2023)\citenamefont
  {Goldman}, \citenamefont {Reddy}, \citenamefont {Paul},\ and\ \citenamefont
  {Fu}}]{fu-cfl-prl23}%
  \BibitemOpen
  \bibfield  {author} {\bibinfo {author} {\bibfnamefont {H.}~\bibnamefont
  {Goldman}}, \bibinfo {author} {\bibfnamefont {A.~P.}\ \bibnamefont {Reddy}},
  \bibinfo {author} {\bibfnamefont {N.}~\bibnamefont {Paul}},\ and\ \bibinfo
  {author} {\bibfnamefont {L.}~\bibnamefont {Fu}},\ }\href
  {https://doi.org/10.1103/PhysRevLett.131.136501} {\bibfield  {journal}
  {\bibinfo  {journal} {Phys. Rev. Lett.}\ }\textbf {\bibinfo {volume} {131}},\
  \bibinfo {pages} {136501} (\bibinfo {year} {2023})}\BibitemShut {NoStop}%
\bibitem [{\citenamefont {Li}()}]{prepare}%
  \BibitemOpen
  \bibfield  {author} {\bibinfo {author} {\bibfnamefont {Q.}~\bibnamefont
  {Li}},\ }\href@noop {} {}\bibinfo {note} {\textit{et al.} manuscript in
  preparation.}\BibitemShut {Stop}%
\bibitem [{\citenamefont {Moon}\ and\ \citenamefont
  {Koshino}(2014)}]{Moon-Koshino-prb2014}%
  \BibitemOpen
  \bibfield  {author} {\bibinfo {author} {\bibfnamefont {P.}~\bibnamefont
  {Moon}}\ and\ \bibinfo {author} {\bibfnamefont {M.}~\bibnamefont {Koshino}},\
  }\href {https://doi.org/10.1103/PhysRevB.90.155406} {\bibfield  {journal}
  {\bibinfo  {journal} {Phys. Rev. B}\ }\textbf {\bibinfo {volume} {90}},\
  \bibinfo {pages} {155406} (\bibinfo {year} {2014})}\BibitemShut {NoStop}%
\bibitem [{\citenamefont {Elias}\ \emph
  {et~al.}(2011{\natexlab{a}})\citenamefont {Elias}, \citenamefont {Gorbachev},
  \citenamefont {Mayorov}, \citenamefont {Morozov}, \citenamefont {Zhukov},
  \citenamefont {Blake}, \citenamefont {Ponomarenko}, \citenamefont
  {Grigorieva}, \citenamefont {Novoselov}, \citenamefont {Guinea} \emph
  {et~al.}}]{elias_natphys2011}%
  \BibitemOpen
  \bibfield  {author} {\bibinfo {author} {\bibfnamefont {D.~C.}\ \bibnamefont
  {Elias}}, \bibinfo {author} {\bibfnamefont {R.}~\bibnamefont {Gorbachev}},
  \bibinfo {author} {\bibfnamefont {A.}~\bibnamefont {Mayorov}}, \bibinfo
  {author} {\bibfnamefont {S.}~\bibnamefont {Morozov}}, \bibinfo {author}
  {\bibfnamefont {A.}~\bibnamefont {Zhukov}}, \bibinfo {author} {\bibfnamefont
  {P.}~\bibnamefont {Blake}}, \bibinfo {author} {\bibfnamefont
  {L.}~\bibnamefont {Ponomarenko}}, \bibinfo {author} {\bibfnamefont
  {I.}~\bibnamefont {Grigorieva}}, \bibinfo {author} {\bibfnamefont
  {K.}~\bibnamefont {Novoselov}}, \bibinfo {author} {\bibfnamefont
  {F.}~\bibnamefont {Guinea}}, \emph {et~al.},\ }\href@noop {} {\bibfield
  {journal} {\bibinfo  {journal} {Nature Physics}\ }\textbf {\bibinfo {volume}
  {7}},\ \bibinfo {pages} {701} (\bibinfo {year}
  {2011}{\natexlab{a}})}\BibitemShut {NoStop}%
\bibitem [{\citenamefont {Vafek}\ and\ \citenamefont
  {Kang}(2020)}]{vafek_prl2020}%
  \BibitemOpen
  \bibfield  {author} {\bibinfo {author} {\bibfnamefont {O.}~\bibnamefont
  {Vafek}}\ and\ \bibinfo {author} {\bibfnamefont {J.}~\bibnamefont {Kang}},\
  }\href {https://doi.org/10.1103/PhysRevLett.125.257602} {\bibfield  {journal}
  {\bibinfo  {journal} {Phys. Rev. Lett.}\ }\textbf {\bibinfo {volume} {125}},\
  \bibinfo {pages} {257602} (\bibinfo {year} {2020})}\BibitemShut {NoStop}%
\bibitem [{\citenamefont {Zhang}\ \emph {et~al.}(2022)\citenamefont {Zhang},
  \citenamefont {Dai},\ and\ \citenamefont {Liu}}]{zhang_prl2022}%
  \BibitemOpen
  \bibfield  {author} {\bibinfo {author} {\bibfnamefont {S.}~\bibnamefont
  {Zhang}}, \bibinfo {author} {\bibfnamefont {X.}~\bibnamefont {Dai}},\ and\
  \bibinfo {author} {\bibfnamefont {J.}~\bibnamefont {Liu}},\ }\href
  {https://doi.org/10.1103/PhysRevLett.128.026403} {\bibfield  {journal}
  {\bibinfo  {journal} {Phys. Rev. Lett.}\ }\textbf {\bibinfo {volume} {128}},\
  \bibinfo {pages} {026403} (\bibinfo {year} {2022})}\BibitemShut {NoStop}%
\bibitem [{\citenamefont {L\"auchli}\ \emph {et~al.}(2013)\citenamefont
  {L\"auchli}, \citenamefont {Liu}, \citenamefont {Bergholtz},\ and\
  \citenamefont {Moessner}}]{moessner-fci-prl13}%
  \BibitemOpen
  \bibfield  {author} {\bibinfo {author} {\bibfnamefont {A.~M.}\ \bibnamefont
  {L\"auchli}}, \bibinfo {author} {\bibfnamefont {Z.}~\bibnamefont {Liu}},
  \bibinfo {author} {\bibfnamefont {E.~J.}\ \bibnamefont {Bergholtz}},\ and\
  \bibinfo {author} {\bibfnamefont {R.}~\bibnamefont {Moessner}},\ }\href
  {https://doi.org/10.1103/PhysRevLett.111.126802} {\bibfield  {journal}
  {\bibinfo  {journal} {Phys. Rev. Lett.}\ }\textbf {\bibinfo {volume} {111}},\
  \bibinfo {pages} {126802} (\bibinfo {year} {2013})}\BibitemShut {NoStop}%
\bibitem [{\citenamefont {Elias}\ \emph
  {et~al.}(2011{\natexlab{b}})\citenamefont {Elias}, \citenamefont {Gorbachev},
  \citenamefont {Mayorov}, \citenamefont {Morozov}, \citenamefont {Zhukov},
  \citenamefont {Blake}, \citenamefont {Ponomarenko}, \citenamefont
  {Grigorieva}, \citenamefont {Novoselov}, \citenamefont {Guinea},\ and\
  \citenamefont {Geim}}]{elias-np11}%
  \BibitemOpen
  \bibfield  {author} {\bibinfo {author} {\bibfnamefont {D.~C.}\ \bibnamefont
  {Elias}}, \bibinfo {author} {\bibfnamefont {R.~V.}\ \bibnamefont
  {Gorbachev}}, \bibinfo {author} {\bibfnamefont {A.~S.}\ \bibnamefont
  {Mayorov}}, \bibinfo {author} {\bibfnamefont {S.~V.}\ \bibnamefont
  {Morozov}}, \bibinfo {author} {\bibfnamefont {A.~A.}\ \bibnamefont {Zhukov}},
  \bibinfo {author} {\bibfnamefont {P.}~\bibnamefont {Blake}}, \bibinfo
  {author} {\bibfnamefont {L.~A.}\ \bibnamefont {Ponomarenko}}, \bibinfo
  {author} {\bibfnamefont {I.~V.}\ \bibnamefont {Grigorieva}}, \bibinfo
  {author} {\bibfnamefont {K.~S.}\ \bibnamefont {Novoselov}}, \bibinfo {author}
  {\bibfnamefont {F.}~\bibnamefont {Guinea}},\ and\ \bibinfo {author}
  {\bibfnamefont {A.~K.}\ \bibnamefont {Geim}},\ }\href
  {https://doi.org/10.1038/nphys2049} {\bibfield  {journal} {\bibinfo
  {journal} {Nature Physics}\ }\textbf {\bibinfo {volume} {7}},\ \bibinfo
  {pages} {701} (\bibinfo {year} {2011}{\natexlab{b}})}\BibitemShut {NoStop}%
\bibitem [{\citenamefont {Kang}\ \emph {et~al.}(2021)\citenamefont {Kang},
  \citenamefont {Bernevig},\ and\ \citenamefont {Vafek}}]{kang-prl21}%
  \BibitemOpen
  \bibfield  {author} {\bibinfo {author} {\bibfnamefont {J.}~\bibnamefont
  {Kang}}, \bibinfo {author} {\bibfnamefont {B.~A.}\ \bibnamefont {Bernevig}},\
  and\ \bibinfo {author} {\bibfnamefont {O.}~\bibnamefont {Vafek}},\ }\href
  {https://doi.org/10.1103/PhysRevLett.127.266402} {\bibfield  {journal}
  {\bibinfo  {journal} {Phys. Rev. Lett.}\ }\textbf {\bibinfo {volume} {127}},\
  \bibinfo {pages} {266402} (\bibinfo {year} {2021})}\BibitemShut {NoStop}%
\end{thebibliography}%

\end{document}